\documentclass[%
 aip,
 amsmath,amssymb,
preprint,%
]{revtex4-1}

\usepackage{subcaption}

\usepackage{graphicx}
\usepackage{dcolumn}
\usepackage{bm}

\usepackage[utf8]{inputenc}
\usepackage[T1]{fontenc}
\usepackage{mathptmx}
\usepackage{etoolbox}

\usepackage{hyperref}
\usepackage{graphicx}
\usepackage{wrapfig}
\usepackage{caption}

\makeatletter
\def\@email#1#2{%
 \endgroup
 \patchcmd{\titleblock@produce}
  {\frontmatter@RRAPformat}
  {\frontmatter@RRAPformat{\produce@RRAP{*#1\href{mailto:#2}{#2}}}\frontmatter@RRAPformat}
  {}{}
}%
\makeatother

\begin{document}

\title[]{Lie Algebra Contractions and Interbasis Expansions on Two-Dimensional Hyperboloid IIB. Non-Subgroup Basis.}
\author{\framebox{G.S.~Pogosyan}}
 \affiliation{Yerevan State University, Yerevan, Armenia}
\author{A.~Yakhno}%
 \email{alexander.yakhno@academicos.udg.mx}
\affiliation{%
Departamento de Matematicas, CUCEI, Universidad de Guadalajara, Mexico
}%

\def\vphi{\varphi}
\def\rphi{\vphi\in[0,2\pi)}
\def\vtheta{\vartheta}
\def\SCPM{{\SCPZ\over\SCPN}}
\def\SCPZ{{\xi^2+\eta^2}}
\def\SCPN{{\xi^2\eta^2}}

\def\ralpha{\alpha\in(\i K',\i K'+2K)}
\def\rPalpha{\alpha\in(\i K',\i K'+K)}
\def\rbeta{\beta\in[0,4K')}
\def\rPbeta{\beta\in(0,K')}
\def\rmu{\mu\in(\i K',\i K'+2K)}
\def\rPmu{\mu\in(\i K',\i K'+K)}
\def\reta{\eta\in[0,4K')}
\def\rPeta{\eta\in(0,K')}

\newdimen\theight
\newcommand{\ds}{\displaystyle}
\newcommand{\be}{\begin{equation}}
\newcommand{\ee}{\end{equation}}
\newcommand{\bi}{\begin{itemize}}
\newcommand{\ei}{\end{itemize}}
\newcommand{\x}{{\ensuremath{\times}}}
\newcommand{\bb}[1]{\makebox[16pt]{{\bf#1}}}
\newtheorem{theorem}{Theorem}
\newtheorem{definition}{Definition}

\newtheorem{lemma}{Lemma}
\newtheorem{comment}{Comment}
\newtheorem{corollary}{Corollary}
\newtheorem{example}{Example}
\newtheorem{examples}{Examples}
\newcommand{\sn}{\mbox{sn}}
\newcommand{\cn}{\mbox{cn}}
\newcommand{\dn}{\mbox{dn}}
\newcommand{\ba}{\begin{array}}
\newcommand{\ea}{\end{array}}
\newcommand{\bea}{\begin{eqnarray}}
\newcommand{\eea}{\end{eqnarray}}
\newcommand{\Res}{\mbox{Res}}
\newcommand{\arcsinh}{\mbox{arsinh}}
\newcommand{\arccosh}{\mbox{arcosh}}
\newcommand{\sign}{\mbox{sign}}
\newcommand{\diag}{\mathop{\mathrm{diag}}}

\def \Column{%
             \vadjust{\setbox0=\hbox{\sevenrm\quad\quad tcol}%
             \theight=\ht0
             \advance\theight by \dp0    \advance\theight by \lineskip
             \kern -\theight \vbox to \theight{\rightline{\rlap{\box0}}%
             \vss}%
             }}%

\catcode`\@=11
\def\qed{\ifhmode\unskip\nobreak\fi\ifmmode\ifinner\else\hskip5\p@\fi\fi
 \hbox{\hskip5\p@\vrule width4\p@ height6\p@ depth1.5\p@\hskip\p@}}
\catcode`@=12 

\def\cents{\hbox{\rm\rlap/c}}
\def\miss{\hbox{\vrule height2pt width 2pt depth0pt}}

\def\vvert{\Vert}                

\def\tcol#1{{\baselineskip=6pt \vcenter{#1}} \Column}

\def\dB{\hbox{{}}}                 
\def\mB#1{\hbox{$#1$}}             
\def\nB#1{\hbox{#1}}               

\date{\today}

\begin{abstract}
The paper describes solutions of the Laplace-Beltrami equation on two-dimensional two-sheeted hyperboloid for three non-subgroup coordinate systems: semi-sircular parabolic, elliptic parabolic and hyperbolic parabolic.
The coefficients of interbasis expansions of solutions in the specified coordinate systems through some subgroup bases are calculated. A contraction procedure for all normalized eigenfunctions in three non-subgroup coordinate systems from the  hyperboloid to the Euclidean plane is realized.  
\end{abstract}

\maketitle

\tableofcontents

\newpage

\section{Introduction}
\label{sec:Introduction}

This paper is a direct continuation of our work \onlinecite{IIA}, where we studied wave functions and interbasis expansions between them in subgroup-type coordinate systems.  The subject of our study is quantum motion described by the Schr\"odinger equation 
\begin{eqnarray}
\label{HE1}
{\cal H} \Psi = - \frac{\Delta_{LB}}{2} \Psi =  {\cal E} \Psi
\end{eqnarray}
on the upper sheet of a two-dimensional two-sheeted hyperboloid $H_2^{+}: u_0^2 - u_1^2 - u_2^2 = R^2$ ($u_0 \ge 0$, $R > 0$) embedded into the pseudo-Euclidean space $E_{2,1}$ with Cartesian coordinates $u_0$, $u_1$, $u_2$. The Laplace-Beltrami (LB) operator $\Delta_{LB}$  in the curvilinear coordinates $(\xi^1, \xi^2)$ has the form
\begin{equation}
\label{ALGEBRA4}
\Delta_{LB} = \frac{1}{\sqrt{g}}\frac{\partial}{\partial \xi^i}
\sqrt{g} g^{ik}\frac{\partial}{\partial \xi^k},
\qquad
g=|\det (g_{ik})|,
\quad
g_{ik}g^{kj} = \delta_{ij}, \quad i, k, j = 1, 2,
\end{equation}
with metric $g_{ik}(\xi)$ and line element $dl^2 = g_{ik}(\xi) d \xi^i d \xi^k$.  The energy spectrum ${\cal E}$  takes the value ${\cal E} = \left(\rho^2 + \frac14\right)/2R^2 > 0$, $\rho \ge 0$\cite{IIA}. The presence of $1/4$ means that the energy cannot be zero, and therefore the particle is in motion all the time.  Instead of the energy ${\cal E}$ and Schr\"odinger  Eq. (\ref{HE1}), we will use  the LB equation everywhere below
\begin{eqnarray}
\label{HE01}
\left(\Delta_{LB} + \frac{\rho^2  + 1/4 }{R^2}\right) \Psi_\rho = 0,
\end{eqnarray}
and the quantum number $\rho$ to label the eigenfunction $\Psi_\rho$.

It is known that the LB equation on $H_2^{+}$ allows separation of variables in nine orthogonal systems of coordinates \cite{OLEV},  so one can obtain nine sets of wave functions corresponding to  each separated system.  Three of  them, namely the subgroup type systems: pseudo-spherical (S), horocyclic (HO)  and equidistant (EQ),  have already been studied in Ref. \onlinecite{IIA} (see references therein).  The next three systems:  semi-circular parabolic (SCP),  elliptic parabolic (EP) and  hyperbolic parabolic (HP) are non-subgroup ones, but also exactly solvable. In all three systems,  separation of variables leads to two equations with singular potentials: trigonometric Rosen-Morse and Pöschl-Teller, hyperbolic centrifugal and algebraic, containing both constants and $(\rho^2 + \frac14)$ as a coupling (the so-called case of “non-separation” of  separation constants).

The Lie algebra $so(2,1)$ corresponding to the isometry group $SO(2,1)$ of $H_2^+$ has the following basis:
\begin{equation}
\label{GEN}
K_1 =   - u_0 \frac{\partial}{\partial u_2} - u_2 \frac{\partial}{\partial u_0},
\qquad
K_2 =   - u_0 \frac{\partial}{\partial u_1} - u_1 \frac{\partial}{\partial u_0},
\qquad
M =  u_1 \frac{\partial}{\partial u_2} - u_2 \frac{\partial}{\partial u_1}.
\end{equation}
The Casimir operator is defined by ${\cal C}= K_1^2 + K_2^2 - M^2$ and is related to the Laplace-Beltrami operator by ${\cal C} = R^2 \Delta_{LB}$. The semi-circular parabolic coordinate system corresponds to the operator  $L^{SCP} = K_1 K_2 + K_2 K_1 + K_2 M + M K_2$, the elliptic parabolic system to $L^{EP} = (K_1 + M)^2 + \gamma K_2^2$, and the hyperbolic parabolic system to $L^{HP} = (K_1 + M)^2 - \gamma K_2^2$, $\gamma > 0$. All eigenfunctions corresponding to these  three systems can be written in closed form as classical special functions: hypergeometric,  Bessel,  MacDonald and Legendre functions.  Some partial results concerning the non-normalized  eigenfunctions of operators $L^{SCP}$, $L^{EP}$  and $L^{HP}$, as well as some examples of interbasis expansions, were presented in Refs. \onlinecite{KAL-MIL1, GROab, GROSCHE-BOOK1, SCP:2024}.

The main objective of this note is to study the eigenfunctions of the LB equation (\ref{HE01})  in three (SCP, EP and HP) non-subgroup coordinate systems, their interbasis expansions over subgroup basis, and analytical contractions of solutions onto the Euclidean plane $E_2$. In Sec. \ref{sec:solutions} we present wave functions for the coordinates listed above and describe the process of their normalization. Note that the case of the HP system is significantly more complicated than the case of SCP and EP coordinates due to the presence of a discrete spectrum. The existence of a discrete spectrum was mentioned in Ref. \onlinecite{KAL-MIL1}, but no solutions were presented. 

In Sec. \ref{sec:INTERBASIS} we relate solutions in non-subgroup coordinates with some  sub-group basis. We separately highlight the case of expansion of the SCP solution through HO  wave functions, whose coefficients are expressed in terms of exponential functions.
Also in Ref. \onlinecite{KAL-MIL1} the expansion of the HP solution through the standard spherical basis was marked as intractable. In Subsection \ref{subsection: HP-EQ} we present the interbasis expansions of HP through EQ system for both discrete and continuous cases.

\section{Non subgroup basis on $H_2^+$ hyperboloid}
\label{sec:solutions}

\subsection{Semi-circular parabolic coordinate system}
\label{sec:SCP}

The wave functions and their properties in SCP coordinates were published in Ref. \onlinecite{SCP:2024}, so we present here only the main results in a compact form.   This system looks like
\begin{eqnarray}
u_0=R{(\SCPZ)^2+4\over8\xi\eta}, \quad
u_1=R{(\SCPZ)^2-4\over8\xi\eta}, \quad
u_2=R{\eta^2-\xi^2\over2\xi\eta}, \qquad \xi,\eta > 0.
\label{sys_scp_xi_eta}
\end{eqnarray}
The LB  equation takes the following form
\begin{equation}
\label{sys_scp_LB}
\frac{\xi^2\eta^2}{\xi^2 + \eta^2}
\left(\frac{\partial^2}{\partial\xi^2} + \frac{\partial^2}
{\partial\eta^2}\right) \Psi(\xi, \eta) = - \left(\rho^2 + \frac{1}{4}\right) \Psi(\xi, \eta),
\end{equation}
and the orthonormal complete set of $\Psi_{\rho A}^{\rm SCP}(\xi,\eta)$ wave functions is defined by intervals for the separation constant $A\in\mathbb{R}\setminus\{0\}$: 
\bea
\label{00-norm_solution_SCP_PSI}
\Psi_{\rho A}^{\rm SCP}(\xi,\eta) = \left\{ 
\Psi^{(1)}_{\rho A}(\xi,\eta), 
\ {\rm if}\  A > 0;\quad  \Psi^{(2)}_{\rho A}(\xi,\eta), \  {\rm if}\  A < 0,
\right.
\eea
where for $A >0$
\begin{equation}
\label{norm_solution_SCP}
\Psi^{(1)}_{\rho A}(\xi,\eta) =  \frac{1}{\pi R}
\sqrt{\frac{\rho \tanh \frac{\pi \rho}{2}}{2}} \sqrt{\xi\eta} \left[ J_{i\rho}\left(\sqrt{|A|}\xi \right)
+  J_{-i\rho}\left(\sqrt{|A|}\xi\right)\right] 
K_{i\rho}\left(\sqrt{|A|}\eta\right),
\end{equation}
and for $A<0$
\begin{equation}
\label{1-norm_solution_SCP}
\Psi^{(2)}_{\rho A}(\xi,\eta) =  \frac{1}{\pi R}
\sqrt{\frac{\rho \tanh \frac{\pi \rho}{2}}{2}}  \sqrt{\xi\eta}\left[ J_{i\rho}\left(\sqrt{|A|}\eta \right)
+  J_{-i\rho}\left(\sqrt{|A|}\eta\right)\right]
K_{i\rho}\left(\sqrt{|A|}\xi\right).
\end{equation}

Let us write the asymptotics of wave functions  $\sqrt{\eta}K_{i\rho}\left(\sqrt{|A|}\eta\right)$ (here we use (12) 7.2.2 and (1) 7.4.2\cite{BE2})
\bea
\label{3-psi1(rho_A)}
-\frac{\pi}{\sinh \pi\rho} \, \frac{\sqrt{\eta}}{|\Gamma(1+i\rho)|}
\sin\left(\rho\ln\frac{ \sqrt{|A|} \eta}{2} + \delta_{\rho}\right),
\qquad
\eta \sim 0;
\\[2mm]
\label{4-psi1(rho_A)}
\sqrt{\frac{\pi}{2\sqrt{|A|}}}
e^{-\sqrt{|A|} \eta},
\qquad
\eta \to \infty,\qquad \delta_{\rho} = {\rm arg} \Gamma(1-i\rho).
\eea
Taking into account the asymptotic (3) 7.13.1\cite{BE2} for Bessel function
we obtain for $\sqrt{\xi} \left[ J_{i\rho}\left(\sqrt{|A|}\xi \right)\right.$  $	\left.+
J_{-i\rho}\left(\sqrt{|A|}\xi\right)\right]$ 
\bea
\label{5-psi1(rho_A)}
 \frac{2 \sqrt{\xi}}{|\Gamma(1+i\rho)|}
\cos\left(\rho \ln \frac{\sqrt{|A|}\xi}{2} + \delta_\rho\right),
\qquad
\xi \sim 0;
\\[2mm]
\label{6-psi1(rho_A)}
\sqrt{\frac{8}{\pi \sqrt{|A|}}}
\cosh\frac{\rho\pi}{2}  \cos\left(\sqrt{|A|} \xi - \frac{\pi}{4}\right),
\qquad
\xi \to \infty.
\eea

\begin{figure}[htbp]
\centering      %
            \includegraphics[width=0.4\textwidth]{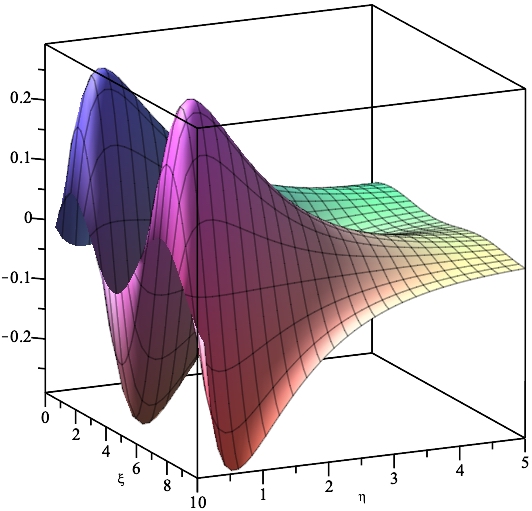} 
\caption{Graphics of the SCP wave function $\Psi^{(1)}_{\rho A}(\xi, \eta)$ for $\rho = 1$, $R = 1$ and $A = 1$.}
\label{figSCP}
\end{figure}

The functions $\Psi^{(1,2)}_{\rho A}(\xi,\eta)$ satisfy the following relations:
\begin{eqnarray}
\label{int_norm_SCP}
 R^2 \int\limits_{0}^{\infty}d\xi\int\limits_{0}^{\infty}d\eta
 \frac{\xi^2 + \eta^2}{\xi^2  \eta^2}
 \Psi^{(1,2)}_{\rho A}(\xi,\eta)
 \Psi^{(1,2) *}_{\rho' A'}(\xi,\eta) = \delta(\rho-\rho ')
 \delta(|A|-|A'|),
\end{eqnarray}
\begin{eqnarray}
\label{00-int_norm_SCP}
R^2 \int\limits_{0}^{\infty}d\rho 
\Biggl\{ \int\limits_{0}^{\infty}  \Psi^{(1)}_{\rho A}(\xi,\eta) \Psi^{(1)*}_{\rho  A}(\xi',\eta') 
d A
&+& 
\int\limits_{-\infty}^{0} \Psi^{(2)}_{\rho A}(\xi,\eta) \Psi^{(2)*}_{\rho  A}(\xi',\eta')  
d A \Biggr\}
\nonumber\\[2mm]
&=&  \frac{\xi^2 \eta^2}{\xi^2+\eta^2} 
\delta(\xi-\xi ') \delta(\eta-\eta').
\end{eqnarray}

The wave functions $\Psi^{(1,2)}_{\rho A}(\xi,\eta)$ are eigenfunctions of the symmetry
operator ${\hat A}$:
\bea
\label{A_A}
{\hat A}(\xi, \eta) \Psi^{(1)}_{\rho A}(\xi,\eta) =
|A| \Psi^{(1)}_{\rho A}(\xi,\eta),
\qquad
{\hat A}(\xi, \eta) \Psi^{(2)}_{\rho A}(\xi,\eta) =
- |A| \Psi^{(2)}_{\rho A}(\xi,\eta),
\eea
\bea
\label{OPERAT-A}
{\hat A}(\xi, \eta) = - \frac{1}{\xi^2+\eta^2}\left(
\xi^2 \frac{\partial^2}{\partial\xi^2} - \eta^2 \frac{\partial^2}
{\partial\eta^2}\right).
\eea

\subsection{Elliptic parabolic coordinate system}
\label{sec:EP}

\begin{figure}[htbp]
\begin{subfigure}[b]{.5\linewidth}
            \centering      %
            \includegraphics[width=0.7\textwidth]{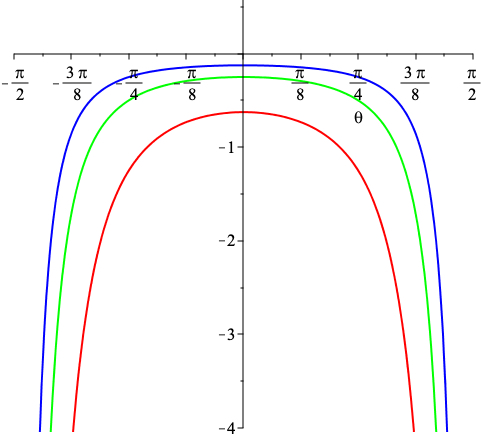} 
             \subcaption{Trigonometric Rosen-Morse potential.
             }
        \end{subfigure}%
        \begin{subfigure}[b]{.5\linewidth}
            \centering      %
           \includegraphics[width=0.8\textwidth]{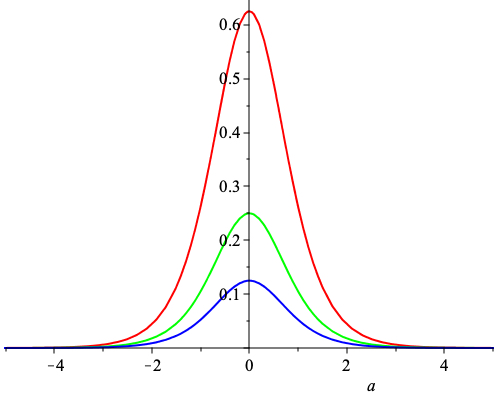}
            \subcaption{P\"oschl-Teller  potential.
            }
        \end{subfigure}%
\caption{Graphics of potentials in separated equations for the EP system for the values  $\rho = 0$ (blue), $\rho = 1/2$ (green) and $\rho = 1$ (red).}
\label{fig1}
\end{figure}

The elliptic parabolic coordinate system has the form
\begin{eqnarray}
  u_0=\frac{R}{\sqrt{\gamma}}{\cosh^2a - \sin^2\theta + \gamma \over 2\cos\theta\cosh a}, \
  u_1=\frac{R}{\sqrt{\gamma}}{\cosh^2a - \sin^2\theta - \gamma \over 2\cos\theta\cosh a}, \
  u_2=R\tan\theta\tanh a. \label{EP}
\end{eqnarray}
The transformation between  Cartesian coordinates of  the ambient space and elliptic parabolic coordinates will be single-valued if $\theta\in(-\pi/2,\pi/2)$, $a \ge 0$. For the LB equation (\ref{HE01}) we get
\be
\label{HELM-EP_hyper-01}
\frac{\cos^2\theta \cosh^2 a}{\cosh^2 a - \cos^2\theta}
\left\{\frac{\partial^2 \Psi}{\partial\theta^2} + \frac{\partial^2 \Psi}{\partial a^2} \right\}
= - \left(\rho^2 + \frac14\right) \Psi.
\ee
Note that the Laplace-Beltrami equation does not depend on the constant $\gamma$, 
so below we set $\gamma=1$.  Separation of variables $\Psi (a, \theta)  = \psi(\theta)\psi(a)$  in Eq. (\ref{HELM-EP_hyper-01}) leads to  two equations with symmetric trigonometric Rosen-Morse $V(\theta) = - (\rho^2 + 1/4)/2\cos^2\theta$ and P\"oschl-Teller  $V(a) = (\rho^2 + 1/4)/{2\cosh^2 a}$ potentials (see Fig. \ref{fig1}):
\bea
\label{cos}
\frac{d^2\psi(\theta)}{d\theta^2} + \left(-\mu^2  + \frac{\rho^2 + 1/4}{\cos^2\theta}\right)\psi (\theta)=0,
\qquad
\frac{d^2\psi(a)}{d a^2} + \left(\mu^2  - \frac{\rho^2 + 1/4}{\cosh^2 a}\right)\psi(a)=0.
\eea
 
These two equations are invariant under the inversion $\mu \to - \mu$, so we take $\mu \geq 0$. The elliptic parabolic separation constant $\mu^2$ has only continuous spectrum.  The general solution of  the  right equation in (\ref{cos}) can be presented as two real linearly independent solutions: 
\bea
\label{COSH}
\psi_{\rho \mu}^{(+)}(a)  &=&  (\cosh a)^{-i\mu} 
{_2F_1}\left(\frac{1}{4} + \frac{i\rho}{2} +
\frac{i\mu}{2}, \frac{1}{4} - \frac{i\rho}{2} + \frac{i\mu}{2};  \frac{1}{2};  \tanh^2 a \right), 
\\[2mm]
\label{00-COSH}
\psi_{\rho \mu}^{(-)}  (a)  &=& (\cosh a)^{-i\mu}  \tanh a   \  {_2F_1}\left(\frac{3}{4} + \frac{i\rho}{2} +
 \frac{i\mu}{2}, \frac{3}{4} - \frac{i\rho}{2} + \frac{i\mu}{2}; \frac{3}{2};  \tanh^2 a  \right).
\eea
The solution of  the left equation in  (\ref{cos})  for the region $\theta \in (-\pi/2,\pi/2)$ can be 
constructed  by  the replacing $a \to i \theta$ in the above functions. Thus, we consider 
\bea
\label{EP_hyper_sol}
\psi_{\rho \mu}^{(+)}(\theta)
&=&  (\cos \theta)^{-i\mu}{_2F_1}\left(\frac{1}{4} + \frac{i\rho}{2} +
\frac{i\mu}{2}, \frac{1}{4} - \frac{i\rho}{2} + \frac{i\mu}{2};  \frac{1}{2};  - \tan^2 \theta \right) 
\\[2mm]
\psi_{\rho \mu}^{(-)}(\theta) &=&  (\cos \theta)^{-i\mu} \tan\theta  \,  {_2F_1}\left(\frac{3}{4} + \frac{i\rho}{2} +
 \frac{i\mu}{2}, \frac{3}{4} - \frac{i\rho}{2} + \frac{i\mu}{2};  \frac{3}{2};  - 
 \tan^2 \theta  \right).
 \label{EP_hyper_sol-00}
\eea
From the definition of  elliptic parabolic coordinates (\ref{EP}) it is clear that the transformations $a \to - a$, $\theta \to -\theta$ leave
the points $(u_0, u_1, u_2)$ of the ambient space fixed. Therefore, one can take functions $\psi_{\rho \mu} (\theta)$ and  $\psi_{\rho \mu} (a)$ with the same parity to form two sets 
\bea
\label{EP-EQUIDIST-02_P}
\Psi_{\rho \mu}^{(\pm)} (a,\theta)
=  N_{\rho \mu}^{(\pm)} \psi_{{\rho}{\mu}}^{(\pm)}(a)\psi_{{\rho}{\mu}}^{(\pm)}(\theta).
\eea
Each of the above sets separately does not form a complete basis due to the following integral 
\begin{eqnarray}
\label{norm_int_EP-ORT}
  \int\limits_{0}^{\infty}da\int\limits_{-\pi/2}^{\pi/2}d\theta
 \frac{\cosh^2 a-\cos^2\theta}{\cosh^2 a\, \cos^2\theta}\,
 \Psi_{\rho \mu}^{(\pm)}(a,\theta)
 \Psi_{\rho '\mu'}^{(\mp)*}(a,\theta) = 0.
\end{eqnarray}
The eigenfunctions $\Psi_{\rho \mu}^{(\pm)} (a,\theta)$  satisfy the following orthonormalization and completeness conditions:
\begin{eqnarray}
\label{norm_int_EP}
  R^2 \int\limits_{0}^{\infty}da\int\limits_{-\pi/2}^{\pi/2}d\theta\,
 \frac{\cosh^2 a-\cos^2\theta}{\cosh^2 a\, \cos^2\theta}\,
 \Psi_{\rho \mu}^{(\pm)}(a,\theta)
 \Psi_{\rho '\mu'}^{(\pm)*}(a,\theta) =  \delta(\rho-\rho')
 \delta(\mu-\mu'),
\end{eqnarray}
\begin{eqnarray}
\label{complete_int_EP}
R^2 \int\limits_{0}^{\infty}d\rho\int\limits_{0}^{\infty}d\mu
\left\{
\Psi_{\rho \mu}^{(+)}(a,\theta) \Psi_{\rho \mu}^{(+)\ast}(a',\theta') 
+   
\Psi_{\rho \mu}^{(-)}(a,\theta)  \Psi_{\rho \mu}^{(-)\ast}(a',\theta') 
\right\}
\nonumber
\\[2mm]
=
\frac{\cosh^2 a\, \cos^2\theta}{\cosh^2 a-\cos^2\theta}
\delta(a-a')  \delta(\theta- \theta'),
\end{eqnarray}
where the normalization constants $N_{\rho \mu}^{(\pm)}$ are as follows
\bea
\label{EP_constant_PLUS}
N_{\rho\mu}^{(+)} 
&=& \frac{\sqrt{\rho\mu \sinh\pi\rho \sinh\pi\mu}}{2\sqrt{2}R\pi^3} 
\left|\Gamma\left(\frac14 + i\frac{\rho + \mu}{2}\right)\right|^2 \left|\Gamma\left(\frac14 + i\frac{\rho - \mu}{2}\right)\right|^2.
\\[2mm]
\label{EP_constant_MINUS}
N_{\rho\mu}^{(-)} 
&=& \sqrt{2}\frac{\sqrt{\rho\mu \sinh\pi\rho \sinh\pi\mu}}{R\pi^3} 
\left|\Gamma\left(\frac34 + i\frac{\rho + \mu}{2}\right)\right|^2 \left|\Gamma\left(\frac34 + i\frac{\rho - \mu}{2}\right)
\right|^2.
\eea
For the proof of the above relations and the asymptotic behavior of the EP solutions, see Appendix \ref{bsection: Completeness of EP}.  In Fig. \ref{fig2} one can observe the oscillations of $\Psi^{(\pm)}_{\rho \mu}(a,\theta)$ functions as $\theta \sim \pm \pi/2$ and $a \to \infty$.

\begin{figure}[htbp]
\begin{subfigure}[b]{.5\linewidth}
            \centering      %
            \includegraphics[width=0.7\textwidth]{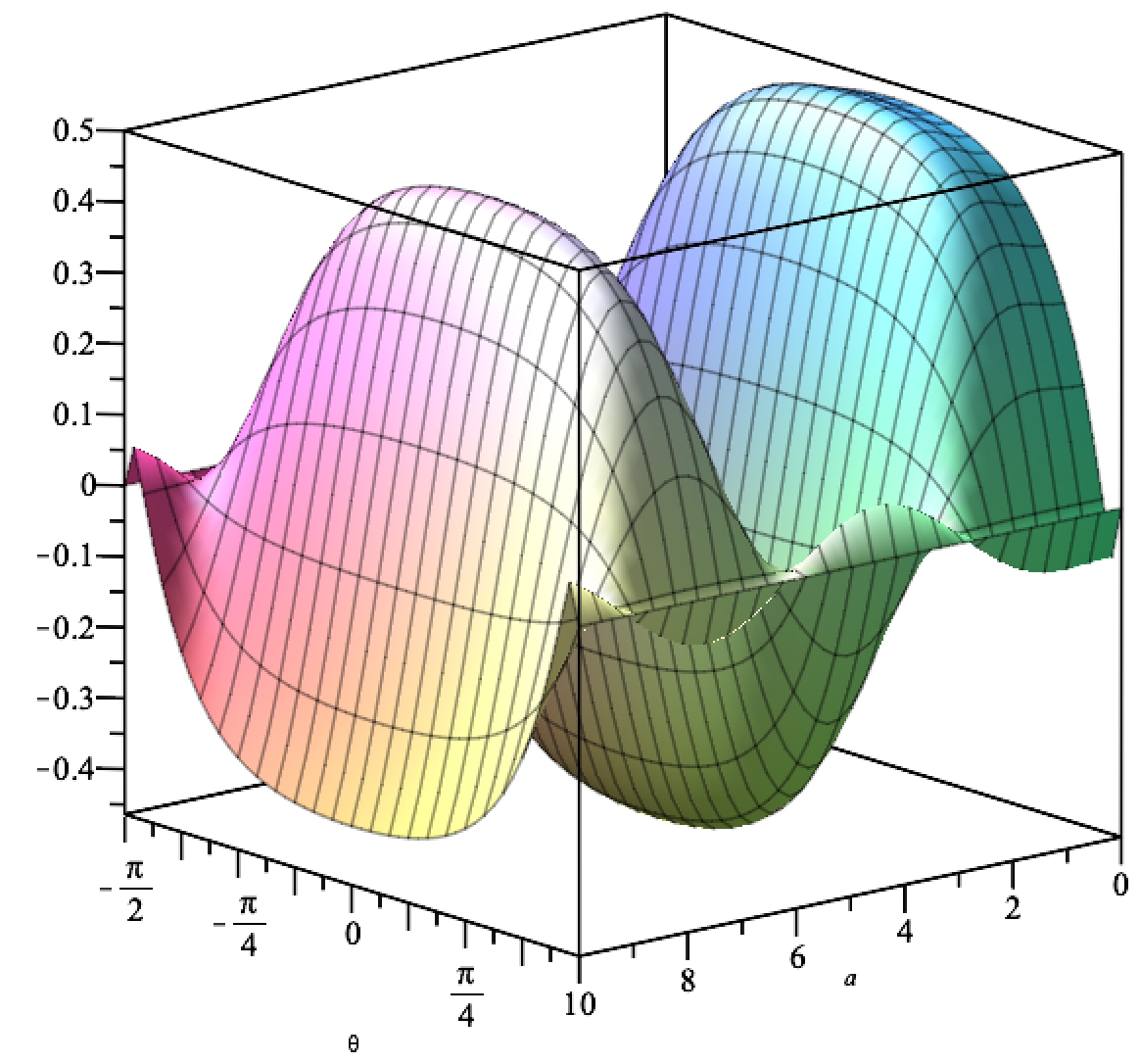} 
             \subcaption{Graphic of $\Psi^{(+)}_{\rho \mu}(a,\theta)$ function.
             }
        \end{subfigure}%
        \begin{subfigure}[b]{.5\linewidth}
            \centering      %
           \includegraphics[width=0.7\textwidth]{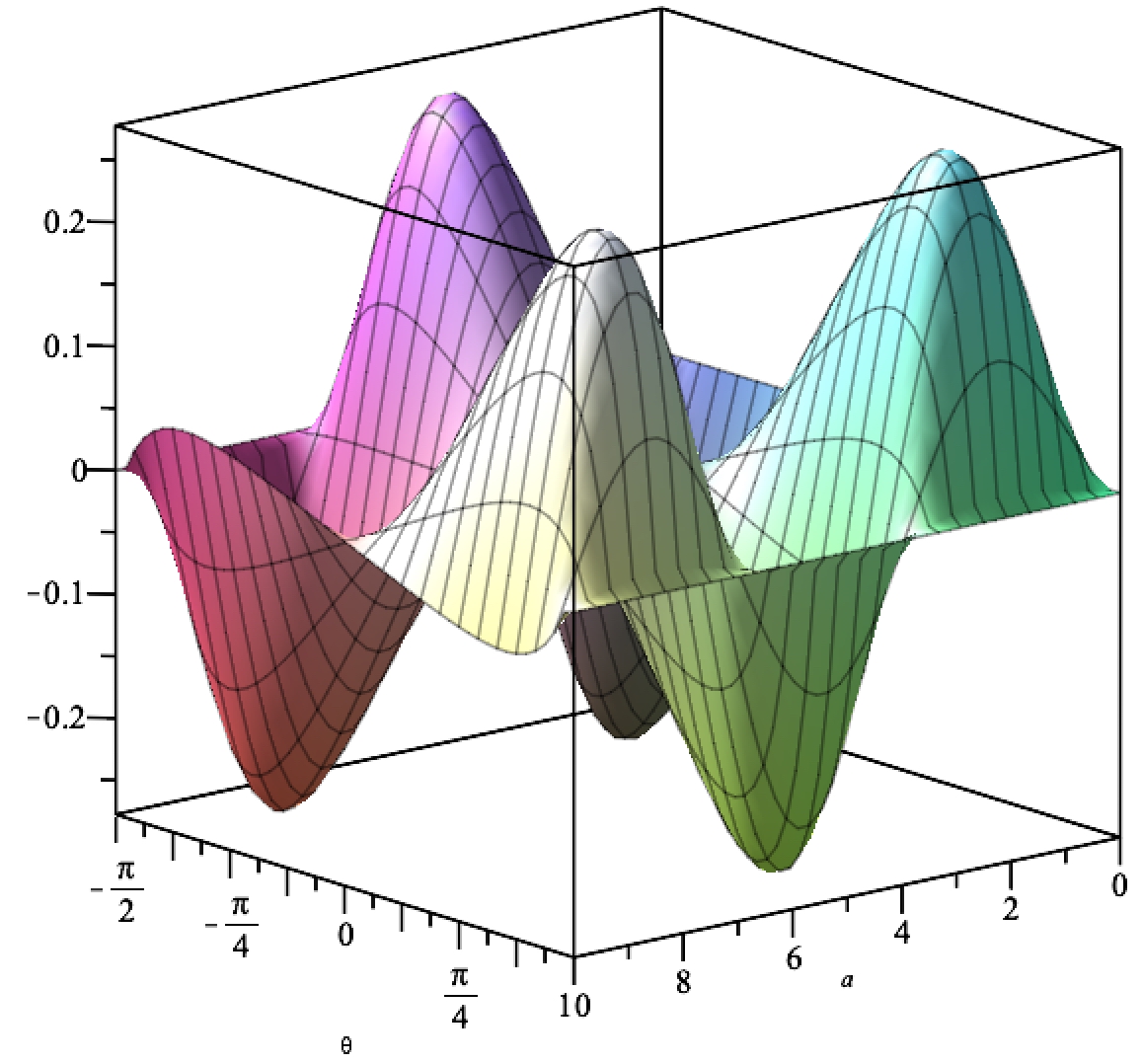}
            \subcaption{Graphic of $\Psi^{(-)}_{\rho \mu}(a,\theta)$ function.
            }
        \end{subfigure}%
\caption{Graphics of EP wave functions for $\rho = 1$, $\mu = 1$ and $R = 1$.}
\label{fig2}
\end{figure}

\subsection{Hyperbolic parabolic coordinate system}

The hyperbolic parabolic system of coordinates is defined by the formulas
\begin{eqnarray}
\label{HP_sys_2SH}
u_0=\frac{R}{\sqrt{\gamma}}{\cosh^2b - \sin^2\theta + \gamma \over 2\sin\theta\sinh b}, \
u_1=\frac{R}{\sqrt{\gamma}}{\cosh^2b - \sin^2\theta - \gamma \over 2\sin\theta\sinh b}, \
u_2=R\cot\theta\coth b,
\end{eqnarray}
with $\theta\in(0,\pi)$, $b>0$, $\gamma>0$.  The Helmholtz equation (\ref{HE01}) takes the form
\bea
\label{2SH_HP_LB}
\frac{\partial^2\Psi}{\partial\theta^2} + \frac{\partial^2\Psi}{\partial b^2} +
\left(\frac14 + \rho^2\right) \left(\frac{1}{\sin^2\theta} + \frac{1}{\sinh^2 b} \right)\Psi = 0.
\eea
Separation of variables  $\Psi (b, \theta)= N_{\rho \varsigma} \psi (b) \psi (\theta)$ yields two
one-dimensional Schr\"odinger equations:
\bea
\label{SINH1}
\frac{d^2\psi}{d b^2} + \left(- \varsigma^2 + \frac{1/4 + \rho^2}{\sinh^2 b}\right)\psi = 0,
\qquad
\frac{d^2\psi}{d\theta^2} + \left(\varsigma^2 + \frac{1/4 + \rho^2}{\sin^2\theta}\right)\psi = 0,
\eea
 which are connected to each other by changing $b\leftrightarrow i\theta$. 
The "energy"  $\sigma(\sigma+1) = \frac14 + \rho^2$
plays the role of a parameter, while the separation constant $\varsigma^2$  generates the spectrum of HP  system. Both equations in (\ref{SINH1}) belong to the class of hypergeometric equations and are therefore exactly solvable (their solution can be written in analytic form). There is a two-parameter family of solution in real form.  
The corresponding eigenvalue problem is singular, for the first equation in (\ref{SINH1}) at $b=0$, and for second one at both ends of the interval $\theta \in (0, \pi)$.
There are two spectra of separation constant:  discrete for  $\varsigma^2 > 0$ and continuous  for $\varsigma^2 < 0$.

The normalization condition takes the form
\bea
\label{Norm_HP}
R^2 \left|N_{\rho \varsigma} \right|^2 \int\limits_{0}^\infty db \int\limits_0^\pi \psi_{\rho \varsigma} (b) \psi^\ast_{\rho' \varsigma'}
(b) \psi_{\rho \varsigma} (\theta) \psi^\ast_{\rho' \varsigma'} (\theta) \left( \frac{1}{\sin^2\theta}
+ \frac{1}{\sinh^2 b}\right)  d\theta = \delta(\rho-\rho') \tilde{\delta}_{\varsigma \varsigma'},
\eea
where $\tilde{\delta}_{\varsigma \varsigma'} = \delta(\varsigma - \varsigma')$ for continuous spectrum, and  
$\tilde{\delta}_{\varsigma \varsigma'} = \delta_{f(\varsigma),f( \varsigma')}$ is the Kronecker delta for discrete values of $\varsigma$, the function $f(\varsigma)$ defines the quantization of $\varsigma$ and will be determined in Sec. \ref{par:norm_discrete_HP}.

\subsubsection{Discrete spectrum}

\paragraph{ The radial equation.}

\begin{figure}[htbp]
\begin{subfigure}[b]{.5\linewidth}
            \centering      %
            \includegraphics[width=0.7\textwidth]{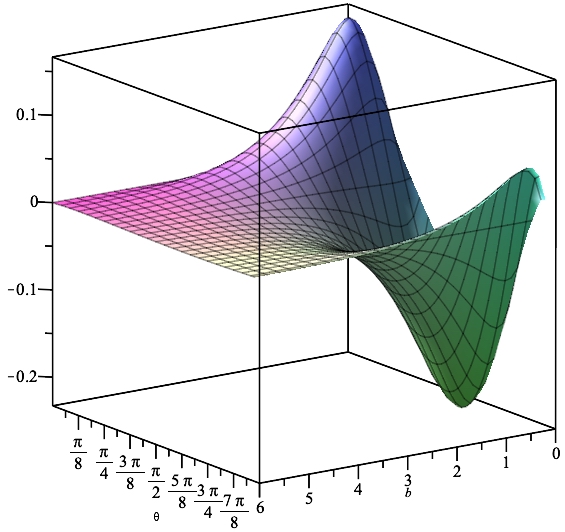} 
             \subcaption{Graphic of real part of $\Psi^{\rm HP}_{\rho \varsigma_n}(b,\theta)$ function.
             }
        \end{subfigure}%
        \begin{subfigure}[b]{.5\linewidth}
            \centering      %
           \includegraphics[width=0.7\textwidth]{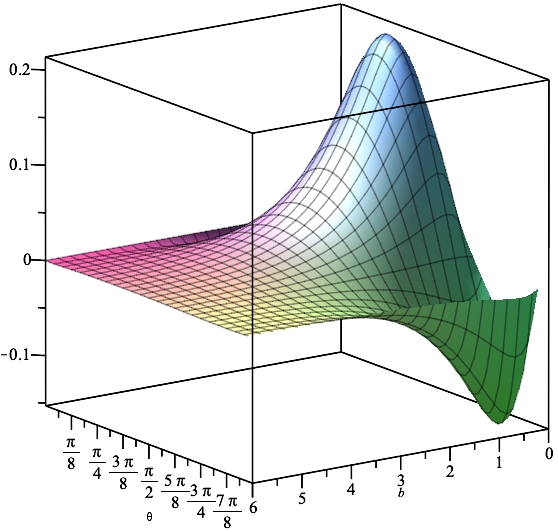}
            \subcaption{Graphic of imaginary part of $\Psi^{\rm HP}_{\rho \varsigma_n}(b,\theta)$ function.
            }
        \end{subfigure}%
\caption{Graphics of HP wave functions for $\rho = 1$, $\varsigma_n = 1$ and $R = 1$ with discrete spectrum.}
\label{figHP_d}
\end{figure}

Let us consider the first equation in (\ref{SINH1}). The following substitution
\bea
\label{SUB1}
y = (1- \cosh b)/2 = - \sinh^2\frac{b}{2},
\qquad
\psi (y) = [y(y-1)]^{1/4 + i\rho/2} w(y)
\eea
transforms the equation into a hypergeometric one
\bea
\label{HYPER-EQ1}
y(1-y) \frac{d^2 w}{d y^2} + \left(1 + i\rho - 2\left[1+ i\rho \right] y \right)
\frac{d w}{d y} - \left(\frac{1}{2} + i\rho + \varsigma \right)\left(\frac{1}{2} + i\rho -
\varsigma \right)w = 0.
\eea
The general solution of left equation in (\ref{SINH1}),  invariant under the change $\rho \to - \rho$ and regular at $b \sim 0$, can be rewritten in real form
\bea
\label{SOLUTION-HP1}
\psi_{\rho \varsigma}  (b)
&=& C (\rho, \varsigma)  \left(\frac{\sinh b}{2}\right)^{\frac12+i\rho}
{_2F_1}\left(\frac{1}{2} + i\rho + \varsigma, \frac{1}{2} + i\rho - \varsigma;  1+i\rho;
 - \sinh^2 \frac{b}{2}\right)
\nonumber \\[2mm]
&+& C(- \rho, \varsigma) \left(\frac{\sinh b}{2}\right)^{\frac12-i\rho}
{_2F_1}\left(\frac{1}{2} - i\rho + \varsigma, \frac{1}{2} - i\rho - \varsigma;  1-i\rho;
 - \sinh^2\frac{b}{2}\right).
\eea
To understand the behavior of the wave function (\ref{SOLUTION-HP1}) as $b \to \infty$
we use transformation $z \to 1/z$ (see (2) from 2.10.\cite{BE1})
\bea
\label{HYPER-1}
{_2F_1}\left(a, b; c; z \right)
=  \frac{\Gamma(c)\Gamma(b-a)}{\Gamma(b)\Gamma(c-a)} (-z)^{-a}\,
{_2F_1}\left(a, c-b; a-b+1, 1/z \right )
\nonumber \\
+
\frac{\Gamma(c)\Gamma(a-b)}{\Gamma(a)\Gamma(c-b)} 
(-z)^{-b}\, {_2F_1}\left(b, c-a; b-a+1, 1/z \right).
\eea
We get for large $b$
\bea
\label{HYPER-2}
{_2F_1} \left(\frac{1}{2} \pm i\rho + \varsigma, \frac{1}{2} \pm i\rho - \varsigma ; 
1 \pm i\rho;  -\sinh^2\frac{b}{2}\right ) \sim 
\frac{ A(\pm \rho,  \varsigma) e^{-\varsigma b} +
A(\pm \rho,  - \varsigma) e^{\varsigma b}}{2\sqrt{\pi} \left(\sinh\frac{b}{2}\right)^{1 \pm 2i\rho} },
\eea
where $A(\rho,  \varsigma) :=  \Gamma(1+i\rho)\Gamma(-\varsigma)/\Gamma(1/2+i\rho-\varsigma)$
and we use formula (15) 1.2.\cite{BE1}
\bea
\label{HYPER-5}
\Gamma(2 z) =  \Gamma(z) \Gamma(z+1/2) \frac{2^{2z-1}}{\sqrt{\pi}}.
\eea
Taking into account (\ref{HYPER-2}) we obtain for the wave functions (\ref{SOLUTION-HP1}) as $b\to \infty$
\bea
\label{HYPER-6}
\psi_{\rho \varsigma} (b)
&\sim&
\frac{1}{2\sqrt{\pi}} \left\{ \left[C(\rho, \varsigma)A(\rho, \varsigma) + C(-\rho, \varsigma)A(-\rho, \varsigma)\right]  e^{-b \varsigma} \right.
\nonumber\\
&+& \left. \left[C(\rho, \varsigma)A(\rho, -\varsigma) + C(-\rho, \varsigma)A(-\rho, - \varsigma)\right]  e^{ b \varsigma}\right\}.
\eea
Choose $\varsigma > 0$, then the square integrability condition requires that the second term in the formula (\ref{HYPER-6}) be  zero, which leads to the formula for the ratio of constants
\bea
\label{HYPER-7}
\frac{C(-\rho, \varsigma)}{C(\rho, \varsigma)} = - \frac{\Gamma(1/2 - i\rho + \varsigma)\Gamma(1 + i\rho)}
{\Gamma(1/2 + i\rho + \varsigma)\Gamma(1-i\rho)},
\eea
and therefore for the asymptotics we have
\bea
\label{HYPER-8}
\psi_{\rho \varsigma} (b) \sim
C(\rho, \varsigma)  \frac{\sqrt{\pi}  \sin(\pi \varsigma)}{\cos\pi(\varsigma - i\rho)} 
 \frac{\Gamma(-\varsigma)}{\Gamma(-i\rho)\Gamma(1/2+i\rho - \varsigma)}
 e^{- \varsigma b}.
\eea
Thus the wave function (\ref{SOLUTION-HP1}) is equal to 
\bea
\label{SOLUTION-HP2}
\psi_{\rho \varsigma}  (b)
&=& C (\rho, \varsigma) 
\left\{ \left(\frac{\sinh b}{2}\right)^{1/2 + i\rho}\,
{_2F_1}\left(\frac{1}{2} + i\rho + \varsigma, \frac{1}{2} + i\rho - \varsigma;  1+i\rho;
 - \sinh^2 \frac{b}{2}\right) \right.
\nonumber \\[2mm]
&-&
\frac{\Gamma(1/2 - i\rho + \varsigma)\Gamma(1+i\rho)}{\Gamma(1/2+i\rho+\varsigma)\Gamma(1-i\rho)}
\left(\frac{\sinh b}{2}\right)^{1/2 -i\rho}
\nonumber \\[2mm]
&\times&
\left. {_2F_1}\left(\frac{1}{2} - i\rho + \varsigma, \frac{1}{2} - i\rho - \varsigma;  1-i\rho;
 - \sinh^2\frac{b}{2}\right) \right\}.
\eea
The above formula can be also rewritten in terms of the Legendre functions of the first kind, using the relation (7) 3.2.\cite{BE1}
\bea
P_{\nu}^{\mu} (z) = \frac{2^{\mu}}{\Gamma(1-\mu)}
(z^2 - 1)^{-\frac{\mu}{2}}
\,
{_2F_1}\left(1 -\mu + \nu, -\mu - \nu;  1-\mu;  \frac{1-z}{2}\right).
\eea
We get
\bea
\label{SOLUTION-HP4}
\psi_{\rho \varsigma}  (b) =
C (\rho, \varsigma) \Gamma(1+i\rho) \sqrt{\frac{\sinh{b}}{2}} 
\left[ P_{-1/2 + \varsigma}^{- i\rho} (\cosh b) -
\frac{\Gamma(1/2-i\rho+\varsigma)}{\Gamma(1/2+i\rho+\varsigma)}
P_{-1/2 + \varsigma}^{i\rho} (\cosh b)\right].
\eea
Taking into account the relation between the Legendre functions of the second and the first kind (4) 3.3.1.\cite{BE1}
\bea
\label{SOLUTION-HP5}
Q_{\nu}^{\mu}(z) \sin\pi\mu = \frac{\pi}{2} e^{i\pi \mu} 
\left[ P_{\nu}^{\mu} (z) - \frac{\Gamma(\nu+\mu+1)}{\Gamma(\nu-\mu+1)}
P_{\nu}^{-\mu} (z)\right],
\eea
we obtain, up to a normalization constant,  the solution of the radial equation 
for $s^2 > 0$ which vanishes for large $b$ (see Fig. \ref{figHP_d}) and is described by the associated Legendre functions (toroidal functions)
\bea
\label{SOLUTION-HP6}
\psi_{\rho \varsigma}  (b) =  \frac{\sqrt{\sinh b}}{\Gamma\left(\frac12+ \varsigma -i\rho\right)} 
 Q_{-1/2 + \varsigma}^{-i\rho}(\cosh b),
\eea
where (see  (36) 3.2.\cite{BE1})
\bea
\label{00-SOLUTION-HP6}
e^{-i \mu \pi} \, Q^{\mu}_{\nu}(z)
&=&  \frac{2^{\nu}}{(z+1)^{1+\nu}}\, \left(\frac{z+1}{z-1}\right)^{\frac{\mu}{2}}
\frac{\Gamma(1+\nu+\mu)\Gamma(1+\nu)}{\Gamma(2+2\nu)}
\nonumber\\[2mm]
&\times&
{_2F_1}\left(1+\nu-\mu, 1+\nu; 2+ 2\nu; \frac{2}{1+z}\right),
\eea
 with asymptotics (21) 3.9.2.\cite{BE1}
\bea
\label{00-ASYMP}
Q^{\mu}_{\nu}(z) \sim \frac{e^{i\mu \pi} \sqrt{\pi}}{(2 z)^{1+\nu}} \, \frac{\Gamma(1+\nu+\mu)}{\Gamma\left(\nu+ \frac32\right)}, \qquad z \to \infty.
\eea
The asymptotic form of the function (\ref{SOLUTION-HP6}) for $b\sim 0$ (here we use transformation $z \to (1-z)/2$,  (32) from 3.2.\cite{BE1}) is
\bea
\label{01-SOLUTION-HP6}
\psi_{\rho \varsigma}  (b) \sim \frac{e^{\pi \rho}}{\sqrt{2}} \left\{\frac{\Gamma(i\rho)}{\Gamma(\frac12 + \varsigma + i\rho)}
\left(\sinh \frac{b}{2}\right)^{\frac12 - i\rho} +
\frac{\Gamma(-i\rho)}{\Gamma(\frac12 + \varsigma - i\rho)}
\left(\sinh \frac{b}{2}\right)^{\frac12 + i\rho}\right\}.
\eea

\paragraph{The angular equation.}
After substituting $b = i \theta$, the radial equation in (\ref{SINH1}) becomes angular.
Accordingly, the variable $z=\cosh b \in [1, \infty)$ in the function $Q_\nu^\mu(z)$ becomes 
$x = \cos\theta \in [-1, 1]$. Taking into account the formulas of the previous paragraph and replacing
$(z-1)$ by $(1-x)e^{\pm i\pi}$ and $(z^2-1)$ with $(1-x^2)e^{\pm i\pi}$, we obtain in the case of discrete spectrum the solution for angular equation in the following form
\bea
\label{SOLUTION-00-HP6}
\psi_{\rho \varsigma}  (\theta) =  \frac{\sqrt{\sin\theta}}{\Gamma(\frac12+ \varsigma -i\rho)}
{\rm Q}_{-1/2 + \varsigma}^{-i\rho}(\cos\theta),
\eea
where now we can use formula (10) 3.4.\cite{BE1}
\bea
\label{SOLUTION-01-HP6}
{\rm Q}_{\nu}^{\mu}(x) &=& \frac{\Gamma(1+\nu+\mu)\Gamma(-\mu)}{2\Gamma(1+\nu-\mu)}
\left(\frac{1-x}{1+x}\right)^{\frac{\mu}{2}}
\,
{_2F_1}\left(-\nu, \nu+1; 1+\mu; \frac{1-x}{2}\right)
\nonumber\\[2mm]
&+&
\frac{ \Gamma(\mu)}{2} \cos(\mu \pi)
\left(\frac{1+x}{1-x}\right)^{\frac{\mu}{2}}
\,
{_2F_1}\left(-\nu, \nu+1; 1-\mu; \frac{1-x}{2}\right)
\eea
to get
\bea
\label{SOLUTION-01-HP6}
&&\psi_{\rho \varsigma}  (\theta) =  \frac{ \sqrt{\sin\theta}}{2}
\Biggl\{\frac{\Gamma(i\rho)}{\Gamma\left(\frac12 + \varsigma + i\rho\right)}
\left(\frac{1-\cos\theta}{1+\cos\theta}\right)^{- \frac{i\rho}{2}}
\,
{_2F_1}\left(\frac12 - \varsigma, \frac12 + \varsigma; 1 - i\rho; \sin^2\frac{\theta}{2}\right)
\nonumber\\[2mm]
&+&
\frac{\Gamma(-i\rho)}{\Gamma\left(\frac12 + \varsigma - i\rho\right)}
\cosh(\pi \rho)
\left(\frac{1+\cos\theta}{1-\cos\theta}\right)^{- \frac{i\rho}{2}}
\,
{_2F_1}\left(\frac12 - \varsigma, \frac12 + \varsigma; 1 + i\rho;  \sin^2\frac{\theta}{2}\right)
\Biggr\}.
\eea
The right equation in (\ref{SINH1}) has two singular points: $\theta = 0$ and $\theta = \pi$.
At the point $\theta = 0$, the asymptotics of the solution are as follows
\bea
\label{SOLUTION-02-HP6}
\psi_{\rho \varsigma}  (\theta) \sim  \frac{1}{\sqrt{2}}
\Biggl\{\frac{\Gamma(i\rho)}{\Gamma\left(\frac12 + \varsigma + i\rho\right)}
\left(\sin\frac{\theta}{2}\right)^{\frac12 - i\rho}
+
\frac{\Gamma(-i\rho)\cosh(\pi \rho )}{\Gamma\left(\frac12 + \varsigma - i\rho\right)}
\left(\sin\frac{\theta}{2}\right)^{\frac12 + i\rho}
\Biggr\}.
\eea
At the point $\theta = \pi$ we have
\bea
\label{SOLUTION-03-HP6}
\psi_{\rho \varsigma}  (\theta) \sim
\frac{- 1}{\sqrt{2}}
\Biggl\{\frac{\Gamma(-i\rho)\sin\pi \varsigma}{\Gamma\left(\frac12 + \varsigma - i\rho\right)}
\left(\cos\frac{\theta}{2}\right)^{\frac12 + i\rho}
+
\frac{\Gamma(i\rho)\sin\pi(s - i\rho)}{\Gamma\left(\frac12 + \varsigma + i\rho\right)}
\left(\cos\frac{\theta}{2}\right)^{\frac12 - i\rho}\Biggl\},
\eea
where we use the analytic continuation formula  (1) 2.10.\cite{BE1}
\bea
 {_2F_1}\left(a, b; c; z \right) &=& A_1\, {_2F_1}\left(a, b; a+b-c+1; 1 - z \right) + \nonumber \\
 &+&  A_2 (1-z)^{c-a-b} {_2F_1}\left(c-a, c-b; c-a-b+1; 1 - z \right), \nonumber
\eea
with
\[
A_1 = \frac{\Gamma(c)\Gamma(c-a-b)}{\Gamma(c-a)\Gamma(c-b)},\ A_2 = \frac{\Gamma(c)\Gamma(a+b-c)}{\Gamma(a)\Gamma(b)}.
\]

The orthonormal HP solution has the form (see Appendix \ref{par:norm_discrete_HP} and Fig. \ref{figHP_d})
\bea
\Psi_{\rho \varsigma_n}^{\rm HP} (b, \theta) = 2 \frac{\sqrt{2\rho \varsigma_n \tanh\pi\rho}}{e^{\pi\rho} \pi^2 R} \frac{\Gamma\left(\frac12 + \varsigma_n +  i\rho\right)}{\Gamma\left(\frac12 + \varsigma_n -  i\rho\right)} \sqrt{\sinh b \sin\theta}Q^{-i\rho}_{-\frac12 + \varsigma_n}(\cosh b) {\rm Q}^{-i\rho}_{-\frac12 + \varsigma_n}(\cos\theta),
\label{HP_discrete_2}
\eea
where $\varsigma_n \in \left\{\varsigma_ 0 + 2n\right\}_{n = 0}^{\infty}$, $\varsigma_0 \in (0, 2]$. 

\subsubsection{Continuous spectrum}

\begin{figure}[htbp]
\begin{subfigure}[b]{.5\linewidth}
            \centering      %
            \includegraphics[width=0.7\textwidth]{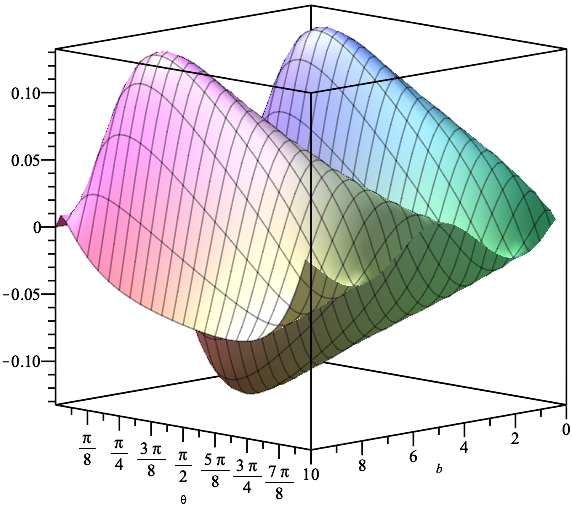} 
             \subcaption{Graphic of real part of $\Psi^{\rm HP}_{\rho \varsigma}(b,\theta)$ function.
             }
        \end{subfigure}%
        \begin{subfigure}[b]{.5\linewidth}
            \centering      %
           \includegraphics[width=0.7\textwidth]{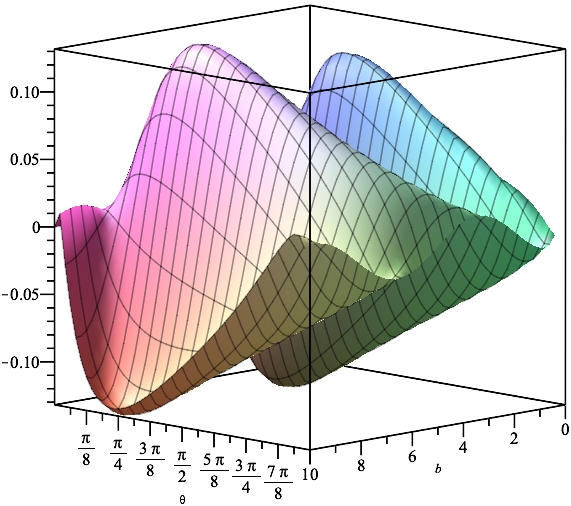}
            \subcaption{Graphic of imaginary part of $\Psi^{\rm HP}_{\rho \varsigma}(b,\theta)$ function.
            }
        \end{subfigure}%
\caption{Graphics of HP wave functions for $\rho = 1$, $\varsigma = 1$ and $R = 1$ with continuous spectrum.}
\label{figHP_c}
\end{figure}

The solution in the case of a continuous spectrum $\varsigma^2 < 0$ can be obtained from the discrete radial and angular wave functions by transforming $\varsigma \to i \varsigma$. Thus, 
\bea
\label{011-SOLUTION-HP6}
\psi_{\rho \varsigma}(b) =  \frac{\sqrt{\sinh b}}{\Gamma\left(\frac12+ i\varsigma -i\rho\right)} 
 Q_{-1/2 + i\varsigma}^{-i\rho}(\cosh b),
\quad
\psi_{\rho \varsigma}  (\theta) =  \frac{\sqrt{\sin\theta}}{\Gamma\left(\frac12 + i\varsigma - i\rho\right)} {\rm Q}_{-1/2 +i \varsigma}^{-i\rho}(\cos\theta).
\eea
The asymptotic form of the solution $\psi_{\rho \varsigma}(b)$ as $b \to \infty$ is
\bea
\label{11-SOLUTION-HP6}
\psi_{\rho \varsigma}  (b) \sim \frac{e^{\pi \rho}\sqrt{\pi}}{\sqrt{2} \Gamma(1+i\varsigma)} 
e^{- i \varsigma b},
\eea
and for $b \sim 0$ it looks as follows (see Fig. \ref{figHP_c})
\bea
\label{12-SOLUTION-HP6}
\psi_{\rho \varsigma}(b) \sim  \frac{e^{\pi \rho}}{\sqrt{2}} \left\{\frac{\Gamma(i\rho)}{\Gamma\left(\frac12+ i\rho + i\varsigma\right)} 
\left(\sinh \frac{b}{2}\right)^{\frac12 -i\rho} +
\frac{\Gamma(-i\rho)}{\Gamma\left(\frac12- i\rho + i\varsigma\right)} 
\left(\sinh \frac{b}{2}\right)^{\frac12 + i\rho}\right\}.
\eea
The asymptotic formulas (\ref{SOLUTION-02-HP6}, (\ref{SOLUTION-03-HP6}) for discrete spectrum solution are valid after the replacing $\varsigma \to i \varsigma$ for continuous spectrum solution. Therefore,
\bea
\label{SOLUTION-02-HP6_C}
\psi_{\rho \varsigma}(\theta) \sim  \frac{1}{\sqrt{2}}
\left\{\frac{\Gamma(i\rho)}{\Gamma\left(\frac12 + i\varsigma + i\rho\right)}
\left(\sin\frac{\theta}{2}\right)^{\frac12 - i\rho}
+
\frac{\Gamma(-i\rho) \cosh(\pi \rho )}{\Gamma\left(\frac12 + i\varsigma - i\rho\right)}
\left(\sin\frac{\theta}{2}\right)^{\frac12 + i\rho}
\right\}, \ \theta \sim 0;
\eea
\bea
\label{SOLUTION-03-HP6_C}
\psi_{\rho \varsigma}  (\theta) \sim
\frac{- i}{\sqrt{2}}
\left\{\frac{\Gamma(-i\rho)\sinh\pi \varsigma}{\Gamma\left(\frac12 + i\varsigma - i\rho\right)}
\left(\cos\frac{\theta}{2}\right)^{\frac12 + i\rho}
+
\frac{\Gamma(i\rho)\sinh\pi(\varsigma - \rho)}{\Gamma\left(\frac12 + i\varsigma + i\rho\right)}
\left(\cos\frac{\theta}{2}\right)^{\frac12 - i\rho}\right\},  \theta \sim \pi.
\eea

 The orthonormal HP solution in the case of a continuous spectrum has the following form (see Appendix \ref{par:norm_continuous_HP} )
\bea
\Psi_{\rho \varsigma}^{\rm HP} (b, \theta) =\frac{2}{\pi R e^{\pi\rho}} \sqrt{\frac{\rho \varsigma \sinh\pi\rho}{\sinh\pi \varsigma (\sinh^2 \pi\rho + \cosh^2 \pi \varsigma) \cosh\pi(\rho - \varsigma)}} \times \nonumber \\
\times \frac{\sqrt{\sinh b \sin\theta}}{\Gamma^2\left(\frac12 - i \rho+ i\varsigma\right)}  Q_{-\frac12 + i \varsigma}^{-i\rho}(\cosh b) 
{\rm Q}_{-\frac12 + i \varsigma}^{-i\rho}(\cos\theta).
\label{02-SOLUTION-00-HP6}
\eea

\section{Interbasis expansions}
\label{sec:INTERBASIS}

\subsection{Connection between SCP and EQ basis}
\label{sec:6.2}

Let us write the expansion of the semi-circular parabolic basis (\ref{norm_solution_SCP}), (\ref{1-norm_solution_SCP}) in terms of equidistant\cite{SCP:2024}
\begin{equation}
\label{SCP-EQUIDIST-01}
\Psi_{\rho A}^{(1,2)} (\xi,\eta) = \int\limits_{-\infty}^{\infty}
 {\cal T}_{\rho A}^{\nu(+)} 
\Psi_{\rho \nu}^{{\rm EQ}(+)}  (\tau_1,\tau_2) d\nu \pm
\int\limits_{-\infty}^{\infty}
{\cal T}_{\rho A}^{\nu(-)} 
\Psi_{\rho \nu}^{{\rm EQ}(-)}  (\tau_1,\tau_2) d\nu,
\end{equation}
with
\bea
\label{SCP-EQUIDIST-09-10}
{\cal T}_{\rho A}^{\nu(\pm)}
= \frac{(|A|/2)^{-\frac{1}{2}+i\nu}}{2^{2i\nu} \sqrt{8\pi}}
\frac{\Gamma\left(\frac{3}{4} - \frac{i\nu}{2}\right)}{\Gamma\left(\frac{3}{4} + \frac{i\nu}{2}\right)}
F^{(\pm)}_{\rho \nu},
\eea
\bea
\label{HORIC-EQUIDIST-006}
F^{(+)}_{\rho \nu} :=
\sqrt{\frac{\Gamma\left(\frac{1}{4} + i \frac{\rho - \nu}{2}\right)
\Gamma\left(\frac{1}{4} - i \frac{\rho + \nu}{2}\right)}
{\Gamma\left(\frac{1}{4} - i \frac{\rho - \nu}{2}\right)
\Gamma\left(\frac{1}{4} + i \frac{\rho + \nu}{2}\right)}},
\quad
 F^{(-)}_{\rho \nu} :=
\sqrt{\frac{\Gamma\left(\frac{3}{4} +  i \frac{\rho - \nu}{2}\right)
\Gamma\left(\frac{3}{4} - i\frac{\rho + \nu}{2}\right)}
{\Gamma\left(\frac{3}{4} - i\frac{\rho - \nu}{2}\right)
\Gamma\left(\frac{3}{4} + i\frac{\rho + \nu}{2}\right)}},
\eea
and functions $\Psi^{{\rm EQ}(\pm)}_{\rho \nu}$ are as follows
\begin{equation}
\label{0-EQUID-EQ1}
\Psi^{{\rm EQ}(\pm)}_{\rho \nu} (\tau_1,\tau_2) = N^{(\pm)}_{\rho \nu} \psi_{\rho \nu}^{(\pm)} (\tau_1) \frac{e^{i\nu\tau_2}}{\sqrt{2\pi}},
\end{equation}
where
\begin{eqnarray}
\label{H2}
\psi_{\rho\nu}^{(+)}(\tau_1) = (\cosh \tau_1)^{ i\nu}
\,
_2F_1 \left(\frac{1}{4} - i\frac{\rho-\nu}{2},
\frac{1}{4}+ i\frac{\rho + \nu}{2};\frac{1}{2}; - \sinh^2 \tau_1 \right),
\\
\label{H3}
\psi_{\rho\nu}^{(-)}(\tau_1)
= \sinh \tau_1 (\cosh \tau_1)^{i\nu}\,
_2F_1 \left(\frac{3}{4} - i\frac{\rho-\nu}{2},
\frac{3}{4} + i\frac{\rho+\nu}{2};\frac{3}{2};  -\sinh^2 \tau_1 \right),
\end{eqnarray}
and
\begin{eqnarray}
\label{EQUIDIS-NORM-01}
N_{\rho \nu}^{(+)} =
\frac{\left|\Gamma\left(\frac{1}{4}
+  i\frac{\rho+\nu}{2}\right) \Gamma\left(\frac{1}{4}+
i\frac{\rho-\nu}{2}\right)\right|}{2\sqrt{\pi^3} R
(\rho \sinh\pi \rho)^{-1/2}},
\quad
N_{\rho \nu}^{(-)} =
\frac{\left|\Gamma\left(\frac{3}{4}
+ i\frac{\rho+\nu}{2}\right)\Gamma\left(\frac{3}{4}+
i\frac{\rho-\nu}{2}\right)\right|}{\sqrt{\pi^3} R (\rho \sinh\pi \rho)^{-1/2}}.
\end{eqnarray}

From (\ref{SCP-EQUIDIST-09-10}) we have
\bea
\label{00-SCP-EQUIDIST-0-14}
{\cal T}_{\rho A}^{\nu(\pm)} {{\cal T}_{\rho A' }^{\nu(\pm) \ast}}
= \frac{e^{i\nu (\ln|A|-\ln|A'|) }}{4\pi \sqrt{|A A'|}},
\eea
and therefore
\be
\label{01-SCP-EQUIDIST-14}
\int\limits_{-\infty}^{\infty} {\cal T}_{\rho A }^{\nu(+)}  {{\cal T}_{\rho A' }^{\nu(+)*}}
d \nu
=
\int\limits_{-\infty}^{\infty} {\cal T}_{\rho A }^{\nu(-)}  {{\cal T}_{\rho A' }^{\nu (-)*}}
d \nu =
\frac{1}{2}\delta(|A| - |A'|).
\ee
Using the fact that ${\cal T}_{\rho A }^{\nu(\pm)}$ are even functions with respect to $A$ it is easy to prove the orthogonality relation
\be
\label{SCP-EQUIDIST-14_2}
\int\limits_{0}^{\infty} {\cal T}_{\rho A }^{\nu(+)}  {{\cal T}_{\rho A}^{\nu' (+)*}}
d A
=
\int\limits_{-\infty}^{0} {\cal T}_{\rho A}^{\nu(-)}  {{\cal T}_{\rho A }^{\nu' (-)*}}
d A =
\frac{1}{2}\delta(\nu - \nu'),
\ee
and the inverse expansion 
\be
\label{SCP-EQUIDIST-15}
\Psi_{\rho \nu}^{{\rm EQ}(\pm)}  (\tau_1,\tau_2)
=
\int\limits_{-\infty}^{\infty}  {{\cal T}_{\rho A' }^{\nu (\pm)*}}
\left\{\theta(A) \,{\Psi_{\rho A}^{(1)} (\xi,\eta)} \pm
\theta(-A)  {\Psi_{\rho A}^{(2)} (\xi, \eta)}\right\}
d A.
\ee
where $\theta(x)$ is a step function 
\be \theta(x) = \left\{ 0,\  {\rm if }\ x<0; 1/2,\ {\rm if }\ x=0; 1, \ {\rm if }\  x > 0. \right.
\ee

\subsection{SCP through HO basis}
\label{subsection:SCP_HO}

The interbasis expansion is of the form
\be
\label{SCP-HOR-01}
\Psi_{\rho A}^{(1,2)} (\xi, \eta) =
\int\limits_{-\infty}^{\infty} {\cal K}^{s(1,2)}_{\rho A}
\Psi_{\rho s}^{\rm HO} (\tilde{x},\tilde{y}) d s,
\ee
where functions $\Psi_{\rho s}^{\rm HO} (\tilde{x},\tilde{y})$ are indicated in Ref. \onlinecite{IIA}
\begin{equation}
\label{Psi_HO}
\Psi^{ \rm HO}_{\rho s}(\tilde{y},\tilde{x}) = N_{\rho s} \sqrt{|s| \tilde{y}} \,
K_{i\rho}(|s|\tilde{y}) \, \frac{e^{i s \tilde{x}}}{\sqrt{2\pi}},
\qquad
N_{\rho s} = \frac{1}{R \pi} \,
\sqrt{\frac{2 \rho \sinh \pi \rho}{|s|}}.
\end{equation}

The SCP and HO coordinates are related as follows
\be
\label{SCP-HOR-02}
\xi^2 = \sqrt{\tilde{x}^2+\tilde{y}^2} - \tilde{x},
\qquad
\eta^2 = \sqrt{\tilde{x}^2+ \tilde{y}^2} + \tilde{x},
\ee
which is similar to the relation between parabolic and Cartesian coordinates in two-dimensional Euclidean space\cite{ARXIV:2025}.
The explicit form of the overlap coefficients ${\cal K}^{s(1,2)}_{\rho A}$ can be determined using the orthogonality of the functions $e^{i s \nu}$, but a  direct calculation of the integral is a difficult problem. We will use an alternative way to calculate the interbasis coefficients using already known expansions, namely, the expansion of SCP wave functions in  equidistant basis with coefficients ${\cal T}^{\nu (\pm)}_{\rho A}$ (\ref{SCP-EQUIDIST-09-10}), and then the expansion of the  equidistant
basis in terms of the horocyclic basis with coefficients ${\cal W}^{\nu(\pm)}_{\rho s}$ (see II(A) \onlinecite{IIA}):
\bea
{\cal W}_{\rho s}^{\nu (+)} =
\frac{(|s|/2)^{i\nu}}{2\sqrt{\pi |s|}}  F^{(+)}_{\rho\nu},
\label{HORIC-EQUIDIST-106}
\quad
{\cal W}_{\rho s}^{\nu(-)}
=
\frac{is}{|s|} \,
\frac{(|s|/2)^{i\nu}}{2 \sqrt{ \pi |s|}}
 F^{(-)}_{\rho\nu},\quad {\cal W}^{s (\pm)}_{\rho \nu}= {\cal W}^{\nu (\pm)\ast}_{\rho s}, \quad s \ne 0,
\eea
where $ F^{(\pm)}_{\rho\nu}$ are the same as in (\ref{HORIC-EQUIDIST-006}).

It can be shown that the coefficients  
${\cal K}^{s(1,2)}_{\rho A}$ are of the form
\be
\label{SCP-HOR-07}
{\cal K}^{s(1,2)}_{\rho A} =
\int\limits_{-\infty}^{\infty} 
{\cal T}^{\nu (+)}_{\rho A }  {{\cal W}^{s(+)}_{\rho \nu}}  d \nu
 \pm 
\int\limits_{-\infty}^{\infty} 
{\cal T}^{\nu (-)}_{\rho A }  {{\cal W}^{s(-)}_{\rho \nu}}  d \nu
\ee
\be
\label{SCP-HOR-08}
=
\frac{1}{4\pi} 
\frac{1}{\sqrt{|A| |s|}} 
\int\limits_{-\infty}^{\infty} 
\left[
\frac{\Gamma(3/4-i\nu/2)}{\Gamma(3/4+i\nu/2)}
\mp
\frac{i s}{|s|} \,
\frac{\Gamma(1/4-i\nu/2)}{\Gamma(1/4+i\nu/2)}
\right]
\left(\frac{|A|}{4 |s|}\right)^{i\nu} d \nu.
\ee
The interbasis coefficients ${\cal K}^{s(1,2)}_{\rho A}$ are calculated in Appendix \ref{sec:APP_1}. The answer is simple and does not depend on the quantum number $\rho$:
\bea
{\cal K}^{s (1,2)}_{\rho A}
= \mp \frac{i}{2\sqrt{\pi}} \frac{e^{\pm i \frac{|A|}{2s} }}{s},
\qquad\qquad
{\cal K}^{s(2)*}_{\rho A} = {\cal K}^{s(1)}_{\rho A}.
\label{SCP-HOR-088}
\eea
Note that the function $\Psi^{(1)}_{\rho A}$ is defined for positive values of $A$, and the function $\Psi^{(2)}_{\rho A}$ is defined for negative values. Therefore, in the decomposition (\ref{SCP-HOR-01}) the coefficients ${\cal K}^{s(1)}_{\rho A}$ are defined for $A > 0$ and ${\cal K}^{s(2)}_{\rho A}$  for $A < 0$. Thus,
\bea
\int\limits_{0}^{\infty}{\cal K}^{s(1)}_{\rho A} {\cal K}^{s'(1)\ast}_{\rho A } dA 
+ \int\limits_{-\infty}^{0} {\cal K}^{s(2)}_{\rho A} {\cal K}^{s' (2)\ast}_{\rho A } dA = \frac{1}{4\pi s s'}\int\limits_{-\infty}^{\infty} e^{i\left(\frac{1}{2s} - \frac{1}{2s'}\right)A} dA = \delta(s - s').
\label{K_dA}
\eea
Next, for $AA' > 0$:
\bea
\int\limits_{-\infty}^{\infty} {\cal K}^{s(1,2)}_{\rho A } 
{\cal K}^{s(1,2)\ast}_{\rho A' } ds =  \frac{1}{\pi}  \int\limits_{0}^{\infty} dt\cos(A - A')t = \delta(A - A').
\label{K_ds}
\eea
Moreover, 
\be
\int\limits_{-\infty}^{\infty} {\cal K}^{s (1,2)}_{\rho A} {\cal K}^{s (2,1)\ast}_{\rho A' } ds = 0,
\ee
because in this case $AA' < 0$, so $\delta(A - A') = 0$.

The relations (\ref{K_dA}) allow us to construct the inverse expansion of the HO basis through the SCP wave functions. Multiplying (\ref{SCP-HOR-01}) for $\Psi^{(1)}_{\rho A}$ by ${\cal K}^{s'(1)\ast}_{\rho A}$ and integrating over $A > 0$, we obtain 
\be
\int\limits_{0}^{\infty} \Psi^{(1)}_{\rho A}(\xi, \eta) {\cal K}^{s' (1)\ast}_{\rho A} dA = \int\limits_{-\infty}^{\infty} ds  \Psi^{\rm HO}_{\rho s}(\tilde{x}, \tilde{y}) \int\limits_{0}^{\infty}  {\cal K}^{s(1)}_{\rho A}{\cal K}^{s' (1)\ast}_{\rho A } dA.
\ee
In the same way we get 
\be
\int\limits_{- \infty}^{0} \Psi^{(2)}_{\rho A}(\xi, \eta) {\cal K}^{s' (2)\ast}_{\rho A} dA = \int\limits_{-\infty}^{\infty} ds  \Psi^{\rm HO}_{\rho s}(\tilde{x}, \tilde{y}) \int\limits_{- \infty}^{0}  {\cal K}^{s(2)}_{\rho A}{\cal K}^{s' (2)\ast}_{\rho A} dA.
\ee
The sum of the two above relations and (\ref{K_dA}) gives the inverse expansion
\be
\Psi^{\rm  HO}_{\rho s}(\tilde{x}, \tilde{y}) = \int\limits_{0}^{\infty} \Psi^{(1)}_{\rho A}(\xi, \eta) {\cal K}^{A (1)}_{\rho s} dA  + \int\limits_{- \infty}^{0} \Psi^{(2)}_{\rho A}(\xi, \eta) {\cal K}^{A (2)}_{\rho s} dA, \quad   {\cal K}^{A (1,2)}_{\rho s} =  {\cal K}^{s (1, 2)\ast}_{\rho A}.
\ee

Let us consider some particular cases. For $\tilde{x} = 0$ we have from (\ref{SCP-HOR-02})  $\xi = \eta = \sqrt{\tilde{y}}$. Then the expression (\ref{SCP-HOR-01}), for example for $\Psi_{\rho A}^{(1)}$, takes the elegant form
\be
\int\limits_{-\infty}^{\infty} \frac{d s}{s} K_{i\rho}(\tilde{y}|s|)e^{i\frac{|A|}{2s}} =
\frac{i\pi}{\cosh \frac{\pi\rho}{2}}\left[J_{i\rho}\left(\sqrt{|A|\tilde{y}}\right) + J_{- i\rho}\left(\sqrt{|A|\tilde{y}}\right)\right] K_{i\rho}\left(\sqrt{|A|\tilde{y}}\right),
\label{VERIFIC}
\ee
instead of the known formula (see 8 from 2.16.15\cite{PRUDNIKOV2} with $\alpha = 0$, $b = |A|/2$, $\nu = i\rho$, $c = \tilde{y}$).

In the case when ${\tilde y} = 1$, introducing the parameter $\alpha \in \mathbb{R}$, ${\tilde x} =: \sinh\alpha$, one can obtain 
\bea
\int\limits_{-\infty}^{\infty} \frac{d s}{s} K_{i\rho}(|s|)
e^{i\frac{|A|}{2s} + i s \sinh\alpha} =
\frac{i\pi}{\cosh \frac{\pi\rho}{2}}
\left[J_{i\rho}\left(\sqrt{|A|e^{-\alpha}}\right) + J_{- i\rho}\left(\sqrt{|A|e^{-\alpha}}\right)\right] K_{i\rho}\left(\sqrt{|A|e^{\alpha}}\right).
\label{VERIFIC-00}
\eea

\subsection{EP basis through PS}
\label{subsection: EP-SPH}

The elliptic parabolic and pseudo-spherical bases for fixed $\rho$ are related by the unitary transformation
\be
\label{EP-SPH-01}
\Psi_{\rho \mu}^{(\pm)} (a,\theta) =
\sum\limits_{m=-\infty}^{\infty} {\cal E}^{m(\pm)}_{\rho \mu}
\Psi_{\rho m}^{\rm  S} (\tau, \varphi),
\ee
where PS wave functions are as follows\cite{IIA}
\begin{equation}
\label{sol_S}
\Psi^{\rm S}_{\rho m}(\tau,\vphi) = N_{\rho m}
P^{|m|}_{-1/2 + i\rho}(\cosh\tau) \frac{e^{i m \vphi}}{\sqrt{2\pi}},
\qquad
N_{\rho m} =  \sqrt{\frac{\rho\sinh \pi\rho}{ \pi  R^2}}
\left| \Gamma\left(\frac{1}{2}-|m| + i\rho \right)\right|.
\end{equation}
The EP coordinate system (\ref{EP}) is expressed through the spherical one as follows (considering $\cosh\tau - \cos\phi\sinh\tau > 0$ and $\gamma = 1$)
\bea
\label{EP-SPH-02A}
\cos^2 \theta = \frac{e^{-\tau}}{\cosh\tau - \sinh\tau \cos\varphi},
\qquad
\cosh^2 a =  \frac{e^\tau}{\cosh\tau - \sinh\tau \cos\varphi}.
\eea
The calculation of the interbasis coefficients is based on the method of 
asymptotics and is practically identical to the calculation of coefficients between pseudo-spherical and equidistant bases\cite{IIA}. Thus, after a long but simple calculation which we omit here, and using the symmetry property of ${_4 F_3}(1)$ polynomials
\bea
\label{EQUIDIST-SPHERIC-22}
{_4 F_3}\left(
\left.\begin{array}{cccc}
- n, \, x, \, y, \, z
\\[2mm]
u, \,   v,  \, w
\end{array}\right| 1 \right)
=
\frac{(v-z)_n (w-z)_n}{(v)_n (w)_n}
\,
{_4 F_3}\left(
\left.\begin{array}{cccc}
- n, \, u-x, \, u-y, \, z
\\[2mm]
u, \,  1-v+z-n, \, 1 - w + z - n
\end{array}\right| 1 \right),
\nonumber
\eea
we write the interbasis coefficients ${\cal E}^{m(\pm)}_{\rho \mu}$ in the form:
\bea
\label{EP-SPH-08}
{\cal E}^{m(+)}_{\rho \mu} &=&
\frac{1}{2\sqrt{\pi^3}}
\frac{\left|\Gamma\left(\frac14 + i\frac{\rho + \mu}{2}\right) 
\Gamma\left(\frac14 + i\frac{\rho - \mu}{2}\right)\right|^2
}
{\Gamma(1/2 - i\rho) |\Gamma(i\mu)|}
\sqrt{\frac{\Gamma(1/2 - i\rho-|m|)}
{\Gamma(1/2 + i\rho-|m|)}}
\nonumber
\\[2mm]
&\times&
_4 F_3\left(
\left.
\begin{array}{cccc}
-|m|, &  \frac{1}{4} - i\frac{\rho+\mu}{2},  \  \frac{1}{4} - i\frac{\rho-\mu}{2}, & |m|
\\[2mm]
\frac{1}{2}, &  \frac{1}{2} - i \rho, &   \frac{1}{2}
\end{array}\right| 1 \right).\eea
and
\bea
\label{EP-SPH-082}
{\cal E}^{m(-)}_{\rho \mu}
&=&
- i m \frac{2}{\sqrt{\pi^3}} \,
\frac{\left|\Gamma\left(\frac34 + i\frac{\rho + \mu}{2}\right)
 \Gamma\left(\frac34 + i\frac{\rho - \mu}{2}\right)\right|^2
}
{\Gamma(3/2 - i\rho) |\Gamma(i\mu)|}
\sqrt{\frac{\Gamma(1/2 - i\rho-|m|)}
{\Gamma(1/2 + i\rho-|m|)}}
\nonumber
\\[2mm]
&\times&
_4 F_3\left(
\left.
\begin{array}{cccc}
1-|m|, &  \frac{3}{4} - i\frac{\rho+\mu}{2},  \  \frac{3}{4} - i\frac{\rho-\mu}{2}, & |m| + 1
\\[2mm]
\frac{3}{2}, &  \frac{3}{2} - i \rho, &   \frac{3}{2}
\end{array}\right| 1 \right).
\eea
From the two formulas given above we have the symmetry relations: ${\cal E}^{- m(\pm)}_{\rho \mu} = \pm {\cal E}^{m(\pm)}_{\rho \mu}$. It follows that
\bea
\label{EP-SPH-083}
\sum_{m=-\infty}^{\infty} {\cal E}^{m(\pm)}_{\rho \mu}  {\cal E}^{m(\mp)*}_{\rho \mu'} = 0. 
\eea

As in the case of  decompositions of equidistant bases over  pseudo-spherical  ones\cite{IIA},  
the overlap coefficients  ${\cal E}^{m(\pm)}_{\rho \mu}$ between the EP and PS basis are expressed through the polynomials $_4F_3(1)$ of the unit argument. Moreover, (\ref{EP-SPH-08}) and (\ref{EP-SPH-082}) are Saalsch\"utz type series and can be expressed in terms of Wilson-Racah polynomials\cite{WILSON,KOEKOEK:2010}:
\bea
\label{EQUIDIST-SPHERIC-25}
W_n (t^2)
&\equiv&
W_n(t^2, \alpha, \beta, \gamma, \delta)
=
(\alpha+\beta)_n  (\alpha + \gamma)_n  (\alpha+\delta)_n
\nonumber
\\[2mm]
&\times&
{_4 F_3}\left(
\left.\begin{array}{cccc} - n,
& \alpha+\beta+\gamma+\delta +n -1,
& \alpha- it, & \alpha+it
\\[2mm]
\alpha+\beta, &   \alpha+\gamma,  &
\alpha+\delta
\end{array}\right| 1 \right), \qquad n=0,1,2,...
\eea
The polynomials  $W_n(t^2)$  are symmetric in the four parameter 
$\alpha$, $\beta$, $\gamma$, $\delta$. If  $\alpha^\ast = \beta$, $\gamma^\ast = \delta$ and $t \in\mathbb{R}$,  then $W_n(t^2)$ is real-valued\cite{Koornwinder:1985}. In the case when  parameters $\alpha$, $\beta$, $\gamma$, $\delta$ have positive real part and non-real parameters occur in conjugate pairs, the polynomials $W_n(t^2)$ are orthogonal on $\mathbb{R}^+$ with respect to the weight function 
$ w(t) := {\left|\Gamma(\alpha  + i t) \Gamma(\beta  + i t) \Gamma(\gamma  + i t) \Gamma(\delta  + i t)\right|^2}
/{|\Gamma(2i t)|^2}$ and form the complete set
\bea
\label{EQUIDIST-SPHERIC-26}
\frac{1}{2\pi}
\int\limits_{0}^{\infty}  W_n (t^2)  W_{n'} (t^2) w(t) d t &=&
n! \, (\alpha + \beta+ \gamma+ \delta +n -1)_n  \Gamma(\alpha+\beta+n) 
\Gamma(\alpha+\gamma+n)
\nonumber
\\[2mm]
&\times&
\frac{\Gamma(\alpha+\delta+n)\Gamma(\beta+\gamma+n)
\Gamma(\beta+\delta+n)
\Gamma(\gamma+\delta+n)}
{\Gamma(\alpha+\beta+\gamma+\delta+2n)}  \delta_{n n'}.
\eea
Comparing formulas (\ref{EP-SPH-08}) and (\ref{EP-SPH-082}) with the definition  
of the Wilson-Racah polynomials   (\ref{EQUIDIST-SPHERIC-25}),  we obtain for $t = \mu/2$:
\bea
\label{EP-SPH-09}
{\cal E}^{m(+)}_{\rho \mu}
&=& 
\frac{1}{2\sqrt{\pi}}
\frac{\left|\Gamma\left(\frac14 + i\frac{\rho + \mu}{2}\right) 
\Gamma\left(\frac14 + i\frac{\rho - \mu}{2}\right)\right|^2
}
{\left|\Gamma(\frac12 + i\rho+ |m|) \Gamma(i\mu)\right|}
\frac{W_{|m|} \left({\mu^2}/{4}\right)}{\Gamma^2\left(\frac12+|m|\right)},
\eea
with $\alpha=\delta = \frac{1}{4} - \frac{i\rho}{2}$, $\beta=\gamma= \frac{1}{4} + \frac{i\rho}{2}$
and 
\bea
\label{EP-SPH-092}
{\cal E}^{m(-)}_{\rho \mu}
&=&
\frac{-i m }{2\sqrt{\pi}}\,
\frac{\left|\Gamma\left(\frac34 + i\frac{\rho + \mu}{2}\right) 
\Gamma\left(\frac34 + i\frac{\rho - \mu}{2}\right)\right|^2
}
{\left|\Gamma(\frac12 + i\rho+ |m|) \Gamma(i\mu)\right|}
\frac{W_{|m|-1} \left({\mu^2}/{4}\right)}{\Gamma^2\left(\frac12+|m|\right)},
\eea
where $\alpha=\delta = \frac{3}{4} - \frac{i\rho}{2}$, $\beta=\gamma= \frac{3}{4} + \frac{i\rho}{2}$. Now one can obtain the orthogonality condition for the coefficients ${\cal E}^{m(\pm)}_{\rho \mu}$  from the orthogonality relations (\ref{EQUIDIST-SPHERIC-26})
\bea
\label{00-E_E_mu}
\int\limits_{ 0}^{\infty} {\cal E}^{m(\pm)}_{\rho \mu} {\cal E}^{m'(\pm)*}_{\rho \mu} d \mu
=  \frac12 \, [ \delta_{m, m'} \pm  \delta_{m,  - m'} ],
\eea
which allows us to write the following expansion of the pseudo-spherical basis in terms of the elliptic parabolic basis
\bea
\label{01-E_E}
\Psi_{\rho m}^{\rm S}(\tau, \vphi) = \int\limits_{0}^{\infty} {\cal E}^{\mu(+)}_{\rho m}   \Psi_{\rho \mu}^{(+)}(a, \theta) 
d \mu
+  \int\limits_{0}^{\infty}  {\cal E}^{\mu(-)}_{\rho m}  \Psi_{\rho \mu}^{(-)}(a, \theta)  d \mu, \quad {\cal E}^{\mu(\pm)}_{\rho m} := {\cal E}^{m(\pm)*}_{\rho \mu}.
\eea

Let us write down the integral representations of the coefficients ${\cal E}^{m(\pm)}_{\rho \mu} $\footnote{The connection between the EP and PS basis is presented as an overlap integral in Ref. \onlinecite{KAL-MIL1}. However, the expression obtained in that paper differs from our result for ${\cal E}^{m(\pm)}_{\rho \mu}$ due to a significant difference in the definition of the EP wave function.}:
\bea
\label{EP-SPH-093}
{\cal E}^{m(+)}_{\rho \mu} 
&=&\frac{ \Gamma\left(\frac12 + i \rho + i\mu\right)}{2^{i\rho} \sqrt{\pi} \left|\Gamma(i\mu)\right|}
\frac{\sqrt{{\Gamma\left(\frac12 - i\rho - |m|\right)}/{\Gamma\left(\frac12 + i\rho - |m|\right)}}}
{\cosh\pi\frac{\rho - \mu}{2} - i \sinh \pi\frac{\rho - \mu}{2}} 
\nonumber
\\ [2mm]
&\times&
\int\limits_{0}^{\pi} {\rm P}^{-i\mu}_{-1/2 + i\rho} (\cos\phi)  
\frac{\cos 2 m \phi}{(\sin\phi)^{\frac12  + i\rho}} d\phi,
\eea
\bea
\label{EP-SPH-094}
{\cal E}^{m(-)}_{\rho \mu} 
&=&
\frac{ \Gamma\left(\frac12 + i \rho + i\mu\right)}{2^{i\rho} \sqrt{\pi} \left|\Gamma(i\mu)\right|}
\frac{\sqrt{{\Gamma\left(\frac12 - i\rho - |m|\right)}/{\Gamma\left(\frac12 + i\rho - |m|\right)}}}
{\cosh\pi\frac{\rho - \mu}{2} - i \sinh \pi\frac{\rho - \mu}{2}}
\nonumber
\\ [2mm]
&\times&
\int\limits_{0}^{\pi} {\rm P}^{-i\mu}_{-1/2 + i\rho} (\cos\phi)  
\frac{\sin 2 m \phi}{(\sin\phi)^{\frac12  + i\rho}} d\phi.
\eea
These relations can be proven using formula (11) 2.8\cite{BE1}
\bea
\label{INT_REP-1}
\cos 2|m|\phi = \sum\limits_{p=0}^{|m|} \frac{(-|m|)_p (|m|)_p}{(1/2)_p p!} (\sin \vphi)^{2p},
\eea
and  (27) 3.12\cite{BE1}
\bea
\label{INT_REP-2}
\int\limits_{0}^{\pi} (\sin t)^{\alpha-1} {\rm P}^{-\mu}_{\nu} (\cos t) d t
=  \frac{2^{-\mu} \pi \Gamma\left(\frac{\alpha + \mu }{2}\right) \Gamma\left(\frac{\alpha - \mu}{2}\right)}
{\Gamma\left(\frac{1 + \alpha + \nu}{2}\right)
\Gamma\left(\frac{\alpha - \nu}{2}\right) \Gamma\left(\frac{\nu + \mu}{2} + 1\right)
\Gamma\left(\frac{1-\nu+\mu}{2}\right)}, \Re(\alpha\pm \mu) > 0
\eea
for the even coefficient. For the odd coefficient,  one can use (12) 2.8\cite{BE1}
\bea
\label{INT_REP-3}
\sin 2|m|\phi = 2 |m|  \sin \phi \cos \phi \sum\limits_{p=0}^{|m|} \frac{(-|m|+1)_p  (|m|+1)_p}{(3/2)_p p!} 
(\sin \phi)^{2p},
\eea
and the following relation between adjacent Legendre functions (12) 3.8\cite{BE1}
\bea
\label{INT_REP-4}
(2\nu +1) x {\rm P}^{\mu}_{\nu} (x)= (\nu-\mu+1) {\rm P}^{\mu}_{\nu+1} (x) + (\nu+\mu)  
{\rm P}^{\mu}_{\nu-1} (x).
\eea

Using integral representations  (\ref{EP-SPH-093}), (\ref{EP-SPH-094}) one can prove the completeness of the overlap coefficients ${\cal E}^{m (\pm)}_{\rho \mu}$.  We have 
\bea
\label{E_E}
\sum\limits_{m = -\infty}^{\infty} {\cal E}^{m(\pm)}_{\rho \mu} {\cal E}^{m(\pm)*}_{\rho \mu' } 
&=&
\frac{1}{\pi \left|\Gamma(i\mu)\Gamma(i\mu')\right|} 
\frac{\Gamma\left(\frac12 + i \rho + i\mu\right)\Gamma\left(\frac12 - i \rho - i\mu'\right)}
{\left(\cosh \pi \frac{\rho - \mu}{2} - i \sinh \pi \frac{\rho - \mu}{2}\right)
\left(\cosh \pi \frac{\rho - \mu'}{2} + i \sinh \pi \frac{\rho - \mu'}{2}\right)}
\nonumber
\\[2mm]
&\times&
\int\limits_{0}^{\pi}  \int\limits_{0}^{\pi}  \frac{d \phi d\phi'  A^{(\pm)}(\phi, \phi')}{(\sin\phi)^{1/2 + i\rho} (\sin\phi')^{1/2 - i\rho}} {\rm P}^{-i\mu}_{-1/2 + i\rho} (\cos\phi) \, {\rm P}^{i\mu'}_{-1/2 - i\rho} (\cos\phi'),  
\eea
where 
\bea
\label{PARAB-POLAR-10}
A^{(+)}(\phi, \phi') 
&:=& \sum\limits_{m = -\infty}^{\infty} \cos 2 m\phi \, \cos 2 m \phi'
=  \frac{\pi}{2}  \delta(\phi - \phi'),
\nonumber
\\[2mm]
A^{(-)}(\phi, \phi') 
&:=&
\sum\limits_{m = -\infty}^{\infty} \sin 2 m\phi \,  \sin 2 m \phi'
= \frac{\pi}{2} \delta(\phi - \phi'),
\nonumber
\eea
which is valid  for $- \pi \leq \vphi - \vphi'  \leq \pi$. Substituting expressions for 
$A^{(\pm)}(\phi, \phi')$ into formula (\ref{E_E}) we  obtain
\bea
\label{00-E_E}
\sum\limits_{m = -\infty}^{\infty} {\cal E}^{m(\pm)}_{\rho \mu} {\cal E}^{m(\pm)*}_{\rho \mu' } 
=
\frac{{\cal I}_{\mu \mu'}}{\left|\Gamma(i\mu)\Gamma(i\mu')\right|} 
\frac{\Gamma\left(\frac12 + i \rho + i\mu\right)\Gamma\left(\frac12 - i \rho - i\mu'\right)}
{\left(\cosh \pi \frac{\rho - \mu}{2} - i \sinh \pi \frac{\rho - \mu}{2}\right)
\left(\cosh \pi \frac{\rho - \mu'}{2} + i \sinh \pi \frac{\rho - \mu'}{2}\right)}
\eea
with
\bea
\label{01-E_E-0}
{\cal I}_{\mu \mu'} := \frac{1}{2} \int\limits_{0}^{\pi} \frac{d \phi}{\sin\phi} {\rm P}^{-i\mu}_{-1/2 + i\rho} (\cos\phi)  {\rm P}^{i\mu'}_{-1/2 - i\rho} (\cos\phi).  
\eea
Let us make the change of variable $\cos\phi = \tanh\tau$ in the above integral and use the orthogonality 
of the Legendre functions (see Ref. \onlinecite{IIA})
\bea
\label{02-E_E}
\frac{1}{2}
\int\limits_{-\infty}^{\infty} 
{\rm P}^{-i\mu}_{-1/2 + i\rho} (\tanh\tau)  {\rm P}^{i\mu'}_{-1/2 - i\rho} (\tanh\tau) d \tau  
= \frac{\sinh^2 \pi\mu + \cosh^2 \pi\rho}{\mu \sinh\pi\mu}  \delta(\mu-\mu').
\eea
We finally have   
\bea
\label{03-E_E}
\sum\limits_{m = -\infty}^{\infty} {\cal E}^{m(\pm)}_{\rho \mu} {\cal E}^{m(\pm)*}_{\rho \mu' }  = \delta(\mu-\mu'),
\eea
which allows us to obtain the decomposition (\ref{EP-SPH-01}) from (\ref{01-E_E}).

\subsection{EP through EQ basis}
\label{subsection:EP_EQ}

Let us consider the interbasis expansion of elliptic parabolic solution $\Psi_{\rho \mu}^{(\pm)} (a,\theta)$ (\ref{EP-EQUIDIST-02_P}) in terms of equidistant functions $\Psi_{\rho\nu}^{{\rm EQ}(\pm)}(\tau_1,\tau_2)$ (\ref{0-EQUID-EQ1})
\begin{equation}
\label{EP-EQUIDIST-01}
\Psi_{\rho \mu}^{(\pm)} (a,\theta) = \int\limits_{-\infty}^{\infty}
{\cal L}_{\rho \mu}^{\nu (\pm)} 
\Psi_{\rho \nu}^{{\rm EQ}(\pm)}  (\tau_1,\tau_2) d\nu.
\end{equation}
These two coordinate systems are related ($\gamma = 1$) as follows
\bea
\label{EP-EQUIDIST-05}
\cos^2 \theta
&=&
\left(\cosh\tau_2 - \sqrt{\sinh^2\tau_2 + \tanh^2\tau_1}\right)
 e^{\tau_2},
\nonumber
\\[2mm]
\cosh^2 a
&=&
\left(\cosh\tau_2 + \sqrt{\sinh^2\tau_2 + \tanh^2\tau_1}\right)
 e^{\tau_2}.
\eea
Passing from elliptic parabolic coordinates $a$, $\theta$ to equidistant coordinates $\tau_1$, $\tau_2$ in the left-hand sides of the expansions (\ref{EP-EQUIDIST-01}) by the formulas (\ref{EP-EQUIDIST-05}), passing to the limit $\tau_1 \sim 0$ in both sides, taking into account that
\bea
\label{EP-EQUIDIST-06}
\cos^2 \theta \sim e^{\tau_2 - |\tau_2|},  
\qquad
\cosh^2 a \sim e^{\tau_2 + |\tau_2|}, 
\eea
and  using the orthogonality property of $e^{i\nu\tau_2}$ in ($-\infty, \infty$),
we obtain
\be
\label{EP-EQUIDIST-07}
{\cal L}_{\rho \mu}^{\nu (+)} = \frac{N_{\rho \mu}^{(+)}}
{\sqrt{2\pi} N_{\rho \nu}^{(+)}}
\int\limits_{-\infty}^{\infty} e^{- i(\nu + \mu)\tau_2} \,
{_2F_1}\left(a_1, b_1; \frac{1}{2}; 1 - e^{- \tau_2 + |\tau_2|} \right) \, {_2F_1}\left(a_1, b_1; \frac{1}{2}; 1 - e^{- \tau_2 - |\tau_2|} \right) d \tau_2,
\ee
with $a_1 := \frac{1}{4} + i(\rho + \mu)/2$, $b_1 := \frac{1}{4} - i(\rho - \mu)/2$. For odd coefficients we have the similar integral representation
\be
\label{EP-EQUIDIST-08}
{\cal L}_{\rho \mu}^{\nu (-)} = \frac{N_{\rho \mu}^{(-)}}
{\sqrt{2\pi} N_{\rho \nu}^{(-)}}
\int\limits_{-\infty}^{\infty} e^{- i(\nu + \mu)\tau_2} \,
{_2F_1}\left(a_2, b_2; \frac{3}{2}; 1 - e^{- \tau_2 + |\tau_2|} \right) \, {_2F_1}\left(a_2, b_2; \frac{3}{2}; 1 - e^{- \tau_2 - |\tau_2|} \right) d \tau_2,
\ee
with $a_2 := \frac{3}{4} + i(\rho + \mu)/2$, $b_2 := \frac{3}{4} - i(\rho - \mu)/2$. 

Let us compute the ${\cal L}_{\rho \mu}^{\nu (+)}$ coefficients. First, we separate the integral (\ref{EP-EQUIDIST-07}) into two integrals over intervals $(0, \infty)$ and $(-\infty, 0)$, then in the second integral making the substitution $\tau_2 \to - \tau_2$, we get
\be
\label{EP-EQUIDIST-10}
{\cal L}_{\rho \mu}^{\nu (+)} = \frac{N_{\rho \mu}^{(+)}}
{\sqrt{2\pi} N_{\rho \nu}^{(+)}} \left\{ J^{(+)}_1(\rho, \mu,\nu)
+ J^{(+)}_2(\rho, \mu,\nu)\right\},
\ee
where
\bea
\label{EP-EQUIDIST-11}
J^{(+)}_{1,2}(\rho, \mu,\nu) :=
\int\limits_{0}^{\infty}  e^{ \mp i(\nu + \mu)\tau_2} \,
{_2F_1}\left(a_1, b_1;\frac{1}{2}; 1 - e^{ \mp 2\tau_2} \right)  d \tau_2.
\eea
Let us make an analytic continuation of the hypergeometric functions (1), (3) from 2.10.\cite{BE1}:
\bea
\label{TRANSFORM-1A}
&&
_2F_1 \left(a,  b;  c;  z \right)
=
\frac{\Gamma(c)\Gamma(c-a-b)}{\Gamma(c-a)\Gamma(c-b)}
\, _2F_1 \left(a,  b;  a+b-c+1; 1-z \right) + 
\\[2mm]
&+&
\frac{\Gamma(c)\Gamma(a+b-c)}{\Gamma(a)\Gamma(b)}
(1-z)^{c-a-b}
\, _2F_1 \left(c - a,  c-b;  c-a-b+1; 1-z \right), \quad z = 1 - e^{- 2\tau_2}, \nonumber
\eea
\bea
\label{TRANSFORM-1B}
&&
_2F_1 \left(a,  b;  c;  z \right)
=
\frac{\Gamma(c)\Gamma(b - a)}{\Gamma(b)\Gamma(c - a)} (1 - z)^{- a}
\, _2F_1 \left(a, c - b;  a - b + 1; (1-z)^{-1} \right) + 
\\[2mm]
&+&
\frac{\Gamma(c)\Gamma(a - b)}{\Gamma(a)\Gamma(c -b)}
(1-z)^{-b}
\, _2F_1 \left(b, c - a; b - a + 1; (1 - z)^{-1} \right), \quad z = 1 - e^{2\tau_2}. \nonumber
\eea
Then from (\ref{EP-EQUIDIST-11}) we obtain:
\bea
J^{(+)}_{1}(\rho, \mu,\nu) = A_1^{(+)} 
\int\limits_{0}^{\infty}  e^{ - i(\nu + \mu)\tau_2} \, {_2F_1}\left(a_1, b_1; 1 + i\mu; e^{ - 2\tau_2} \right)  d \tau_2 +  \nonumber \\
+ A_2^{(+)} 
\int\limits_{0}^{\infty}  e^{ - i(\nu - \mu)\tau_2} \, {_2F_1}\left(a_1^\ast, b_1^\ast; 1 - i\mu; e^{ - 2\tau_2} \right)  d \tau_2 =: A_1^{(+)} J^{(+)}_{11} +  A_2^{(+)} J^{(+)}_{12},
\eea 
\bea
A_1^{(+)} := \frac{\Gamma(1/2)\Gamma(- i\mu)}{\Gamma(a_1^\ast)\Gamma(b_1^\ast)}, \quad A_2^{(+)} := \frac{\Gamma(1/2)\Gamma(i\mu)}{\Gamma(a_1)\Gamma(b_1)},
\eea
and
\bea
J^{(+)}_{2}(\rho, \mu,\nu) = B_1^{(+)} 
\int\limits_{0}^{\infty}  e^{ -\tau_2/2 + i(\nu - \rho)\tau_2} \, {_2F_1}\left(a_1, b_1^\ast; 1 + i\rho; e^{ - 2\tau_2} \right)  d \tau_2 + \nonumber \\
+ B_2^{(+)} 
\int\limits_{0}^{\infty}  e^{- \tau_2/2 + i(\nu + \rho)\tau_2} \, {_2F_1}\left(a_1^\ast, b_1; 1 - i\rho; e^{ - 2\tau_2} \right)  d \tau_2 =: B_1^{(+)} J^{(+)}_{21} + B_2^{(+)} J^{(+)}_{22},
\eea 
with
\bea
B_1^{(+)} := \frac{\Gamma(1/2)\Gamma(- i\rho)}{\Gamma(a_1^\ast)\Gamma(b_1)}, \quad B_2^{(+)} = \frac{\Gamma(1/2)\Gamma(i\rho)}{\Gamma(a_1)\Gamma(b_1^\ast)}.
\eea
Representing the hypergeometric functions in the four integrals given above in series with indices $n_{1,2,3,4}$ respectively, we have:
\bea
J^{(+)}_{11} = \sum\limits_{n_1 = 0}^{\infty} \frac{(a_1)_{n_1} (b_1)_{n_1}}{(1 + i\mu)_{n_1} n_1!}\int\limits_{0}^{\infty}  e^{[-2n_1 - i(\nu + \mu)]\tau_2} d\tau_2 = 
\nonumber\\
= \pi\delta(\nu + \mu) - \frac{i}{\nu + \mu}\, {_3 F_2}\left(
\left.
\begin{array}{ccc}
a_1, & b_1, &   i\frac{\nu + \mu}{2}
\\[2mm]
1 + i\mu, &  1 + i\frac{\nu + \mu}{2} &
\end{array}\right| 1 \right),
\eea
\bea
J^{(+)}_{12} =  \pi\delta(\nu - \mu) - \frac{i}{\nu - \mu}\, {_3 F_2}\left(
\left.
\begin{array}{ccc}
a_1^\ast, & b_1^\ast, &   i\frac{\nu - \mu}{2}
\\[2mm]
1 - i\mu, &  1 + i\frac{\nu - \mu}{2} &
\end{array}\right| 1 \right),
\eea
\bea
J^{(+)}_{21} = \frac{1}{\frac{1}{2} + i(\rho - \nu)}\, {_3 F_2}\left(
\left.
\begin{array}{ccc}
a_1, & b_1^\ast, & \frac{1}{4} +  i\frac{\rho - \nu}{2}
\\[2mm]
1 + i\rho, &  \frac{5}{4} + i\frac{\rho - \nu}{2} &
\end{array}\right| 1 \right),
\eea
\bea
J^{(+)}_{22} = \frac{1}{\frac{1}{2} - i(\rho + \nu)}\, {_3 F_2}\left(
\left.
\begin{array}{ccc}
a_1^\ast, & b_1, & \frac{1}{4} -  i\frac{\rho + \nu}{2}
\\[2mm]
1 - i\rho, &  \frac{5}{4} - i\frac{\rho + \nu}{2} &
\end{array}\right| 1 \right),
\eea
where we use formulas\cite{MADELUNG:1957}:
\be
\label{EP-EQUIDIST-13}
\delta_{\pm} (z) = \int\limits_{0}^{\infty} e^{\pm 2\pi i z s}
d s = \frac{1}{2} \delta (z) \mp \frac{1}{2\pi i z}.
\ee
Finally, 
\bea
{\cal L}_{\rho \mu}^{\nu (+)} = \frac{N_{\rho \mu}^{(+)}}
{\sqrt{2\pi} N_{\rho \nu}^{(+)}}\left\{ A_1^{(+)}J^{(+)}_{11}  + A_2^{(+)} J^{(+)}_{12} + B_1^{(+)} J^{(+)}_{21} + B_2^{(+)} J^{(+)}_{22} \right\}.
\eea

In the same way, from (\ref{EP-EQUIDIST-08}) we get
\bea
{\cal L}_{\rho \mu}^{\nu (-)} = \frac{N_{\rho \mu}^{(-)}}
{\sqrt{2\pi} N_{\rho \nu}^{(-)}}\left\{ A_1^{(-)}J^{(-)}_{11}  + A_2^{(-)} J^{(-)}_{12} + B_1^{(-)} J^{(-)}_{21} + B_2^{(-)} J^{(-)}_{22} \right\},
\eea
where
\bea
J^{(-)}_{11} = \pi\delta(\nu + \mu) - \frac{i}{\nu + \mu}\, {_3 F_2}\left(
\left.
\begin{array}{ccc}
a_2, & b_2, &   i\frac{\nu + \mu}{2}
\\[2mm]
1 + i\mu, &  1 + i\frac{\nu + \mu}{2} &
\end{array}\right| 1 \right),
\eea
\bea
J^{(-)}_{12} =  \pi\delta(\nu - \mu) - \frac{i}{\nu - \mu}\, {_3 F_2}\left(
\left.
\begin{array}{ccc}
a_2^\ast, & b_2^\ast, &   i\frac{\nu - \mu}{2}
\\[2mm]
1 - i\mu, &  1 + i\frac{\nu - \mu}{2} &
\end{array}\right| 1 \right),
\eea
\bea
J^{(-)}_{21} = \frac{1}{\frac{3}{2} + i(\rho - \nu)}\, {_3 F_2}\left(
\left.
\begin{array}{ccc}
a_2, & b_2^\ast, & \frac{3}{4} +  i\frac{\rho - \nu}{2}
\\[2mm]
1 + i\rho, &  \frac{7}{4} + i\frac{\rho - \nu}{2} &
\end{array}\right| 1 \right),
\eea
\bea
J^{(-)}_{22} = \frac{1}{\frac{3}{2} - i(\rho + \nu)}\, {_3 F_2}\left(
\left.
\begin{array}{ccc}
a_2^\ast, & b_2, & \frac{3}{4} -  i\frac{\rho + \nu}{2}
\\[2mm]
1 - i\rho, &  \frac{7}{4} - i\frac{\rho + \nu}{2} &
\end{array}\right| 1 \right),
\eea
and
\bea
A_1^{(-)} = \frac{\Gamma(3/2)\Gamma(- i\mu)}{\Gamma(a_2^\ast)\Gamma(b_2^\ast)}, \quad A_2^{(-)} = \frac{\Gamma(3/2)\Gamma(i\mu)}{\Gamma(a_2)\Gamma(b_2)},\nonumber\\
B_1^{(-)} = \frac{\Gamma(3/2)\Gamma(- i\rho)}{\Gamma(a_2^\ast)\Gamma(b_2)}, \quad B_2^{(-)} = \frac{\Gamma(3/2)\Gamma(i\rho)}{\Gamma(a_2)\Gamma(b_2^\ast)}.
\eea

\subsection{Expansions between  HP and EQ basis}
\label{subsection: HP-EQ}

\subsubsection{Discrete spectrum}
\label{subsection: HP-EQ discrete}

The interbasis expansion for the HP wave functions in the case of a discrete spectrum is as follows
\bea
\label{HP-EQ-01}
\Psi_{\rho \varsigma_n}^{\rm HP} (b, \theta) =
\int\limits_{-\infty}^{\infty} {\cal A}^{\nu (+)}_{\rho \varsigma_n }
\Psi_{\rho \nu}^{ {\rm EQ} (+)} (\tau_1, \tau_2) d\nu  +
\int\limits_{-\infty}^{\infty} {\cal A}^{\nu (-)}_{\rho \varsigma_n}
\Psi_{\rho \nu}^{{\rm EQ} (-)} (\tau_1, \tau_2) d\nu,
\eea
where functions $\Psi_{\rho \nu}^{{\rm EQ} (\pm)}(\tau_1, \tau_2)$  are given by (\ref{0-EQUID-EQ1}) 
and the HP basis is (\ref{HP_discrete_2}). 
The equidistant coordinate system is related to the hyperbolic parabolic coordinates in this way
\bea
\label{COOR-HP-EQ-01}
\cos^2 \theta = \frac{u_0- \sqrt{u_1^2+ R^2}}{u_0-u_1}
&=&
\frac{\cosh\tau_1 \cosh\tau_2 - \sqrt{\cosh^2\tau_1 \sinh^2\tau_2 + 1}}{ e^{-\tau_2} \cosh\tau_1} ,
\nonumber
\\[3mm]
\cosh^2 b =  \frac{u_0 + \sqrt{u_1^2+ R^2}}{u_0-u_1}
&=&
\frac{\cosh\tau_1 \cosh\tau_2 + \sqrt{\cosh^2\tau_1 \sinh^2\tau_2 + 1}}{ e^{-\tau_2} \cosh\tau_1 }.
\nonumber
\eea

To start the calculation we note that the inversion $\tau_1 \to - \tau_1$ (or $u_2 \to - u_2$) 
for EQ system corresponds to the transformation $\theta \to \pi - \theta$ for HP system. For equidistant wave functions we have $\Psi_{\rho \nu}^{{\rm EQ}(\pm)} (- \tau_1, \tau_2) = \pm \Psi_{\rho \nu}^{{\rm EQ}(\pm)} (\tau_1, \tau_2)$.
Thus, making the transformations $\tau_1 \to - \tau_1$ in the both sides of
interbasis expansions (\ref{HP-EQ-01}), we can rewrite it as
\bea
\label{11-HP-EQ-01}
\Psi^{\rm HP}_{\rho \varsigma_n} (b, \pi - \theta) =
\int\limits_{-\infty}^{\infty} {\cal A}^{\nu (+)}_{\rho \varsigma_n}
\Psi_{\rho \nu}^{{\rm EQ} (+)} (\tau_1, \tau_2) d\nu  -
\int\limits_{-\infty}^{\infty} {\cal A}^{\nu (-)}_{\rho \varsigma_n}
\Psi_{\rho \nu}^{{\rm EQ} (-)} (\tau_1, \tau_2) d\nu.
\eea
Comparing formulas (\ref{HP-EQ-01}) and (\ref{11-HP-EQ-01}
and using orthogonality of functions $e^{i\nu\tau_2}$, we obtain from
(\ref{HP-EQ-01}) and (\ref{11-HP-EQ-01}) 
\bea
\label{ARHO}
{\cal A}_{\rho \varsigma_n }^{\nu (\pm)} \psi_{\rho \nu}^{(\pm)}(\tau_1)
&=&
\frac{N_{\rho \varsigma_n}^{\rm d}}{2\sqrt{2\pi}\, N_{\rho\nu}^{(\pm)}}
\int\limits_{-\infty}^{\infty}
\sqrt{\sinh b \sin\theta}
Q_{-1/2 +  \varsigma_n}^{-i\rho}(\cosh b)
\nonumber
\\[2mm]
&\times&
\left\{{\rm Q}_{-1/2 + \varsigma_n}^{-i\rho}(\cos\theta) \pm {\rm Q}_{-1/2 +  \varsigma_n}^{-i\rho}(-\cos\theta)\right\}
e^{-i\nu\tau_2} d\tau_2,
\eea
where $N_{\rho \varsigma_n}^{\rm d}$ is given by (\ref{N_rho_s_D}). To calculate the integrals in (\ref{ARHO}) we take the following steps. First, the formula (12) 3.4. \cite{BE1} for
${\rm Q}_{\nu}^{\mu}(x)$ 
\bea
\label{100-ARHO}
\frac{(1-x^2)^{\frac{\mu}{2}}}{2^{\mu} \pi^{3/2}}
{\rm Q}_{\nu}^{\mu}(x)
&=&
x \cot\left( \pi\frac{\nu+\mu}{2}\right) \frac{\
{_2F_1}(\frac12 -\frac{\nu}{2}-\frac{\mu}{2},  1+\frac{\nu}{2}-\frac{\mu}{2} ;
\frac32; x^2)}
{\Gamma\left(\frac12 +\frac{\nu}{2}-\frac{\mu}{2}\right) \Gamma\left(- \frac{\nu}{2}-\frac{\mu}{2}\right)}
\nonumber
\\[2mm]
&-&
\tan\left(\pi \frac{\nu+\mu}{2}\right) \frac{
\,
{_2F_1}(- \frac{\nu}{2}-\frac{\mu}{2},  \frac12+\frac{\nu}{2}-\frac{\mu}{2}; 
\frac12;  x^2)}
{2 \Gamma\left(1 +\frac{\nu}{2}-\frac{\mu}{2}\right)\Gamma\left(\frac12-\frac{\nu}{2}-\frac{\mu}{2}\right)}
\eea
gives us 
\bea
\label{101-ARHO}
{\rm Q}_{-1/2 + \varsigma_n}^{-i\rho}(\cos\theta)
&+&
{\rm Q}_{-1/2 +  \varsigma_n}^{-i\rho}(-\cos\theta)
=
\frac{\pi^{\frac32}}{2^{i\rho}}
\frac{\tan\left(\frac{\pi}{4} - \frac{\varsigma_n - i\rho}{2}\pi\right)}
{\Gamma\left(\frac34 + \frac{\varsigma_n + i\rho}{2}\right)\Gamma\left(\frac34 - \frac{\varsigma_n - i\rho}{2}\right)}
\nonumber
\\[2mm]
&\times&
(\sin\theta)^{i\rho} \, {_2F_1}\left(\frac{1}{4} - \frac{\varsigma_n - i\rho}{2},  \frac14 + \frac{\varsigma_n + i\rho}{2}; 
\frac12;  \cos^2\theta \right),
\\[2mm]
\label{102-ARHO}
{\rm Q}_{-1/2 + \varsigma_n}^{-i\rho}(\cos\theta)
&-&
{\rm Q}_{-1/2 +  \varsigma_n}^{-i\rho}(-\cos\theta)
=
-  \frac{\pi^{\frac32}}{2^{-1 + i\rho}}
\frac{\cot\left(\frac{\pi}{4} - \frac{\varsigma_n - i\rho}{2}\pi \right)}
{\Gamma\left(\frac14 + \frac{\varsigma_n + i\rho}{2}\right)\Gamma\left(\frac14 - \frac{\varsigma_n - i\rho}{2}\right)}
\nonumber
\\[2mm]
&\times&
\cos\theta  (\sin\theta)^{i\rho} \, 
{_2F_1}\left(\frac{3}{4} - \frac{\varsigma_n - i\rho}{2},  \frac34 + \frac{\varsigma_n + i\rho}{2}; 
\frac32;  \cos^2\theta \right).
\eea

Now we can pass to the limit $\tau_1 \sim 0$ on both sides of relations
(\ref{ARHO}). Having previously noted that from formula (\ref{COOR-HP-EQ-01}) one can obtain
\bea
\label{103-ARHO}
\cos^2 \theta \sim \tau_1^2 \frac{e^{\tau_2}}{2\cosh\tau_2},
\qquad
\cosh^2 b \sim 1+e^{2\tau_2} = 2 e^{\tau_2}\cosh\tau_2,
\eea
we get
\bea
\label{104-ARHO}
{\cal A}_{\rho \varsigma_n }^{\nu (+)}
=
\frac{N_{\rho \varsigma_n}^{\rm d}}{N_{\rho \nu}^{(+)}}
\frac{{\pi}}{2^{\frac32 + i\rho}}
\frac{\tan\left(\frac{\pi}{4} - \frac{\varsigma_n - i\rho}{2}\pi\right)}
{\Gamma\left(\frac34 + \frac{\varsigma_n + i\rho}{2}\right)\Gamma\left(\frac34 - \frac{\varsigma_n - i\rho}{2}\right)}
\int\limits_{-\infty}^{\infty}
Q_{- \frac12 +  \varsigma_n}^{-i\rho}\left(\sqrt{1 + e^{2\tau_2}}\right)
e^{\left(\frac12 - i\nu\right)\tau_2}  d\tau_2,
\\[2mm]
{\cal A}_{\rho \varsigma_n }^{\nu (-)}
=
\frac{-N_{\rho \varsigma_n}^{\rm d}}{N_{\rho \nu}^{(-)}}
\frac{{\pi}}{2^{1 + i\rho}}
\frac{\cot\left(\frac{\pi}{4} - \frac{\varsigma_n - i\rho}{2}\pi\right)}
{\Gamma\left(\frac14 + \frac{\varsigma_n + i\rho}{2}\right)\Gamma\left(\frac14 - \frac{\varsigma_n - i\rho}{2}\right)}
\int\limits_{-\infty}^{\infty}
Q_{- \frac12 +  \varsigma_n}^{-i\rho}\left(\sqrt{1+e^{2\tau_2}}\right)
\frac{e^{(1 - i\nu)\tau_2}}{\sqrt{\cosh\tau_2}} d\tau_2.  \label{105-ARHO}
\eea
Substituting $\sinh t = e^{\tau_2}$ into Eqs. (\ref{104-ARHO}), (\ref{105-ARHO}) and using formula (2) 3.8.\cite{BE1}
\bea
\label{106-ARHO}
(2\nu+1) z Q_{\nu}^{\mu} (z) = (\nu-\mu+1) Q_{\nu+1}^{\mu} (z)
+ (\nu+\mu) Q_{\nu-1}^{\mu} (z),
\eea
we transform the corresponding integrals as follows
\bea
\label{107-ARHO}
{\cal A}_{\rho \varsigma_n }^{\nu (+)}
&=&
\frac{N_{\rho \varsigma_n}^{\rm d}}{N_{\rho \nu}^{(+)}}
\frac{{\pi}}{2^{\frac32 + i\rho}}
\frac{\tan\left(\frac{\pi}{4} - \frac{\varsigma_n - i\rho}{2}\pi\right)}
{\Gamma\left(\frac34 + \frac{\varsigma_n + i\rho}{2}\right)\Gamma\left(\frac34 - \frac{\varsigma_n - i\rho}{2}\right)}
\nonumber
\\[2mm]
&\times&
\int\limits_{0}^{\infty}
\left[
\frac{\frac12 + \varsigma_n + i\rho}{2\varsigma_n}
Q_{\frac12 +  \varsigma_n}^{-i\rho}(\cosh t)
-
\frac{\frac12 - \varsigma_n + i\rho}{2\varsigma_n}
Q_{-\frac32 +  \varsigma_n}^{-i\rho}(\cosh t)
\right]
\frac{d t}{(\sinh t)^{\frac12 + i\nu}},
\nonumber
\\[2mm]
\label{108-ARHO_00}
{\cal A}_{\rho \varsigma_n }^{\nu (-)}
&=&
- \frac{N_{\rho \varsigma_n}^{\rm d}}{N_{\rho \nu}^{(-)}}
\frac{{\pi}}{2^{\frac12 + i\rho}}
\frac{\cot\left(\frac{\pi}{4} - \frac{\varsigma_n - i\rho}{2}\pi\right)}
{\Gamma\left(\frac14 + \frac{\varsigma_n + i\rho}{2}\right)\Gamma\left(\frac14 - \frac{\varsigma_n - i\rho}{2}\right)}
\int\limits_{0}^{\infty}
Q_{- \frac12 +  \varsigma_n}^{-i\rho}(\cosh t)
(\sinh t)^{\frac12 - i\nu}  d t.
\eea
The integrals in the above expressions are well known and can be calculated using formula (29) 3.12.\cite{BE1}
\bea
\label{108-ARHO}
\int\limits_{0}^{\infty}
Q_{\nu}^{\mu}(\cosh t)
(\sinh t)^{\alpha-1}  d t
&=&
\frac{e^{i\mu \pi}}{2^{\alpha-\mu}}
\Gamma\left(\frac{\alpha}{2}- \frac{\mu}{2}\right)
\Gamma\left(\frac{\alpha}{2}+ \frac{\mu}{2}\right)
\frac{\Gamma\left(\frac12+ \frac{\nu}{2}+ \frac{\mu}{2}\right)
\Gamma\left(1+ \frac{\nu}{2}- \frac{\alpha}{2}\right)}
{\Gamma\left(1+ \frac{\nu}{2}- \frac{\mu}{2}\right)
\Gamma\left(\frac12+ \frac{\nu}{2}+ \frac{\alpha}{2}\right)},
\nonumber
\\[2mm]
&&  \Re(\alpha\pm \mu)>0,
\qquad
\Re(\nu-\alpha+2)>0.
\eea
We get 
\bea
\label{109-ARHO}
{\cal A}_{\rho \varsigma_n }^{\nu (+)}
=
2^{\frac12 + i\nu}\sqrt{\frac{\varsigma_n}{\pi \cosh\pi\rho}} \frac{\Gamma\left(\frac{\varsigma_n + i\nu}{2}\right) }{\Gamma\left(1 + \frac{\varsigma_n - i\nu}{2}\right)}
\sin\left(\frac{\pi}{4} - \frac{\varsigma_n - i\rho}{2}\pi\right)
F^{(+)}_{\rho \nu},
\label{110-ARHO}
\\
{\cal A}_{\rho \varsigma_n }^{\nu (-)}
=
-2^{- \frac12 + i\nu}\sqrt{\frac{\varsigma_n}{\pi \cosh\pi\rho}} \frac{\Gamma\left(\frac{\varsigma_n + i\nu}{2}\right) }{\Gamma\left(1 + \frac{\varsigma_n - i\nu}{2}\right)}
\cos\left(\frac{\pi}{4} - \frac{\varsigma_n - i\rho}{2}\pi\right)
F^{(-)}_{\rho \nu},
\label{110-ARHO_m}
\eea
where $F^{(\pm)}_{\rho \nu}$ as in (\ref{HORIC-EQUIDIST-006}).


\subsubsection{Continuous spectrum}
\label{subsection: HP-EQ con}

The interbasis expansion of the HP wave function (\ref{02-SOLUTION-00-HP6}) is of  the form
\bea
\label{HPc-EQ-01}
\Psi_{\rho \varsigma}^{\rm HP} (b, \theta) =
\int\limits_{-\infty}^{\infty} {\cal B}^{\nu (+)}_{\rho \varsigma }
\Psi_{\rho \nu}^{{\rm EQ} (+)} (\tau_1, \tau_2) d\nu  +
\int\limits_{-\infty}^{\infty} {\cal B}^{\nu (-)}_{\rho \varsigma }
\Psi_{\rho \nu}^{{\rm EQ} (-)} (\tau_1, \tau_2) d\nu.
\eea
Repeating analysis as in the case of discrete spectrum, described in the above subsection, with the change $\varsigma_n \to i\varsigma$ and $N_{\rho \varsigma_n}^{\rm d} \to N_{\rho \varsigma}^{\rm c} $, we obtain
\bea
\label{107-BRHO}
{\cal B}_{\rho \varsigma }^{\nu (+)}
&=&
\frac{N_{\rho \varsigma}^{\rm c}}{N_{\rho \nu}^{(+)}}
\frac{{\pi}}{2^{\frac32 + i\rho}}
\frac{\tan\left(\frac{\pi}{4} - \frac{i \varsigma - i\rho}{2}\pi\right)}
{\Gamma\left(\frac34 + \frac{i \varsigma + i\rho}{2}\right)\Gamma\left(\frac34 - \frac{i \varsigma - i\rho}{2}\right)}
\nonumber
\\[2mm]
&\times&
\int\limits_{0}^{\infty}
\left[
\frac{\frac12 + i\varsigma + i\rho}{2i\varsigma}
Q_{\frac12 +  i\varsigma}^{-i\rho}(\cosh t)
-
\frac{\frac12 - i\varsigma + i\rho}{2i\varsigma}
Q_{-\frac32 +  i\varsigma}^{-i\rho}(\cosh t)
\right]
\frac{d t}{(\sinh t)^{\frac12 + i\nu}},
\nonumber
\\[2mm]
\label{108-BRHO}
{\cal B}_{\rho \varsigma }^{\nu (-)}
&=&
- \frac{N_{\rho \varsigma}^{\rm c}}{N_{\rho \nu}^{(-)}}
\frac{{\pi}}{2^{\frac12 + i\rho}}
\frac{\cot\left(\frac{\pi}{4} - \frac{i\varsigma - i\rho}{2}\pi\right)}
{\Gamma\left(\frac14 + \frac{i\varsigma + i\rho}{2}\right)\Gamma\left(\frac14 - \frac{i\varsigma - i\rho}{2}\right)}
\int\limits_{0}^{\infty}
Q_{-\frac12 +  i\varsigma}^{-i\rho}(\cosh t)
(\sinh t)^{\frac12 - i\nu}  d t.
\eea
The first restriction in (\ref{108-ARHO}) is satisfied for all integrals in (\ref{108-BRHO}), but the second one is not valid for two integrals:
\be
\label{FI}
\int\limits_{0}^{\infty} Q_{-3/2 +  i\varsigma}^{-i\rho}(\cosh t) (\sinh t)^{- \frac12 - i\nu} dt
\ee
with $\alpha = 1/2 - i\nu$, and for
\be
\label{SI}
\int\limits_{0}^{\infty} Q_{-1/2 +  i\varsigma}^{-i\rho}(\cosh t) (\sinh t)^{\frac12 - i\nu} dt,
\ee
with $\alpha = 3/2 - i\nu$.

Let us consider $S := \varsigma + i\epsilon$, $\epsilon > 0$. Then for integral (\ref{FI}) $\nu = -3/2 + \epsilon + iS$ and $\Re(\nu - \alpha + 2) = \epsilon > 0$. For integral (\ref{SI}) we have $\nu = -1/2 + \epsilon + iS$ and once again $\Re(\nu - \alpha + 2) = \epsilon > 0$. Thus, we obtain
\bea
\int\limits_{0}^{\infty} Q_{-\frac32 + \epsilon +  iS}^{-i\rho}(\cosh t) (\sinh t)^{- \frac12 - i\nu} dt = \frac{e^{\pi\rho}}{2^{\frac12 - i\nu + i\rho}} \Gamma\left(\frac14 + i \frac{\rho - \nu}{2} \right) \Gamma\left(\frac14 - i \frac{\rho + \nu}{2} \right) \times \nonumber \\
\times \frac{\Gamma\left(- \frac14 - i\frac{\rho - S}{2} + \frac{\epsilon}{2}\right)}{\Gamma\left(\frac14 + i\frac{\rho + S}{2} + \frac{\epsilon}{2}\right)} \frac{\Gamma\left(i\frac{S + \nu}{2} + \frac{\epsilon}{2}\right)}{\Gamma\left(i\frac{S - \nu}{2} + \frac{\epsilon}{2}\right)},
\eea
\bea
\int\limits_{0}^{\infty} Q_{-\frac12 + \epsilon +  iS}^{-i\rho}(\cosh t) (\sinh t)^{\frac12 - i\nu} dt = \frac{e^{\pi\rho}}{2^{\frac32 - i\nu + i\rho}} \Gamma\left(\frac34 + i \frac{\rho - \nu}{2} \right) \Gamma\left(\frac34 - i \frac{\rho + \nu}{2} \right) \times \nonumber \\
\times \frac{\Gamma\left(\frac14 - i\frac{\rho - S}{2} + \frac{\epsilon}{2}\right)}{\Gamma\left(\frac34 + i\frac{\rho + S}{2} + \frac{\epsilon}{2}\right)} \frac{\Gamma\left(i\frac{S + \nu}{2} + \frac{\epsilon}{2}\right)}{\Gamma\left(1 + i\frac{S - \nu}{2} + \frac{\epsilon}{2}\right)}.
\eea
Since there is no singularity on the right side of the two above expressions in the limit $\epsilon\sim 0$, we can conclude that the coefficients ${\cal B}^{\nu (\pm)}_{\rho \varsigma}$ can be obtained from (\ref{108-ARHO_00}) by replacing $\varsigma_n \to i\varsigma$, $N_{\rho \varsigma_n}^{\rm d} \to N_{\rho \varsigma}^{\rm c}$. Finally, 
\bea
{\cal B}_{\rho \varsigma }^{\nu (+)} = 2^{i\nu}\sqrt{\frac{\pi \varsigma}{\sinh\pi \varsigma}} \frac{\Gamma\left(i \frac{\varsigma + \nu}{2}\right) }{\Gamma\left(1 + i\frac{\varsigma - \nu}{2}\right)}
\frac{\left(\sinh^2\pi\rho + \cosh^2\pi \varsigma\right)^{-\frac12}}{\Gamma\left(\frac12 - i\rho + i\varsigma\right)\Gamma\left(\frac12 + i\rho + i\varsigma\right)} \nonumber \\
\times 
\sin\left(\frac{\pi}{4} +i \frac{\rho - \varsigma}{2}\pi\right)
\frac{F^{(+)}_{\rho \nu}}{\sqrt{\cosh\pi(\rho - \varsigma)}},
\label{B_P}
\eea
\bea
{\cal B}_{\rho \varsigma }^{\nu(-)} = - \frac{2^{i\nu}}{2}\sqrt{\frac{\pi \varsigma}{\sinh\pi \varsigma}} \frac{\Gamma\left(i \frac{\varsigma + \nu}{2}\right) }{\Gamma\left(1 + i\frac{\varsigma - \nu}{2}\right)} \frac{\left(\sinh^2\pi\rho + \cosh^2\pi \varsigma\right)^{-\frac12}}{\Gamma\left(\frac12 - i\rho + i\varsigma\right)\Gamma\left(\frac12 + i\rho + i\varsigma\right)}
 \nonumber \\
\times
\cos\left(\frac{\pi}{4} +i \frac{\rho - \varsigma}{2}\pi\right)
\frac{F^{(-)}_{\rho \nu}}{\sqrt{\cosh\pi(\rho - \varsigma)}}.
\label{B_M}
\eea

\section{Contractions}
\label{sec:Contractions}

Details of the analytic contraction procedure can be found in numerous articles and books. Since this work is a continuation of Ref. \onlinecite{IIA}, we will adhere to the notations adopted there. See also Ref. \onlinecite{POG-YAKH4}, where contractions of  coordinate systems are shown.

\subsection{Contractions in SCP coordinates}
\label{sec:Contractions_SCP}

For rotated semi-circular parabolic system\cite{POG-YAKH4}
\bea
\label{sys_scp_xi_eta_rot}
u_0 &=& R\frac {\left(\eta^2 + \xi^2\right)^2 + 4}{8 \xi \eta},
\nonumber
\\[2mm]
u_1^\prime &=& R \frac{\sqrt{2}}{2}\left(\frac {{\eta}^{2}-{\xi}^{2}}{2\xi\eta}
+ \frac {\left(\eta^2 + \xi^2\right)^2 - 4}{8 \xi \eta}  \right),
\\[2mm]
u_2^\prime &=& R \frac{\sqrt{2}}{2}\left( \frac {{\eta}^{2}-{\xi}^{2}}{2\xi\eta}
- \frac {\left(\eta^2 + \xi^2\right)^2 - 4}{8 \xi \eta}  \right),
\nonumber
\eea
in the limit $R\to\infty$ we have
\begin{equation}
\label{contr_SCP}
\eta^2 \to 1 + \sqrt{2}\frac{x}{R}, \qquad
\xi^2 \to 1 + \sqrt{2}\frac{y}{R}.
\end{equation}
Laplace-Beltrami operator in Eq. (\ref{sys_scp_LB})  with the above limits goes to Laplacian $\Delta = \partial^2_{xx} + \partial^2_{yy}$ in Cartesian coordinates on Euclidean plane, therefore we take $\rho \sim k R$, where $k^2 = k_1^2 + k_2^2$ is a separation constant in Laplace equation $\Delta \Psi(x,y) + k^2 \Psi = 0$. From (\ref{OPERAT-A}), taking into account (\ref{contr_SCP}), we have $ {\hat{A}(\xi, \eta)}/{R^2} + \Delta_{LB} \sim 2\partial^2_{xx}$, so from the left Eq. (\ref{A_A}) we get for $A > 0$
\be
\frac{A - \rho^2 - 1/4}{2R^2} \sim -k_1^2,
\ee
where $k_1^2$ is a separation constant for contracted equation $X''(x) + k_1^2 X = 0$. Therefore $A \sim R^2 (k_2^2 - k_1^2)$, $|k_2| > |k_1|$. In the same way, from the right Eq. (\ref{A_A}) we obtain the same limit for constant $A < 0$ when $|k_2| < |k_1|$.

To analyze the contractions of the basis functions, let us use known asymptotic
relations for the Bessel function of pure imaginary index (3.14.2\cite{MAGNUS}, $p/z$ is fixed as $p,z\to\infty$)
\begin{equation}
2\pi J_{\pm ip}(z) \sim \frac{\sqrt{2\pi}}{(p^2 + z^2)^{1/4}}
\exp\left[\pm i\left(\sqrt{p^2 + z^2} -  p\, \arcsinh\frac{p}{z} -
\frac{\pi}{4} \right)\right] e^{\frac{\pi p}{2}},
\end{equation}
and for the MacDonald function let us use relation (3.14.2\cite{MAGNUS}, see also (19) 7.13.2\cite{BE2})
\begin{equation}
\label{K_inu}
K_{i\nu}(z) \sim \frac{\sqrt{2\pi}}{(\nu^2 - z^2)^{1/4}}
\sin\left(\frac{\pi}{4}
- \sqrt{\nu^2 - z^2} + \nu\, \arccosh \frac{\nu}{z} \right) e^{-\frac{\pi\nu}{2}},
\end{equation}
for $\nu > z > 0$. Then we obtain ($|k_2| > |k_1|$):
\begin{eqnarray}
J_{\pm i\rho}\left(\sqrt{|A|}\xi\right) &\sim& J_{ikR}\left(R \sqrt{k_2^2-k_1^2}
\sqrt{1+\sqrt{2} \frac{y}{R}} \right)
\nonumber\\[2mm]
&\sim& \frac{e^{\frac{\pi}{2} kR} 2^{-\frac{1}{4}} }{\sqrt{2\pi |k_2| R }}
\exp\left[\pm i\left(|k_2| y + R \left[\sqrt{2} |k_2| - k\, \arcsinh\frac{k}{\sqrt{k_2^2-k_1^2}}
\right] -  \frac{\pi}{4} \right)\right],
\end{eqnarray}
\begin{eqnarray}
K_{i\rho}(\sqrt{|A|}\eta) & \sim& K_{ikR} \left(R \sqrt{k_2^2-k_1^2} \sqrt{1+\sqrt{2}
\frac{x}{R}} \right)
\nonumber\\[2mm]
&\sim&
\frac{2^\frac{1}{4}\sqrt{\pi}e^{- \frac{\pi}{2}kR}}
{\sqrt{R |k_1|}} \sin\left(\frac{\pi}{4}  -  x |k_1|  -  R \left[\sqrt{2} |k_1| -
k\, \arccosh\frac{k}{\sqrt{k_2^2-k_1^2}}
\right] \right).
\end{eqnarray}
Thus, for large $R$ we have for the functions (\ref{norm_solution_SCP})
\begin{eqnarray}
\label{contr_sol_SCP}
\Psi_{\rho A}^{(1)}  (\xi,\eta) \sim  \frac{-\sqrt{k}}{\pi R \sqrt{ |k_1 k_2| R}}
\sin\left[|k_1| x + \delta_1(k_1,k_2, R)\right]
\cos\left[|k_2| y + \delta_2(k_1,k_2, R) \right], \ |k_2| > |k_1|, 
\end{eqnarray}
where 
\bea
\delta_{1,2}(k_1,k_2, R) := \sqrt{2} R|k_{1,2}| - Rk\, \arccosh\frac{k}{\sqrt{k_2^2 - k_1^2}} - \pi/4. 
\eea
The contraction of the wave function $\Psi_{\rho A}^{(2)}(\xi,\eta)$ (\ref{1-norm_solution_SCP}) up to the interchange $x \leftrightarrow y$ and $k_1 \leftrightarrow k_2$ coincides with the formula (\ref{contr_sol_SCP}). Therefore
\begin{eqnarray}
\label{contr_sol_SCP_2}
\Psi_{\rho A}^{(2)}  (\xi,\eta) \sim  \frac{- \sqrt{k}}{\pi R \sqrt{ |k_1 k_2| R}}
\sin\left[|k_2| y + \delta_1(k_2,k_1, R)\right]
\cos\left[|k_1| x + \delta_2(k_2,k_1, R) \right],\ |k_1| > |k_2|.
\end{eqnarray}

\subsection{Elliptic parabolic basis  to parabolic}

In the contraction limit $R \to \infty$ we have from (\ref{EP}) with $\gamma = 1$
\begin{equation}
\cos^2\theta \sim 1 - \frac{\eta^2}{R},
\qquad
\cosh^2 a \sim 1 + \frac{\xi^2}{R},
\end{equation}
where $(\xi, \eta)$ are the parabolic coordinates $x=(\xi^2 -\eta^2)/2$, $y = \xi \eta$, $\xi \ge 0$,  $\eta \in \mathbb{R}$ on Euclidean plane $E_2$.  Let us take $\mu \sim \kappa R$, $\rho \sim kR$, $\kappa, k > 0$ in Eqs. (\ref{cos}) and introduce a new constant $\lambda := R(\kappa^2 - k^2)$. Then the contraction procedure as $R\to\infty$ leads to two equations for parabolic-cylinder functions:
\begin{eqnarray}
\left(
\frac{d^2}{d\xi^2} + k^2 \xi^2 + \lambda
\right)
\Phi(\xi) = 0,
\qquad
\left(
\frac{d^2}{d\eta^2} + k^2 \eta^2 -\lambda
\right)
\Phi(\eta) = 0.
\end{eqnarray}
 Taking into account that $\mu \sim \kappa R  \sim k R + \frac{\lambda}{2k}$ and the asymptotic formulas for gamma functions at the large values of variable (see 1.18.\cite{BE1} (2), (4) and (6)):
\bea
\label{Gamma_limit}
\left|\Gamma(x+iy)\right| \exp\left(\frac{\pi |y|}{2}\right) |y|^{\frac{1}{2}-x} \sim
\sqrt{2\pi},
\quad
\frac{\Gamma(z + \alpha)}{\Gamma(z + \beta)} \sim z^{\alpha - \beta},
\quad
\Gamma(z) \sim \sqrt{2\pi} \frac{e^{\left(z - \frac12\right)\ln z}}{e^z},
\eea
 we obtain from (\ref{EP_constant_PLUS}) and (\ref{EP_constant_MINUS}) the following asymptotics for normalization constants:
\bea
\label{EP-SPH-112}
N_{\rho \mu}^{(+)} \sim \frac{\sqrt{2}}{4\pi^2}  \sqrt{\frac{k}{R}} 
\left|\Gamma\left(\frac{1}{4} + \frac{i\lambda}{4k}\right)\right|^2,
\qquad
N_{\rho \mu}^{(-)}  \sim \frac{\sqrt{2k^3 R}}{\pi^2} 
\left|\Gamma\left(\frac{3}{4} + \frac{i\lambda}{4k}\right)\right|^2,
\eea
and for (\ref{EP-EQUIDIST-02_P})
\begin{eqnarray}
\label{EP-SPH-00-110}
\Psi_{\rho \mu}^{(+)}
\sim \sqrt{\frac{k}{R}} \,
\left|\Gamma\left(\frac{1}{4} + \frac{i\lambda}{4k}\right)\right|^2
\, \frac{e^{\frac{i k}{2} (\eta^2 - \xi^2)} }{2\sqrt{2}\, \pi^2}
{_{1}F_{1}\left(\frac{1}{4} + \frac{i \lambda}{4k}; \frac{1}{2};  ik \xi^2\right)}
{_{1}F_{1}\left(\frac{1}{4} + \frac{i \lambda}{4k}; \frac{1}{2};  - ik \eta^2\right)},
\end{eqnarray}
\begin{eqnarray}
\label{EP-SPH-01-111}
\Psi_{\rho \mu}^{(-)}
\sim
\xi\eta \sqrt{\frac{2k^3}{R}} 
\left|\Gamma\left(\frac{3}{4} + \frac{i\lambda}{4k}\right)\right|^2
 \frac{e^{\frac{i k}{2} (\eta^2 - \xi^2)} }{\pi^2}
{_{1}F_{1}\left(\frac{3}{4} + \frac{i \lambda}{4k}; \frac{3}{2};  ik \xi^2\right)}
{_{1}F_{1}\left(\frac{3}{4} + \frac{i \lambda}{4k}; \frac{3}{2};  - ik \eta^2\right)}.
\end{eqnarray}
Taking into account that $ {_{1}F_{1}}\left(a; b; z\right) = e^z {_{1}F_{1}}\left(b - a; b; - z\right)$ ((7) from 6.3\cite{BE1})  for (\ref{EP-SPH-00-110}), (\ref{EP-SPH-01-111}) we get
\bea
\label{EP-SPH-00-11}
\lim_{R\rightarrow\infty} \sqrt{\frac{R}{k}} \,\,
\Psi_{\rho \mu}^{(\pm)}(a, \theta) = \,\Psi^{(\pm)}_{k \beta} (\xi, \eta),
\eea
which coincides with the solution for parabolic coordinates on the Euclidean plane\cite{ARXIV:2025} for  $\lambda = 2\beta$.
The factor $\sqrt{k/R}$ in (\ref{EP-SPH-00-11}) is obtained from contraction of the normalization integrals
\be
\delta(\rho - \rho') \delta(\mu - \mu') \sim \frac{\delta(k - k') \delta(\kappa - \kappa')}{R^2} \sim \frac{k }{R}\delta(k - k') \delta(\beta - \beta').
\ee

Now, looking at expansion (\ref{EP-SPH-01}) and taking into account that 
\be
\frac{1}{\Gamma(1/2 - i\rho)}
\sqrt{\frac{\Gamma(1/2 - i\rho-|m|)}
{\Gamma(1/2 + i\rho-|m|)}} \sim \frac{(-i)^{-|m|} e^{\pi k R/2}}{\sqrt{2\pi}},
\ee
we get for (\ref{EP-SPH-08})
\bea
\label{INTER-EP-SPH-01}
{\cal E}^{m(+)}_{\rho \mu}
\sim
\frac{i^{|m|}}{2\pi\sqrt{\pi}} \left|\Gamma\left(\frac{1}{4} + \frac{i\lambda}{4k}\right)\right|^2 \,
 {_3}F_2\left(
\left.
\begin{array}{ccc}
-|m|, &  \frac{1}{4} + i\frac{\lambda}{4k},  &   |m|
\\[2mm]
\frac{1}{2},  &   \frac{1}{2}
\end{array}\right| 1 \right).
\eea
Thus, from the limit of $\Psi_{\rho m}^{\rm S}(\tau, \varphi)$ eigenfunction on $H^{+}_2$ to the wave function $\Psi_{km}(r, \vphi)$ in polar coordinates on $E_2$ (see Refs. \onlinecite{IIA} and \onlinecite{ARXIV:2025})
\begin{eqnarray}
\label{CONTR-FUNCTION1}
\lim_{R\rightarrow\infty}\sqrt{R}\, \Psi_{\rho m}^{\rm S}(\tau, \varphi)
= (-1)^{|m|} \,  \sqrt{k} \, J_{|m|}(kr) \,
\frac{e^{im\varphi}} {\sqrt{2\pi}} =  (-1)^{|m|}  \Psi_{km}(r, \vphi), 
\end{eqnarray}
using (\ref{EP-SPH-00-110}), (\ref{EP-SPH-01}) and (\ref{INTER-EP-SPH-01}) one can obtain the expansion 
\be
\label{PARAB-POLAR-01}
\Psi^{ (+)}_{k \beta} (\xi, \eta)  =  \sum\limits_{m = -\infty}^{\infty} {\cal W}^{(+)}_{k \beta m}
\Psi_{km}(r, \phi)
\ee
for parabolic even and polar bases on Euclidean plain with coefficients
\bea
\label{PARAB-POLAR-02}
{\cal W}^{(+)}_{k \beta m} 
&=& \frac{(- i)^{|m|}}{2\sqrt{\pi^3 k}}
\, \left|\Gamma\left(\frac14 + \frac{i \beta}{2 k}\right)\right|^2 \,
{_3 F_2}
\left(
\left.\begin{array}{cccc}
- |m|  &  |m| & \frac14 + \frac{i\beta}{2 k}
\\[2mm]
\frac12, & \frac12  & 
\end{array}\right| 1
\right).
\eea
By analogy,
\be
\frac{1}{\Gamma(3/2 - i\rho)}
\sqrt{\frac{\Gamma(1/2 - i\rho-|m|)}
{\Gamma(1/2 + i\rho-|m|)}} \sim \frac{(-i)^{-1 - |m|} e^{\pi k R/2}}{kR\sqrt{2\pi}},
\ee
therefore
\bea
\label{INTER-EP-SPH-02}
{\cal E}^{m(-)}_{\rho \mu}
\sim
2m \frac{(-i)^{- |m|}}{\pi\sqrt{\pi}} \left|\Gamma\left(\frac{3}{4} + \frac{i\lambda}{4k}\right)\right|^2 \,
 {_3}F_2\left(
\left.
\begin{array}{ccc}
1-|m|, &  \frac{3}{4} + i\frac{\lambda}{4k},  &   |m|+1
\\[2mm]
\frac{3}{2},  &   \frac{3}{2}
\end{array}\right| 1 \right).
\eea
Thus from (\ref{EP-SPH-01}) we obtain expansion for parabolic odd and polar basis on Euclidean plain with coefficients
\bea
\label{PARAB-POLAR-03}
{\cal W}^{(-)}_{k \beta m} 
&=& 2m \frac{(- i)^{|m|}}{\sqrt{\pi^3 k}}
\, \left|\Gamma\left(\frac34 + \frac{i \beta}{2 k}\right)\right|^2 \,
{_3 F_2}
\left(
\left.\begin{array}{cccc}
1 - |m|  &  1 + |m| & \frac34 + \frac{i\beta}{2 k}
\\[2mm]
\frac32, & \frac32  & 
\end{array}\right| 1
\right).
\eea

\subsection{Elliptic parabolic basis to Cartesian}
\label{sec:Elliptic-parabolic_basis_to_Cartesian}

For variables from (\ref{EP}) in the contraction limit $R\to \infty$, when $\gamma \ne 1$ we have\cite{POG-YAKH4}:
\bea
\cos^2\theta \to \gamma\left(1 + 2\frac{x}{R}\right),
\quad
 \cosh^2 a \to  1 - \frac{\gamma}{\gamma - 1}\frac{y^2}{R^2}, \ \gamma\in(0,1);
\label{sys_ep_xi}
\\
 \cos^2\theta \to 1 - \frac{\gamma}{\gamma - 1}\frac{y^2}{R^2},
\quad
 \cosh^2 a \to \gamma\left(1 + 2\frac{x}{R}\right), \ \gamma >1,
\label{sys_ep_xi_2}
\eea
where $(x,y)$ are Cartesian coordinates on Euclidean plane. Consider the first case $\gamma\in(0,1)$. Eqs. (\ref{cos}) as $R\to \infty$ go to
\bea
 \left(
 \frac{d^2}{dx^2} + \frac{\gamma}{1-\gamma}\left[\frac{\rho^2}{\gamma R^2}  - \frac{\mu^2}{R^2} \right]
 \right)\Psi(x) =0, \quad
 \left(
 \frac{d^2}{dy^2} + \frac{\gamma}{1-\gamma}\left[-\frac{\rho^2}{R^2}  + \frac{\mu^2}{R^2} \right]
 \right)\Psi(y) =0.
 \eea
The case $\gamma >1$ gives the same equations in the contraction limit and is not considered here.

For the simplicity we take $\gamma = 1/2$, therefore $\rho \sim k R$, $\mu \sim R \sqrt{k^2 + k_2^2}$, where $k = \sqrt{k_1^2 + k_2^2}$ is a separation constant in Helmholtz equation $\Delta \Psi(x,y) + k^2 \Psi = 0$ in Cartesian coordinates on $E_2$. For variables $\theta$, $a$ in contraction limit $R\to\infty$ we have
\bea
\theta \sim \frac{\pi}{4} - \frac{x}{R},\quad  x\in\left(- \frac{\pi R}{4}, \frac{3\pi R}{4}\right),\qquad
\cosh^2 a \sim 1 + \frac{y^2}{R^2}.
\label{VAR_LIMITS}
\eea
 For normalization constants (\ref{EP_constant_PLUS}), (\ref{EP_constant_MINUS}) one can obtain 
\bea
N_{\rho\mu}^{(+)} \sim \frac{\left[2k\sqrt{k^2 + k_2^2}\right]^{1/2}}{|k_2|\pi R}e^{\frac{\pi R}{2}\left(k-\sqrt{k^2 + k_2^2}\right)} =: \tilde{N}_{k k_2}^{(+)}, \\
N_{\rho\mu}^{(-)} \sim |k_2| R \frac{\left[2k\sqrt{k^2 + k_2^2}\right]^{1/2}}{\pi}e^{\frac{\pi R}{2}\left(k-\sqrt{k^2 + k_2^2}\right)} =: \tilde{N}_{k k_2}^{(-)}.
\eea

For the radial part of the even solution (\ref{COSH}), using relations (\ref{VAR_LIMITS}) and that
\be
{_2F_1}\left(a,b;c;\frac{z}{ab}\right) \sim {_0F_1}\left(;c; z\right), \qquad a,b \to \infty, 
\label{2F1_INFINITY}
\ee
we obtain $\psi_{\rho \mu}^{(+)}(a) \sim \cos k_2 y $. For the angular part of the even solution (\ref{EP_hyper_sol}) we need to calculate the asymptotics of $\psi_{\rho \mu}^{(+)}\left({\pi}/{4} - {x}/{R}\right)$. We know, that  the leading terms of the expansion have the form
\be
\Psi_{\rho \mu}^{(+)}(a,\theta) \sim \tilde{N}^{(+)}_{k k_2} \psi_{\rho \mu}^{(+)}\left(\frac{\pi}{4} - \frac{x}{R}\right) \cos k_2 y \sim \left(A^{(+)}(R) e^{ik_1 x} + B^{(+)}(R) e^{- ik_1 x}\right) \cos |k_2| y, 
\label{PSI_THETA}
\ee
where $A^{(+)}$, $B^{(+)}$ are some constants. To determine them, let  $x\sim 0$, then $\theta\sim\pi/4$,
\bea
\psi_{\rho \mu}^{(+)}\left(\frac{\pi}{4}\right) \sim 2^{\frac{iR\sqrt{k^2 + k_2^2}}{2}}\, {_2F_1}\left(\frac14 + iR\frac{k+\sqrt{k^2 + k_2^2}}{2},\frac14 - iR\frac{k - \sqrt{k^2 + k_2^2}}{2}; \frac12; -1\right) \nonumber \\
\sim 2^{\frac{iR\sqrt{k^2 + k_2^2}}{2}}\, {_0F_1}\left(; \frac12; \frac{R^2 k_2^2}{4}\right) \sim 2^{ \frac{iR\sqrt{k^2 + k_2^2}}{2}} \frac{e^{|k_2| R}}{2},
\eea
and
\bea
\tilde{N}^{(+)}_{k k_2} \psi_{\rho \mu}^{(+)}\left(\frac{\pi}{4}\right) \sim A^{(+)}  + B^{(+)}.
\eea
The derivative of both sides of (\ref{PSI_THETA}) with respect to $x$, evaluated at the same point $x \sim 0$, leads to
\be
\tilde{N}^{(+)}_{k k_2} \psi_{\rho \mu}^{(+)\prime}\left(\frac{\pi}{4}\right) \sim  i k_1\left(A^{(+)}  - B^{(+)}\right).
\ee
Taking into account that ${_2F_1}^\prime(a, b; c; z)  = ab/c\, {_2F_1}(a + 1, b + 1 ; c + 1; z)$, we obtain
\be
 \psi_{\rho \mu}^{(+)\prime}\left(\frac{\pi}{4}\right) \sim - 2^{\frac{iR\sqrt{k^2 + k_2^2}}{2}} e^{|k_2| R}\left(|k_2| + \frac{i}{2} \sqrt{k^2 + k_2^2}\right).
\ee
Therefore
\bea
\Psi_{\rho \mu}^{(+)}(a,\theta) \sim \frac{\left[k\sqrt{k^2 + k_2^2}\right]^{1/2}}{\pi}e^{\frac{\pi R}{2}  \left(k-\sqrt{k^2 + k_2^2}\right) + |k_2|R}\, 2^{\frac{1 + iR\sqrt{k^2 + k_2^2}}{2}} \nonumber \\
\times\left[\frac{\cos k_1 x}{2} -  \left(|k_2| + \frac{i}{2}\sqrt{k^2 + k_2^2}\right)\frac{\sin k_1 x}{k_1}\right] \frac{\cos|k_2|y}{R|k_2|}.
\eea

By analogy, the radial part of the odd solution (\ref{00-COSH}) contracts as follows
\bea
\psi_{\rho \mu}^{(-)}  (a) \sim \frac{y}{R}\, {_0F_1}\left(;\frac{3}{2};-\frac{y^2 k_2^2}{4}\right) 
  \sim \frac{1}{R |k_2|}\sin |k_2| y.
\eea
For the angular odd solution (\ref{EP_hyper_sol-00}) we obtain
\bea
\label{EP_sim2}
\Psi_	{\rho\mu}^{(-)}(a,\theta) \sim \tilde{N}_{k k_2}^{(-)}\Psi_{\rho\mu}^{(-)}\left(\frac{\pi}{4} - \frac{x}{R}\right) \frac{\sin |k_2| y}{R |k_2|} \sim \left(A^{(-)} e^{i k_1 x} + B^{(-)} e^{-i k_1 x}\right) \sin |k_2| y.
\eea
If $x\sim 0$, then 
\bea
\psi_{\rho \mu}^{(-)}\left(\frac{\pi}{4}\right) \sim  2^{ \frac{iR\sqrt{k^2 + k_2^2}}{2}} \frac{e^{|k_2| R}}{2 R |k_2|},\
 \psi_{\rho \mu}^{(-)\prime}\left(\frac{\pi}{4}\right) \sim - 2^{\frac{iR\sqrt{k^2 + k_2^2}}{2}}  \frac{e^{|k_2| R}}{R |k_2|}\left(|k_2| + \frac{i}{2} \sqrt{k^2 + k_2^2}\right),
\eea
and finally
\bea
\Psi_	{\rho\mu}^{(-)}(a,\theta)  \sim \frac{\left[k\sqrt{k^2 + k_2^2}\right]^{1/2}}{\pi}e^{\frac{\pi R}{2}  \left(k-\sqrt{k^2 + k_2^2}\right) + |k_2|R}\, 2^{\frac{1 + iR\sqrt{k^2 + k_2^2}}{2}} \nonumber \\
\times\left[\frac{\cos k_1 x}{2} -  \left(|k_2| + \frac{i}{2}\sqrt{k^2 + k_2^2}\right)\frac{\sin k_1 x}{k_1}\right] \frac{\sin|k_2|y}{R|k_2|}.
\eea


\subsection{Hyperbolic parabolic basis to Cartesian}
\label{sec:Hyperbolic-parabolic_basis_to_Cartesian}
 For the variables from (\ref{HP_sys_2SH}) in contraction limit  $R\to \infty$ we have\cite{POG-YAKH4}
\begin{equation}
\label{CONTR:HPtoC}
 \sin^2 \theta \sim 1 - \frac{\gamma}{\gamma + 1}\frac{y^2}{R^2}, \quad \sinh^2 b \sim \gamma\left(1 + 2\frac{x}{R} \right),
\end{equation}
where $(x,y)$ are Cartesian coordinates on Euclidean plane. Then equations (\ref{SINH1}) contract to the following ones:
\begin{equation}
\label{HPxy}
 \left(
 \frac{d^2}{dx^2} + \frac{\gamma}{\gamma+1}\left[\frac{\rho^2}{\gamma R^2}  - \frac{\varsigma^2}{R^2} \right]
 \right)\Psi(x) =0,\
 \left(
 \frac{d^2}{dy^2} + \frac{\gamma}{\gamma+1}\left[\frac{\rho^2}{R^2}  + \frac{\varsigma^2}{R^2} \right]
 \right)\Psi(y) =0.
 \end{equation}
For the simplicity we take $\gamma = 1$. When $\rho \sim k R$, $\varsigma_n \sim R\sqrt{k_2^2 - k_1^2 }$ ($|k_2|>|k_1|$) for discrete spectrum and $\varsigma \sim R\sqrt{k_1^2 - k_2^2 }$ ($|k_1|>|k_2|$) for continuos one, Eqs. (\ref{HPxy}) go to $X''(x) + k_1^2 X = 0$, $Y''(y) + k_2^2 Y = 0$ respectively on $E_2$.

\paragraph{Contraction for discrete spectrum.} Consider the angular part of the solution (\ref{HP_discrete_2})
\be
  \psi_{\rho s_n}^{HP}(\theta) := \sqrt{\sin\theta} {\rm Q}^{-i\rho}_{-\frac12 + \varsigma_n}(\cos\theta)
\ee
in contraction limit $\cos\theta \sim y/\sqrt{2}R$, therefore, using the representation (\ref{100-ARHO}), the asymptotics of the gamma-functions (\ref{Gamma_limit}) and (\ref{2F1_INFINITY}) we obtain 
\bea
\psi_{\rho s_n}^{HP}(\theta) \sim \frac{\sqrt{\pi} e^{\frac{i\pi}{4} +\frac{\pi k R}{2} } }{2^{\frac74} \sqrt{|k_2| R}} \left(\frac{e}{R|k_2|\sqrt{2}}\right)^{ikR}\left(\frac{ik - \sqrt{k_2^2 - k_1^2 } }{ik + \sqrt{k_2^2 - k_1^2 }}\right)^{\frac{R\sqrt{k_2^2 - k_1^2 }}{2}} e^{i |k_2| y}.
\label{HP_rho_sn_THETA}
\eea
For the constant (\ref{N_rho_s_D}) we obtain
\bea
N_{\rho s_n}^{\rm d} \sim  \frac{2\sqrt{2k\sqrt{k_2^2 - k_1^2}}}{\pi^2 e^{\pi k R}} \left(\frac{2 k_2^2 R^2}{e^2}\right)^{ikR} \left(\frac{\sqrt{k_2^2 - k_1^2} + ik}{\sqrt{k_2^2 - k_1^2} - ik}\right)^{{R\sqrt{k_2^2 - k_1^2 }}}.
\eea
The radial part 
\be
\psi^{HP}_{\rho s_n}(b) := \sqrt{\sinh b}Q^{-i\rho}_{-\frac12 + \varsigma_n}(\cosh b) 
\ee
in the contraction limit with $\cosh b \sim \sqrt{2}\left(1 + x/2R\right)$ should have an expansion of the form
\be
\psi^{HP}_{\rho s_n}(b) \sim A e^{i k_1 x} + B e^{- i k_1 x},
\label{HP_b}
\ee
and as $x \sim 0$
\be
 \psi^{HP}_{\rho s_n}(b) \sim Q^{-i kR}_{-\frac12 + R\sqrt{k_2^2 - k_1^2}}\left(\sqrt{2}\right) \sim A + B. 
\ee
Further calculation of asymptotics of (\ref{00-SOLUTION-HP6}) gives
\be
\frac{e^{\pi k R} 2^{-\frac12 - R\sqrt{k_2^2 - k_1^2}} \sqrt{\pi}  \left({e}/{R}\right)^{ikR}  }{\sqrt{R}\left(1 + \sqrt{2}\right)^{\frac12 + R\sqrt{k_2^2 - k_1^2} + ikR}} \frac{\left(\sqrt{k_2^2 - k_1^2} - ik\right)^{R\sqrt{k_2^2 - k_1^2} - ikR}}{\left(k_2^2 - k_1^2\right)^{\frac14 + R\sqrt{k_2^2 - k_1^2}/2}} e^{R\frac{\sqrt{k_2^2 - k_1^2} + ik}{1+\sqrt{2}}} \sim A + B.
\label{EQ1}
\ee
The derivative of (\ref{HP_b}) with respect to $x$, as $x\sim0$ leads to
\bea
 \frac{A + B}{1 + \sqrt{2}} \left( ik\sqrt{2}  - {\sqrt{k_2^2 - k_1^2}}\right)   \sim ik_1(A - B). 
\label{EQ2}
\eea
Eqs. (\ref{EQ1}), (\ref{EQ2}) allow us to express the constants $A$, $B$ and finally obtain the contraction of the  HP solution.

\paragraph{Contraction for continuous spectrum.} The only difference in the contraction of the HP solution (\ref{02-SOLUTION-00-HP6}) compared to $\Psi^{\rm HP}_{\rho\varsigma_n}(b,\theta)$ is the change of inequality $|k_2| > |k_1|$ to $|k_1| > |k_2|$ and a different normalization constant. Therefore, all expressions of the previous  paragraph can be repeated by replacing $\sqrt{k_2^2 - k_1^2}$ with $i\sqrt{k_1^2 - k_2^2}$. For example, from (\ref{HP_rho_sn_THETA}) we get
\bea
\sqrt{\sin\theta}{\rm Q}_{-\frac12 + i \varsigma}^{-i\rho}(\cos\theta) \sim \frac{\sqrt{\pi} e^{\frac{i\pi}{4} +\frac{\pi k R}{2} } }{2^{\frac74} \sqrt{|k_2| R}} \left(\frac{e}{R|k_2|\sqrt{2}}\right)^{ikR}\left(\frac{k - \sqrt{k_1^2 - k_2^2 } }{k + \sqrt{k_1^2 - k_2^2 }}\right)^{\frac{iR\sqrt{k_1^2 - k_2^2 }}{2}} e^{i |k_2| y}.
\eea
Asymptotics of $N_{\rho s}^{\rm c}$ (\ref{N_rho_s_c}) 
\bea
N_{\rho s}^{\rm c} \sim \frac{2^\frac32}{\pi^2} e^{\pi R \left(\sqrt{k_1^2 - k_2^2} -k\right)} \left[eR\left(k + \sqrt{k_1^2 - k_2^2} \right)\right]^{2iR\left(k + \sqrt{k_1^2 - k_2^2} \right)}
\eea
completes the contraction procedure.


\section{Conclusions}

The paper describes solutions of the Laplace-Beltrami equation on two-dimensional two-sheeted hyperboloid for three non-subgroup coordinate systems: semi-sircular parabolic, elliptic parabolic and hyperbolic parabolic. The behavior of wave functions near singular points is analyzed and all normalization constants are calculated. Eigenfunctions in the hyperbolic parabolic system with a discrete spectrum are presented for the first time.

The coefficients of interbasis expansions of solutions in the specified coordinate systems through some subgroup bases are calculated. These expansions allow one to generalize some integral representations for special functions. In most cases, the expansion coefficients are expressed through gamma functions. Let us note the simplest form of the expansion coefficients of the semi-sircular parabolic functions on the horocyclic basis, expressed in exponential functions. The most complex form has the expansion of the elliptic parabolic system  through the equidistant system, containing delta functions. The expansion of the eigenfunctions of the elliptic parabolic system through the functions of the pseudo-spheric basis uses Wilson-Racah polynomials. The integral representations found for them made it possible to prove the completeness of these coefficients.

A contraction procedure for all normalized eigenfunctions in three non-subgroup coordinate systems from the  hyperboloid to the Euclidean plane is realized. Contraction of solutions from the hyperboloid to the Euclidean plane is not direct and obvious. The obtained results mainly contain phases in which the contraction parameter $R$ is present. Only in the case of contraction of the elliptic parabolic basis to the parabolic one on the plane we were able to find the limits of both the wave functions and the corresponding interbasis expansion. These difficulties are related to the choice of solutions on the hyperboloid and to the non-subgroup nature of the coordinate systems under consideration. 

\section*{APPENDIX}

\subsection{Orthonormality and completeness of EP wave functions}
\label{bsection: Completeness of EP}

Let us calculate the integrals in (\ref{norm_int_EP}). First, from the left differential equation (\ref{cos}) it follows that
\bea
\label{1-norm_int_EP}
\int\limits_{-\pi/2}^{\pi/2}\,
\left(\mu_1^2 - \mu_2^2  + \frac{\rho_2^2 - \rho_1^2}{\cos^2\theta}\right)
\psi_{\rho_1 \mu_1}^{(\pm)}(\theta)
\psi_{\rho_2 \mu_2}^{(\pm)}(\theta)\, d\theta = 
\nonumber
\\[2mm]
=
\left\{\psi_{{\rho_2}{\mu_2}}^{(\pm)}(\theta)
\frac{d \psi_{{\rho_1}{\mu_1}}^{(\pm)}(\theta)}{d \theta}
- \psi_{\rho_1 \mu_1}^{(\pm)}(\theta)
\frac{d \psi_{\rho_2 \mu_2}^{(\pm)
}(\theta)}{d \theta}\right\}
\Biggr|_{-\pi/2}^{\pi/2}.
\eea
Applying the transformation $z \to 1/(1-z)$ for $\psi^{(\pm)}_{\rho\mu}(\theta)$ in hypergeometric functions
(\ref{EP-EQUIDIST-02_P}) according to (3) 2.10\cite{BE1} we get the asymptotics as $\theta \sim \pi/2$
\bea
\psi_{\rho \mu}^{(+)}(\theta)
\sim
 \frac{\sqrt{\pi} \, \Gamma(-i\rho) \, (\cos\theta)^{1/2+i\rho}}
{\Gamma\left(\frac14 - i\frac{\rho + \mu}{2}\right)
\Gamma\left(\frac14 - i\frac{\rho - \mu}{2}\right)}
+
\frac{\sqrt{\pi} \, \Gamma(i\rho) \, (\cos\theta)^{1/2-i\rho}}
{\Gamma\left(\frac14 + i\frac{\rho + \mu}{2}\right)
\Gamma\left(\frac14 + i\frac{\rho - \mu}{2}\right)},\nonumber\\[3mm]
\psi_{\rho \mu}^{(-)}(\theta)
\sim
 \frac{\sqrt{\pi} \, \Gamma(-i\rho) \, (\cos\theta)^{1/2+i\rho}}
{2\Gamma\left(\frac34 - i\frac{\rho + \mu}{2}\right)
\Gamma\left(\frac34 - i\frac{\rho - \mu}{2}\right)}
+
\frac{\sqrt{\pi} \, \Gamma(i\rho) \, (\cos\theta)^{1/2-i\rho}}
{2\Gamma\left(\frac34 + i\frac{\rho + \mu}{2}\right)
\Gamma\left(\frac34 + i\frac{\rho - \mu}{2}\right)}.
\label{02-norm_int_EP}
\eea
Let us divide expression (\ref{1-norm_int_EP}) by $\rho_2^2 - \rho_1^2$,
\bea
&&\int\limits_{-\pi/2}^{\pi/2}\,
\left(\frac{\mu_1^2 - \mu_2^2}{\rho_2^2 - \rho_1^2} + \frac{1}{\cos^2\theta}\right)\psi_{\rho_1 \mu_1}^{(\pm)}(\theta)
\psi_{\rho_2 \mu_2}^{(\pm)}(\theta)\, d\theta = \nonumber\\ 
&&
= \frac{1}{\rho_2^2 - \rho_1^2}\left\{\psi_{{\rho_2}{\mu_2}}^{(\pm)}(\theta)
\frac{d \psi_{{\rho_1}{\mu_1}}^{(\pm)}(\theta)}{d \theta}
- \psi_{\rho_1 \mu_1}^{(\pm)}(\theta)
\frac{d \psi_{\rho_2 \mu_2}^{(\pm)
}(\theta)}{d \theta}\right\}
\Biggr|_{- \frac{\pi}{2} + \epsilon}^{\frac{\pi}{2} - \epsilon} =: A^{\mu_1\mu_2 (\pm)}_{\rho_1\rho_2}.
\label{1-norm_int_EP_n}
\eea
Substituting (\ref{02-norm_int_EP}) and its derivative into (\ref{1-norm_int_EP_n}) and considering $\sin\epsilon \sim \epsilon$, we obtain for the even solution 
\bea
 &A^{\mu_1\mu_2 (+)}_{\rho_1\rho_2} = 2 i \pi \lim\limits_{\epsilon \to 0}
\left\{ \frac{\Gamma\left( - i\rho_1\right) \Gamma\left( - i\rho_2\right) e^{i(\rho_1 + \rho_2)\ln\epsilon}}{(\rho_2 + \rho_1)
\Gamma\left(\frac14 - i \frac{\rho_1 + \mu_1}{2}\right)
\Gamma\left(\frac14 - i\frac{\rho_1 - \mu_1}{2}\right)
\Gamma\left(\frac14 - i\frac{\rho_2 + \mu_2}{2}\right)
\Gamma\left(\frac14 - i\frac{\rho_2 - \mu_2}{2}\right)}
\right. + \nonumber \\
& + 
\left. \frac{\Gamma\left(i\rho_1\right) \Gamma\left( - i\rho_2\right) e^{ - i(\rho_1 - \rho_2)\ln\epsilon}}{(\rho_2 - \rho_1)
\Gamma\left(\frac14 + i \frac{\rho_1 + \mu_1}{2}\right)
\Gamma\left(\frac14 + i\frac{\rho_1 - \mu_1}{2}\right)
\Gamma\left(\frac14 - i\frac{\rho_2 + \mu_2}{2}\right)
\Gamma\left(\frac14 - i\frac{\rho_2 - \mu_2}{2}\right)}
\right. - 
\nonumber \\
& - 
\left. \frac{\Gamma\left(- i\rho_1\right) \Gamma\left(i\rho_2\right) e^{i(\rho_1 - \rho_2)\ln\epsilon}}{(\rho_2 - \rho_1)
\Gamma\left(\frac14 - i \frac{\rho_1 + \mu_1}{2}\right)
\Gamma\left(\frac14 - i\frac{\rho_1 - \mu_1}{2}\right)
\Gamma\left(\frac14 + i\frac{\rho_2 + \mu_2}{2}\right)
\Gamma\left(\frac14 + i\frac{\rho_2 - \mu_2}{2}\right)}
\right. - \nonumber \\
& - \left. \frac{\Gamma\left(i\rho_1\right) \Gamma\left(i\rho_2\right) e^{- i(\rho_1 + \rho_2)\ln\epsilon}}{(\rho_2 + \rho_1)
\Gamma\left(\frac14 + i \frac{\rho_1 + \mu_1}{2}\right)
\Gamma\left(\frac14 + i\frac{\rho_1 - \mu_1}{2}\right)
\Gamma\left(\frac14 + i\frac{\rho_2 + \mu_2}{2}\right)
\Gamma\left(\frac14 + i\frac{\rho_2 - \mu_2}{2}\right)}
\right\},
 \label{I10} 
\eea
and similarly the expression for $A^{\mu_1\mu_2 (-)}_{\rho_1\rho_2}$. Considering that $e^{\pm i(\rho_1 - \rho_2)\ln\epsilon}/(\rho_1 - \rho_2 ) \sim \mp i\pi\delta(\rho_1 - \rho_2)$ as $\epsilon \sim 0$ and that $\rho_{1,2} > 0$, we finally obtain
\bea
 \label{I1} 
& A^{\mu_1\mu_2 (+)}_{\rho_1\rho_2} = 
 \frac{2\pi^3 \delta(\rho_1 - \rho_2)}{\rho_1\sinh\pi\rho_1}
\left\{ \frac{1}{
\Gamma\left(\frac14 + i \frac{\rho_1 + \mu_1}{2}\right)
\Gamma\left(\frac14 + i\frac{\rho_1 - \mu_1}{2}\right)
\Gamma\left(\frac14 - i\frac{\rho_1 + \mu_2}{2}\right)
\Gamma\left(\frac14 - i\frac{\rho_1 - \mu_2}{2}\right)}
\right. + \\
&\left.
+ \{\mu_1\leftrightarrow \mu_2 \}
\right\},
\nonumber
\eea
\bea
\label{2-norm_int_EP}
& A^{\mu_1\mu_2 (-)}_{\rho_1\rho_2} = 
 \frac{\pi^3 \delta(\rho_1 - \rho_2)}{2\rho_1\sinh\pi\rho_1}
\left\{ \frac{1}{
\Gamma\left(\frac34 + i \frac{\rho_1 + \mu_1}{2}\right)
\Gamma\left(\frac34 + i\frac{\rho_1 - \mu_1}{2}\right)
\Gamma\left(\frac34 - i\frac{\rho_1 + \mu_2}{2}\right)
\Gamma\left(\frac34 - i\frac{\rho_1 - \mu_2}{2}\right)}
\right. + \\
&\left.
+ \{\mu_1\leftrightarrow \mu_2 \}
\right\}.
\nonumber
\eea

From the right Eq. in (\ref{cos}) we have the following integral
\bea
\label{6-norm_int_EP}
&&
\int\limits_{0}^{\infty}
\left(\mu_2^2 - \mu_1^2 + \frac{\rho_1^2-\rho_2^2}{\cosh^2 a}\right)
\psi_{\rho_1 \mu_1}^{(\pm)}(a)
\psi_{\rho_2 \mu_2}^{(\pm)}(a) d a
\nonumber
\\[2mm]
&=&
\left\{\psi_{{\rho_2}{\mu_2}}^{(\pm)}(a)
\frac{d \psi_{{\rho_1}{\mu_1}}^{(\pm)}(a)}{d a}
- \psi_{\rho_1 \mu_1}^{(\pm)}(a)
\frac{d \psi_{\rho_2 \mu_2}^{(\pm)}(a)}{d a}\right\}
\Biggr|_{0}^{\infty}.
\eea
The asymptotic behavior of the wave functions $\psi_{\rho \mu}^{(\pm)}(a)$ (\ref{COSH}), (\ref{00-COSH}) as $a\to \infty$ (transforming $z \to 1 - z$ due to (1) from  2.10\cite{BE1}) is as follows
\bea
\label{06-norm_int_EP}
\psi_{\rho \mu}^{(+)}(a)
&\sim&
\frac{\sqrt{\pi} \, \Gamma(-i\mu)  e^{-i\mu a}}
{2^{-i\mu}\Gamma\left(\frac14 - i\frac{\rho + \mu}{2}\right)
\Gamma\left(\frac14 + i\frac{\rho - \mu}{2}\right)}
+
\frac{\sqrt{\pi} \, \Gamma(i\mu)  e^{i\mu a}}
{2^{i\mu}\Gamma\left(\frac14 + i\frac{\rho + \mu}{2}\right)
\Gamma\left(\frac14 - i\frac{\rho - \mu}{2}\right)},
 \nonumber\\[3mm]
\psi_{\rho \mu}^{(-)}(a)
&\sim& \frac{\sqrt{\pi} \, \Gamma(-i\mu)  e^{-i\mu a}}
{2^{1 - i\mu}\Gamma\left(\frac34 - i\frac{\rho + \mu}{2}\right)
\Gamma\left(\frac34 + i\frac{\rho - \mu}{2}\right)}
+
\frac{\sqrt{\pi} \, \Gamma(i\mu)  e^{i\mu a}}
{2^{1 + i\mu}\Gamma\left(\frac34 + i\frac{\rho + \mu}{2}\right)
\Gamma\left(\frac34 - i\frac{\rho - \mu}{2}\right)}.
\eea
Dividing (\ref{6-norm_int_EP}) by $\mu_2^2 - \mu_1^2$, we get
\bea
\label{7-norm_int_EP}
&&\int\limits_{0}^{\infty}\left(1 + \frac{\rho_1^2 - \rho_2^2}{(\mu_2^2 - \mu_1^2)\cosh^2 a}\right)
\psi_{\rho_1 \mu_1}^{(\pm)}(a)
\psi_{\rho_2 \mu_2}^{(\pm)}(a)\, d a = 
\nonumber\\ 
&=&
\frac{1}{{\mu_2^2 - \mu_1^2}
}
\left.\left\{\psi_{{\rho_2}{\mu_2}}^{(\pm)}(a)
\frac{d \psi_{{\rho_1}{\mu_1}}^{(\pm)}(a)}{d a}
- \psi_{\rho_1 \mu_1}^{(\pm)}(a)
\frac{d \psi_{\rho_2 \mu_2}^{(\pm)}(a)}{d a}\right\}
\right|_{\epsilon}^{\frac{1}{\epsilon}} =: B^{\mu_1 \mu_2(\pm)}_{\rho_1\rho_2}.
\eea
Substituting (\ref{06-norm_int_EP}) to the right side of the above expression and taking into account that point $a = \epsilon \sim 0$ does not contribute to the integral, we obtain
\bea
&B^{\mu_1 \mu_2(+)}_{\rho_1\rho_2} =
i\pi \lim\limits_{\epsilon \to 0}
\Biggl\{
\frac{\Gamma(i\mu_1)\Gamma(-i\mu_2) e^{i(\mu_1-\mu_2)/\epsilon}}{(\mu_2 - \mu_1)2^{i(\mu_1 - \mu_2)}
\Gamma\left(\frac14 - i\frac{\rho_2 + \mu_2}{2}\right)
\Gamma\left(\frac14 + i\frac{\rho_2 - \mu_2}{2}\right)
\Gamma\left(\frac14 + i\frac{\rho_1 + \mu_1}{2}\right)
\Gamma\left(\frac14 - i\frac{\rho_1 - \mu_1}{2}\right) }
\nonumber
\\[2mm]
&-
\frac{\Gamma(-i\mu_1)\Gamma(i\mu_2)}{2^{-i(\mu_1 - \mu_2)}
\Gamma\left(\frac14 - i\frac{\rho_1 + \mu_1}{2}\right)
\Gamma\left(\frac14 + i\frac{\rho_1 - \mu_1}{2}\right)
\Gamma\left(\frac14 + i\frac{\rho_2 + \mu_2}{2}\right)
\Gamma\left(\frac14 - i\frac{\rho_2 - \mu_2}{2}\right) }
\,\, \frac{e^{-i(\mu_1-\mu_2)/\epsilon}}{\mu_2 - \mu_1}
\nonumber
\\[2mm]
&+
\frac{\Gamma(- i\mu_1)\Gamma(- i\mu_2)}{2^{- i(\mu_1 + \mu_2)}
\Gamma\left(\frac14 - i\frac{\rho_1 + \mu_1}{2}\right)
\Gamma\left(\frac14 + i\frac{\rho_1 - \mu_1}{2}\right)
\Gamma\left(\frac14 - i\frac{\rho_2 + \mu_2}{2}\right)
\Gamma\left(\frac14 + i\frac{\rho_2 - \mu_2}{2}\right) }
\,\, \frac{e^{- i(\mu_1 + \mu_2)/\epsilon}}{\mu_2 + \mu_1}
\nonumber
\\[2mm]
&-
\frac{\Gamma(i\mu_1)\Gamma(i\mu_2)}{2^{i(\mu_1 + \mu_2)}
\Gamma\left(\frac14 + i\frac{\rho_1 + \mu_1}{2}\right)
\Gamma\left(\frac14 - i\frac{\rho_1 - \mu_1}{2}\right)
\Gamma\left(\frac14 + i\frac{\rho_2 + \mu_2}{2}\right)
\Gamma\left(\frac14 - i\frac{\rho_2 - \mu_2}{2}\right) }
\,\, \frac{e^{i(\mu_1 + \mu_2)/\epsilon}}{\mu_2 + \mu_1}
\Biggr\},
\eea
and a similar expression for $B^{\mu_1 \mu_2(-)}_{\rho_1\rho_2}$. Considering that $e^{\pm i(\mu_1 - \mu_2)/\epsilon}/(\mu_1 - \mu_2)  \sim \pm i \pi \delta(\mu_1 - \mu_2)$ as $\epsilon \sim 0$ and taking into account that $\mu_1, \mu_2 > 0$, we get for the even function
\bea
\label{10-norm_int_EP_even}
&&B^{\mu_1 \mu_2(+)}_{\rho_1\rho_2} = \frac{\pi^3\delta(\mu_1 - \mu_2)}{\mu_1 \sinh\pi \mu_1} \times
\nonumber
\\[3mm]
&\times& 
\left\{
\frac{1}
{\Gamma\left(\frac14 - i\frac{\rho_1 + \mu_1}{2}\right)
\Gamma\left(\frac14 + i\frac{\rho_1 - \mu_1}{2}\right)
\Gamma\left(\frac14 + i\frac{\rho_2 + \mu_1}{2}\right)
\Gamma\left(\frac14 - i\frac{\rho_2 - \mu_1}{2}\right)}
+
\left\{\rho_1 \leftrightarrow \rho_2\right\}
\right\},
\eea
and for the odd function
\bea
\label{10-norm_int_EP_odd}
&&B^{\mu_1 \mu_2(-)}_{\rho_1\rho_2} = \frac{\pi^3\delta(\mu_1 - \mu_2)}{4 \mu_1 \sinh\pi \mu_1} \times
\nonumber
\\[3mm]
&\times& 
\left\{
\frac{1}
{\Gamma\left(\frac34 - i\frac{\rho_1 + \mu_1}{2}\right)
\Gamma\left(\frac34 + i\frac{\rho_1 - \mu_1}{2}\right)
\Gamma\left(\frac34 + i\frac{\rho_2 + \mu_1}{2}\right)
\Gamma\left(\frac34 - i\frac{\rho_2 - \mu_1}{2}\right)}
+
\left\{\rho_1 \leftrightarrow \rho_2\right\}
\right\}.
\eea
From (\ref{1-norm_int_EP_n}), (\ref{7-norm_int_EP}) and (\ref{I1}), (\ref{10-norm_int_EP_even}) it follows that
\bea
&\int\limits_{-\pi/2}^{\pi/2}\,
 \frac{d\theta }{\cos^2\theta}\psi_{\rho_1 \mu_1}^{(+)}(\theta)
\psi_{\rho_2 \mu_2}^{(+)}(\theta)\int\limits_{0}^{\infty}
\psi_{\rho_1 \mu_1}^{(+)}(a)
\psi_{\rho_2 \mu_2}^{(+)}(a)\, d a = \frac{8\pi^6}{\rho_1\mu_1\sinh\pi\rho_1\sinh\pi\mu_1} \times \nonumber
\\ 
&\times \frac{\delta(\rho_1 - \rho_2)\delta(\mu_1 - \mu_2)}{\left|\Gamma\left(\frac14 + i\frac{\rho_1 + \mu_1}{2}\right)\right|^4 \left|\Gamma\left(\frac14 + i\frac{\rho_1 - \mu_1}{2}\right)\right|^4} + 
\int\limits_{-\pi/2}^{\pi/2}\,
\psi_{\rho_1 \mu_1}^{(+)}(\theta)
\psi_{\rho_2 \mu_2}^{(+)}(\theta)\int\limits_{0}^{\infty}  \frac{d a}{\cosh^2 a}
\psi_{\rho_1 \mu_1}^{(+)}(a)
\psi_{\rho_2 \mu_2}^{(+)}(a),
\eea
where we use the property $z\delta(z) = 0$. Returning to (\ref{norm_int_EP}) and taking into account the above expression, we get the normalization constant
\be
\label{EP_constant_PLUS_A}
N_{\rho\mu}^{(+)} = \frac{\sqrt{\rho\mu \sinh\pi\rho \sinh\pi\mu}}{2\sqrt{2}R\pi^3} \left|\Gamma\left(\frac14 + i\frac{\rho + \mu}{2}\right)\right|^2 \left|\Gamma\left(\frac14 + i\frac{\rho - \mu}{2}\right)\right|^2.
\ee
Similarly, from (\ref{1-norm_int_EP_n}), (\ref{7-norm_int_EP}) and (\ref{2-norm_int_EP}), (\ref{10-norm_int_EP_odd}) we obtain
\bea
&\int\limits_{-\pi/2}^{\pi/2}\,
 \frac{d\theta }{\cos^2\theta}\psi_{\rho_1 \mu_1}^{(-)}(\theta)
\psi_{\rho_2 \mu_2}^{(-)}(\theta)\int\limits_{0}^{\infty}
\psi_{\rho_1 \mu_1}^{(-)}(a)
\psi_{\rho_2 \mu_2}^{(-)}(a)\, d a = \frac{\pi^6}{2 \rho_1\mu_1\sinh\pi\rho_1\sinh\pi\mu_1} \times \nonumber
\\ 
&\times \frac{\delta(\rho_1 - \rho_2)\delta(\mu_1 - \mu_2)}{\left|\Gamma\left(\frac34 + i\frac{\rho_1 + \mu_1}{2}\right)\right|^4 \left|\Gamma\left(\frac34 + i\frac{\rho_1 - \mu_1}{2}\right)\right|^4} + 
\int\limits_{-\pi/2}^{\pi/2}\,
\psi_{\rho_1 \mu_1}^{(-)}(\theta)
\psi_{\rho_2 \mu_2}^{(-)}(\theta)\int\limits_{0}^{\infty}  \frac{d a}{\cosh^2 a}
\psi_{\rho_1 \mu_1}^{(-)}(a)
\psi_{\rho_2 \mu_2}^{(-)}(a),
\eea
therefore
\be
\label{EP_constant_MINUS_A}
N_{\rho\mu}^{(-)} =\sqrt{2}\frac{\sqrt{\rho\mu \sinh\pi\rho \sinh\pi\mu}}{R\pi^3} \left|\Gamma\left(\frac34 + i\frac{\rho + \mu}{2}\right)\right|^2 \left|\Gamma\left(\frac34 + i\frac{\rho - \mu}{2}\right)\right|^2.
\ee

To prove the completeness condition (\ref{complete_int_EP}) of the EP basis we use the completeness of the PS  basis\cite{IIA}, the expansion (\ref{EP-SPH-01}) and the property (\ref{00-E_E_mu}). Indeed,
\bea
&&
R^2
\int\limits_{0}^{\infty} d \rho \int\limits_{0}^{\infty} d \mu 
\left[\Psi^{(+)}_{\rho \mu}(a, \theta)\Psi^{(+)\ast}_{\rho \mu}(a', \theta') 
+ \Psi^{(-)}_{\rho \mu}(a, \theta)\Psi^{(-)\ast}_{\rho \mu}(a', \theta')\right]
\nonumber\\[2mm] 
&=&   
R^2\sum_{m, m' = - \infty}^{\infty}
\int\limits_{0}^{\infty} d \rho \int\limits_{0}^{\infty} d \mu 
\Biggl[ {\cal E}^{m(+)}_{\rho \mu } {\cal E}^{m' (+)*}_{\rho \mu }  + 
{\cal E}^{m(-)}_{\rho \mu} {\cal E}^{m'(-)*}_{\rho \mu}\Biggr]
 \Psi^{S}_{\rho m } (\tau, \vphi) \Psi^{S*} _{\rho m' } (\tau', \vphi')
\nonumber\\[2mm] 
&=&
R^2 \sum_{m = - \infty}^{\infty} 
\int\limits_{0}^{\infty} d \rho   \Psi^{S}_{\rho m } (\tau, \vphi) \Psi^{S*} _{\rho m' } (\tau', \vphi')
= (\sinh \tau)^{-1} \delta(\tau-\tau') \delta(\vphi - \vphi').
\eea
Using the relation $\cosh a = e^\tau \cos\theta$, we obtain
\bea
\delta(a-a')  \delta(\theta- \theta') = \frac{\sinh a}{\cos\theta} \frac{|\sin\theta|}{e^{\tau}}\delta(\tau - \tau')  \delta(\cos\theta - \cos\theta').
\eea 
Further simplification leads to the equality 
\bea
\frac{\cosh^2 a\, \cos^2\theta}{\cosh^2 a-\cos^2\theta}
\delta(a-a')  \delta(\theta- \theta') = (\sinh \tau)^{-1} \delta(\tau-\tau') \delta(\vphi - \vphi'),
\eea
and finally to expression (\ref{complete_int_EP}).

\subsection{Orthonormality of HP wave functions}

\paragraph{ Orthogonality relation in discrete spectrum.}
\label{par:norm_discrete_HP}

For further calculation of normalization integrals we use the formula (which is easily deduced from the left Eq. (\ref{SINH1}))
\bea
\label{00-SOLUTION-HP10}
\int\limits_{0}^{\infty} \Biggl\{
\varsigma_1^2 - \varsigma_2^2
+\frac{\rho_2^2 - \rho_1^2}{\sinh^2 b }\Biggr\} 
\psi_{\rho_1 \varsigma_1}(b) \, \psi_{\rho_2 \varsigma_2}^{\ast}(b) d b
=
\left\{ \psi_{\rho_2 \varsigma_2}^{\ast}
\frac{d \psi_{\rho_1 \varsigma_1}}{d b}
-  \psi_{\rho_1 \varsigma_1}\frac{d \psi_{\rho_2 \varsigma_2}^{\ast}}{d b}
\right\}\Biggl|_{0}^{\infty} =: B_\infty - B_0.
\eea
Substituting the asymptotics (\ref{01-SOLUTION-HP6}) of the function $\psi_{\rho s}(b)$ and its derivatives yields $B_\infty = 0$, and
\bea
&B_0(\rho_1, \rho_2) := \lim_{b\to 0} \left( \psi_{\rho_2 \varsigma_2}^{\ast}
 \psi_{\rho_1 \varsigma_1}^\prime -  \psi_{\rho_1 \varsigma_1} \psi_{\rho_2 \varsigma_2}^{\ast\prime} \right) = \nonumber \\
& = \frac{i e^{\pi(\rho_1 + \rho_2)}}{4}\lim_{b\to 0} 
 \left[
 - \frac{\Gamma(i\rho_1) \Gamma(i\rho_2) (\rho_1 - \rho_2)}{\Gamma(1/2 + i\rho_1 +\varsigma_1) \Gamma(1/2 + i\rho_2 + \varsigma_2)} \left(\sinh\frac{b}{2}\right)^{-i(\rho_1 + \rho_2)} + \right.\nonumber\\
& +  \frac{\pi \Gamma(i\rho_2) (\rho_1 + \rho_2)}{\Gamma(i\rho_1) \rho_1\sinh\pi\rho_1 \Gamma(1/2 - i\rho_1 +\varsigma_1) \Gamma(1/2 + i\rho_2 + \varsigma_2)} \left(\sinh\frac{b}{2}\right)^{i(\rho_1 - \rho_2)} - 
 \nonumber \\
&-  \frac{\pi \Gamma(i\rho_1) (\rho_1 + \rho_2)}{\Gamma(i\rho_2) \rho_2\sinh\pi\rho_2 \Gamma(1/2 + i\rho_1 + \varsigma_1) \Gamma(1/2 - i\rho_2 + \varsigma_2)} \left(\sinh\frac{b}{2}\right)^{- i(\rho_1 - \rho_2)} + \nonumber \\
&\left. 
+  \frac{\pi^2 (\rho_1 - \rho_2)}{\Gamma(i\rho_1) \Gamma(i\rho_2) \Gamma(1/2 - i\rho_1 +\varsigma_1) \Gamma(1/2 - i\rho_2 + \varsigma_2)  \rho_1 \rho_2 \sinh\pi\rho_1  \sinh\pi\rho_2} \left(\sinh\frac{b}{2}\right)^{i(\rho_1 + \rho_2)} 
 \right].
 \label{B0_DISCR}
\eea
Let us note, that 
\bea
\label{DB}
& B_0(\rho_1, \rho_1) = \frac{i\pi e^{2\pi\rho_1}}{2\sinh\pi\rho_1}\left[\frac{1}{\Gamma(\frac{1}{2}- i\rho_1 +\varsigma_1) \Gamma(\frac{1}{2} + i\rho_1 + \varsigma_2)} - \frac{1}{\Gamma(\frac{1}{2} + i\rho_1 + \varsigma_1) \Gamma(\frac{1}{2} - i\rho_1 + \varsigma_2)}\right] = \nonumber \\
& = - (\varsigma_1^2 - \varsigma_2^2) \int\limits_{0}^{\infty}
\psi_{\rho_1 \varsigma_1}(b) \, \psi_{\rho_1 \varsigma_2}^{\ast}(b)
\, d b,
\eea
and 
\bea
\label{B0_R1R2}
&\frac{B_0(\rho_1, \rho_2)}{\rho_2^2 - \rho_1^2} = -\frac{\pi^2 e^{2\pi\rho_1} \delta(\rho_1 - \rho_2)}{4\rho_1\sin\pi\rho_1}\left[\frac{1}{\Gamma(\frac{1}{2}- i\rho_1 +\varsigma_1) \Gamma(\frac{1}{2} + i\rho_1 + \varsigma_2)} + 
\right.
\nonumber \\ 
& \left.
 + \frac{1}{\Gamma(\frac{1}{2} + i\rho_1 + \varsigma_1) \Gamma(\frac{1}{2} - i\rho_1 + \varsigma_2)}\right].
\eea
Finally, from (\ref{00-SOLUTION-HP10}) we have
\be
\label{INT_B}
\int\limits_{0}^{\infty} \frac{db}{\sinh^2 b} \psi_{\rho_1 \varsigma_1}(b) \psi_{\rho_2 \varsigma_2}^{\ast}(b)  
 = - \frac{B_0(\rho_1, \rho_2)}{\rho_2^2 - \rho_1^2} - \frac{\varsigma_1^2 - \varsigma_2^2}{\rho_2^2 - \rho_1^2} \int\limits_{0}^{\infty} \psi_{\rho_1 \varsigma_1}(b)  \psi_{\rho_2 \varsigma_2}^{\ast}(b)\,d b.
\ee

From the right Eq. (\ref{SINH1}) we have
\bea
\label{101-SOLUTION-HP10_0}
\int\limits_{0}^{\pi}\, \left( \varsigma_2^2 - \varsigma_1^2 + \frac{\rho_2^2 - \rho_1^2}{\sin^2\theta} \right)
\psi_{\rho_1 \varsigma_1}(\theta)  \psi_{\rho_2 \varsigma_2}^{\ast}(\theta)
d \theta
=
\left\{\psi_{\rho_2 \varsigma_2}^{\ast}
\frac{d \psi_{\rho_1 \varsigma_1}}{d \theta}
-  \psi_{\rho_1 \varsigma_1}\frac{d \psi_{\rho_2 \varsigma_2}^{\ast}}{d \theta}
\right\}\Biggl|_{0}^{\pi} =: \Theta_\pi - \Theta_0,
\eea
where
\bea
&\Theta_\pi(\rho_1, \rho_2) := \lim_{\theta\to\pi} \left( \psi_{\rho_2 \varsigma_2}^{\ast}
 \psi_{\rho_1 \varsigma_1}^\prime -  \psi_{\rho_1 \varsigma_1} \psi_{\rho_2 \varsigma_2}^{\ast\prime} \right) =  \nonumber \\
 &= - \frac{i\pi (\rho_1 + \rho_2) }{4\rho_1 \sinh\pi\rho_1} \frac{\Gamma(i\rho_2)}{\Gamma(i\rho_1)} \frac{\sin\pi \varsigma_1 \sin\pi \varsigma_2}{\Gamma_{-\rho_1, \varsigma_1} \Gamma_{\rho_2, \varsigma_2}} \lim_{\theta \to \pi} \left(\cos\frac{\theta}{2}\right)^{i(\rho_1 - \rho_2)} + \nonumber \\
 &+ \frac{i\pi (\rho_1 + \rho_2) }{4\rho_2 \sinh\pi\rho_2} \frac{\Gamma(i\rho_1)}{\Gamma(i\rho_2)} \frac{\sin\pi(\varsigma_1 - i\rho_1) \sin\pi(\varsigma_2 + i\rho_2)}{\Gamma_{\rho_1, \varsigma_1} \Gamma_{ - \rho_2, \varsigma_2}} \lim_{\theta \to \pi} \left(\cos\frac{\theta}{2}\right)^{- i(\rho_1 - \rho_2)} + \nonumber \\
 &+ \frac{i(\rho_1 - \rho_2) \Gamma(i\rho_1) \Gamma(i\rho_2)}{4\Gamma_{\rho_1, \varsigma_1} \Gamma_{\rho_2, \varsigma_2}}\sin\pi(\varsigma_1 - i\rho_1) \sin\pi \varsigma_2 \lim_{\theta \to \pi} \left(\cos\frac{\theta}{2}\right)^{- i(\rho_1 + \rho_2)} -
 \label{THETA_PI}\\
 &- \frac{i\pi^2(\rho_1 - \rho_2) \sin\pi \varsigma_1 \sin\pi(\varsigma_2 + i\rho_2)}{4\rho_1 \rho_2\sinh\pi\rho_1 \sinh\pi\rho_2 \Gamma(i\rho_1)\Gamma(i\rho_2)\Gamma_{- \rho_1, \varsigma_1} \Gamma_{- \rho_2, \varsigma_2}} \lim_{\theta \to \pi} \left(\cos\frac{\theta}{2}\right)^{i(\rho_1 + \rho_2)}, \nonumber
 \eea
 here we denote  $ \Gamma_{\pm\rho, \varsigma} := \Gamma\left(\frac{1}{2} \pm i\rho + \varsigma\right)$, and 
  \bea
 &\Theta_0(\rho_1, \rho_2) := \lim_{\theta\to 0} \left( \psi_{\rho_2 \varsigma_2}^{\ast}
 \psi_{\rho_1 \varsigma_1}^\prime -  \psi_{\rho_1 \varsigma_1} \psi_{\rho_2 \varsigma_2}^{\ast\prime} \right) = 
 \nonumber \\
 &=  \frac{i\pi (\rho_1 + \rho_2) }{4\rho_1 \sinh\pi\rho_1} \frac{\Gamma(i\rho_2)}{\Gamma(i\rho_1)} \frac{\cosh\pi\rho_1 \cosh\pi\rho_2}{\Gamma_{-\rho_1, \varsigma_1} \Gamma_{\rho_2, \varsigma_2}} \lim_{\theta \to 0} \left(\sin\frac{\theta}{2}\right)^{i(\rho_1 - \rho_2)} - \nonumber \\
 &- \frac{i\pi (\rho_1 + \rho_2) }{4\rho_2 \sinh\pi\rho_2} \frac{\Gamma(i\rho_1)}{\Gamma(i\rho_2)} \frac{1}{\Gamma_{\rho_1, \varsigma_1} \Gamma_{ - \rho_2, \varsigma_2}} \lim_{\theta \to 0} \left(\sin\frac{\theta}{2}\right)^{- i(\rho_1 - \rho_2)} - \nonumber \\
 &- \frac{i(\rho_1 - \rho_2) \Gamma(i\rho_1) \Gamma(i\rho_2)}{4\Gamma_{\rho_1, \varsigma_1} \Gamma_{\rho_2, \varsigma_2}}\cosh\pi\rho_2 \lim_{\theta \to 0} \left(\sin\frac{\theta}{2}\right)^{- i(\rho_1 + \rho_2)} +
 \label{THETA_0} \\
 &+ \frac{i\pi^2(\rho_1 - \rho_2) \cosh\pi\rho_1}{4\rho_1 \rho_2\sinh\pi\rho_1 \sinh\pi\rho_2 \Gamma(i\rho_1)\Gamma(i\rho_2)\Gamma_{- \rho_1, \varsigma_1} \Gamma_{- \rho_2, \varsigma_2}} \lim_{\theta \to 0} \left(\sin\frac{\theta}{2}\right)^{i(\rho_1 + \rho_2)}, \nonumber
 \eea
with properties:
\bea
\label{T_PI_T_0}
\Theta_\pi(\rho_1, \rho_1) - \Theta_0(\rho_1, \rho_1) = \frac{i\pi}{2\sinh\pi\rho_1}
\left\{
\frac{1 + \sin\pi(\varsigma_1 - i\rho_1)\sin\pi(\varsigma_2 + i\rho_1)}{\Gamma\left(\frac12 + \varsigma_1 + i\rho_1\right)\Gamma\left(\frac12 + \varsigma_2 - i\rho_1\right)} 
- \right. \nonumber \\[2mm]
\left.
- 
 \frac{\cosh^2\pi\rho_1 + \sin\pi \varsigma_1 \sin\pi \varsigma_2}{\Gamma\left(\frac12 + \varsigma_1 - i\rho_1\right)\Gamma\left(\frac12 + \varsigma_2 + i\rho_1\right)} 
\right\} =\left( \varsigma_2^2 - \varsigma_1^2 \right) \int\limits_{0}^{\pi}
\psi_{\rho_1 \varsigma_1}(\theta) \, \psi_{\rho_1 \varsigma_2}^{\ast}(\theta)
\, d \theta,
\eea
and
\bea
\label{DELTA_THETA}
\frac{\Theta_\pi - \Theta_0}{\rho_2^2 - \rho_1^2} = \frac{\pi^2 \delta(\rho_1 - \rho_2)}{4\rho_1\sinh\pi\rho_1}
\left\{
\frac{1 + \sin\pi(\varsigma_1 - i\rho_1)\sin\pi(\varsigma_2 + i\rho_1)}{\Gamma\left(\frac12 + \varsigma_1 + i\rho_1\right)\Gamma\left(\frac12 + \varsigma_2 - i\rho_1\right)} 
+ \right. \nonumber \\[2mm]
\left.
+ 
 \frac{\cosh^2\pi\rho_1 + \sin\pi \varsigma_1 \sin\pi \varsigma_2}{\Gamma\left(\frac12 + \varsigma_1 - i\rho_1\right)\Gamma\left(\frac12 + \varsigma_2 + i\rho_1\right)} 
\right\}.
\eea

From (\ref{101-SOLUTION-HP10_0}) we obtain
\be
\label{INT_THETA}
\int\limits_{0}^{\pi}\ \frac{d \theta}{\sin^2\theta}
\psi_{\rho_1 \varsigma_1}(\theta)  \psi_{\rho_2 \varsigma_2}^{\ast}(\theta) = \frac{\Theta_\pi - \Theta_0}{\rho_2^2 - \rho_1^2} - \frac{\varsigma_2^2 - \varsigma_1^2}{\rho_2^2 - \rho_1^2} \int\limits_{0}^{\pi} \psi_{\rho_1 \varsigma_1}(\theta)  \psi_{\rho_2 \varsigma_2}^{\ast}(\theta) d\theta.
\ee
Let us multiply (\ref{INT_B})  by $\int\limits_{0}^{\pi} \psi_{\rho_1 \varsigma_1}(\theta)\psi_{\rho_2 \varsigma_2}^{\ast}(\theta) d\theta$ and sum to (\ref{INT_THETA}), multiplied by $\int\limits_{0}^{\infty} \psi_{\rho_1 \varsigma_1}(b) \psi_{\rho_2 \varsigma_2}^{\ast}(b)\,d b$. We obtain
\bea
\int\limits_{0}^{\infty} 
db \int\limits_{0}^{\pi}\ d \theta \left( \frac{1}{\sin^2\theta} + \frac{1}{\sinh^2 b} \right) \psi_{\rho_1 \varsigma_1}(b) \psi_{\rho_1 \varsigma_1}(\theta) \, \psi_{\rho_2 \varsigma_2}^{\ast}(b) \psi_{\rho_2 \varsigma_2}^{\ast}(\theta) =
\nonumber
\eea
\bea
\label{BTHETA}
= - \frac{B_0(\rho_1, \rho_2)}{\rho_2^2 - \rho_1^2} \int\limits_{0}^{\pi} \psi_{\rho_1 \varsigma_1}(\theta)  \psi_{\rho_2 \varsigma_2}^{\ast}(\theta) d\theta +
 \frac{\Theta_\pi - \Theta_0}{\rho_2^2 - \rho_1^2} \int\limits_{0}^{\infty} \psi_{\rho_1 \varsigma_1}(b) \psi_{\rho_2 \varsigma_2}^{\ast}(b) db.  
\eea
Taking into account (\ref{B0_R1R2}), (\ref{T_PI_T_0}) and  (\ref{DELTA_THETA}), (\ref{DB}) we get 
\bea
\label{psi1_psi2}
\int\limits_{0}^{\infty} 
db \int\limits_{0}^{\pi}\ d \theta \left( \frac{1}{\sin^2\theta} + \frac{1}{\sinh^2 b} \right) \psi_{\rho_1 \varsigma_1}(b) \psi_{\rho_1 \varsigma_1}(\theta)  \psi_{\rho_2 \varsigma_2}^{\ast}(b) \psi_{\rho_2 \varsigma_2}^{\ast}(\theta) 
\nonumber \\
= 
\frac{i\pi^3 e^{2\pi\rho_1} \delta(\rho_1 - \rho_2)}{4\rho_1\sinh\pi\rho_1 \left|\Gamma\left(\frac12 + \varsigma_1 + i\rho_1\right)\right|^2 
\left|\Gamma\left(\frac12 + \varsigma_2 + i\rho_1\right)\right|^2}
\nonumber \\ 
\times \frac{\sinh\pi\rho_1\left[\cos\pi(\varsigma_2 - \varsigma_1) - 1\right] - i\cosh\pi\rho_1\sin\pi(\varsigma_2 - \varsigma_1)}{\varsigma_2^2 - \varsigma_1^2}.
\eea
If $\varsigma_2 \ne \varsigma_1$, to get the orthogonality of functions  $\psi_{\rho \varsigma}(b) \psi_{\rho \varsigma}(\theta)$ we choose $\varsigma_2 - \varsigma_1 = 2m$, $m\in\mathbb{Z}\setminus\left\{0\right\}$. In the limit $\varsigma_2 \to \varsigma_1$, we obtain
\be
\label{NN}
 \frac{\sinh\pi\rho_1\left[\cos\pi(\varsigma_2 - \varsigma_1) - 1\right] - i\cosh\pi\rho_1\sin\pi(\varsigma_2 - \varsigma_1)}{\varsigma_2^2 - \varsigma_1^2}  \to \frac{- i\pi}{2 \varsigma_1} \cosh\pi\rho_1.
\ee
The values of $m$ are restricted due to inequality $\varsigma > 0$. Therefore, we can associate the discrete spectrum of $\varsigma$ with the natural quantum number $n$:
\be
\varsigma_0 < \varsigma_0 + 2 < ... < \varsigma_0 + 2n < ... 
\ee
starting from the minimum positive fixed value $\varsigma_0 \in (0, 2]$, i.e. $\varsigma_n \in \left\{\varsigma_ 0 + 2n\right\}_{n = 0}^{\infty}$. Thus, the orthogonal basis is formed by the wave functions $\Psi_{\rho \varsigma_n}(b, \theta)$.

Coming back to (\ref{Norm_HP}), using (\ref{psi1_psi2}) and  (\ref{NN}), we have for the discrete spectrum $\varsigma_n$
\bea
\label{00-Norm_HP}
&&R^2 \left|N_{\rho \varsigma_n} \right|^2 \int\limits_{0}^\infty db \int\limits_0^\pi \psi_{\rho \varsigma_n} (b) \psi_{\rho' \varsigma_n'}^\ast
(b) \psi_{\rho \varsigma_n} (\theta) \psi_{\rho' \varsigma_n'}^\ast (\theta) \left( \frac{1}{\sin^2\theta}
+ \frac{1}{\sinh^2 b}\right)  d\theta = \nonumber
\\[2mm]
&=& R^2 \left|N_{\rho \varsigma_n} \right|^2 \frac{\pi^4 e^{2 \pi\rho}}{8 \rho \varsigma_n \tanh\pi\rho}
\frac{\delta(\rho - \rho')}{\left|\Gamma(\frac12 + \varsigma_n + i\rho)\right|^4} \delta_{\varsigma_n, \varsigma_n'}.
\eea
Therefore,  the orthonormal solution has the form
\bea
\label{01-SOLUTION-00-HP6}
\Psi_{\rho \varsigma_n}^{\rm HP} (b, \theta) =   N_{ \rho \varsigma_n}^{\rm d} \sqrt{\sinh b \sin\theta}
Q_{-\frac12 + \varsigma_n}^{-i\rho}(\cosh b) \, {\rm Q}_{-\frac12 + \varsigma_n}^{-i\rho}(\cos\theta),
\eea
where
\be
\label{N_rho_s_D}
 N_{ \rho \varsigma_n}^{\rm d} := 2 \frac{\sqrt{2\rho \varsigma_n \tanh\pi\rho}}{e^{\pi\rho} \pi^2 R} \frac{\Gamma\left(\frac12 + \varsigma_n +  i\rho\right)}{\Gamma\left(\frac12 + \varsigma_n -  i\rho\right)}.
\ee

\paragraph{ Orthogonality relation in continuous spectrum.}
\label{par:norm_continuous_HP} 
The corresponding integral (\ref{00-SOLUTION-HP10}) can be taken in the form
\bea
\label{02-SOLUTION-HP10}
\int\limits_{0}^{\infty}
\psi_{\rho_1 \varsigma_1}(b)  \psi_{\rho_2 \varsigma_2}^{\ast}(b)
\, d b
= \left. A^{\varsigma_1 \varsigma_2}_{\rho_1 \rho_2} \right|_{0}^{\infty} - \frac{\rho_2^2 - \rho_1^2}{\varsigma_2^2 - \varsigma_1^2} \int\limits_{0}^{\infty} \frac{d b}{\sinh^2 b} 
\psi_{\rho_1 \varsigma_1}(b) \, \psi_{\rho_2 \varsigma_2}^{\ast}(b),
\eea
where we denote
\be
A^{\varsigma_1 \varsigma_2}_{\rho_1 \rho_2} := \frac{1}{\varsigma_2^2 - \varsigma_1^2}
\left\{\psi_{\rho_2 \varsigma_2}^{\ast}
\frac{d \psi_{\rho_1 \varsigma_1}}{d b}
-  \psi_{\rho_1 \varsigma_1}\frac{d \psi_{\rho_2 \varsigma_2}^{\ast}}{d b}
\right\}.
\ee
Substituting (\ref{11-SOLUTION-HP6}), (\ref{12-SOLUTION-HP6}) and its derivatives into the above relation yields 
\be
 \lim_{b\to\infty} A^{\varsigma_1 \varsigma_2}_{\rho_1 \rho_2} = - \frac{i\pi}{2} \frac{e^{\pi(\rho_1 + \rho_2)}}{\varsigma_1 \varsigma_2 \Gamma(i \varsigma_1)\Gamma(-i \varsigma_2)}\lim_{b\to\infty} \frac{e^{- i(\varsigma_1 - \varsigma_2)b}}{\varsigma_2 - \varsigma_1},
\ee
and $\lim\limits_{b\to 0} A^{\varsigma_1 \varsigma_2}_{\rho_1 \rho_2} = 0$, because $e^{i(\rho_1 - \rho_2)\ln\left(\sinh b/2 \right)} \sim i\pi(\rho_2 - \rho_1) \delta(\rho_1 - \rho_2) \sim 0$ as $b\sim 0$.

Finally, we obtain
\bea
\left. A^{\varsigma_1 \varsigma_2}_{\rho_1 \rho_2} \right|_{0}^{\infty}  = \frac{\pi^2 e^{\pi(\rho_1 + \rho_2)} \sinh\pi \varsigma_1}{2\varsigma_1} 
\delta(\varsigma_1-\varsigma_2),
\eea
where we use the relation $e^{- i(\varsigma_1 - \varsigma_2)b}/(\varsigma_2 - \varsigma_1) \sim i\pi\delta(\varsigma_1 - \varsigma_2)$ as $b\to\infty$.

From the right Eq. in (\ref{SINH1}) (with the change $s \to is$) we get
\bea
\label{00-SOLUTION-HP111}
\int\limits_{0}^{\pi} \frac{d\theta}{\sin^2\theta}
\psi_{\rho_1 \varsigma_1}(\theta)  \psi_{\rho_2 \varsigma_2}^{\ast}(\theta)
= \left. B^{\varsigma_1 \varsigma_2}_{\rho_1\rho_2} \right|_{0}^{\pi} - \frac{\varsigma_1^2 - \varsigma_2^2}{\rho_2^2 - \rho_1^2} \int\limits_{0}^{\pi}
\psi_{\rho_1 \varsigma_1}(\theta)  \psi_{\rho_2 \varsigma_2}^{\ast}(\theta) d\theta,
\eea
where we introduce
\bea
B^{\varsigma_1 \varsigma_2}_{\rho_1\rho_2} := \frac{1}{\rho_2^2 - \rho_1^2}
\left\{ \psi_{\rho_2 \varsigma_2}^{\ast}
\frac{d \psi_{\rho_1 \varsigma_1}}{d \theta}
-  \psi_{\rho_1 \varsigma_1}\frac{d \psi_{\rho_2 \varsigma_2}^{\ast}}{d \theta}
\right\}.
\label{BBB}
\eea
Substituting the relations (\ref{SOLUTION-02-HP6_C}), (\ref{SOLUTION-03-HP6_C}) and its derivatives into (\ref{BBB}) yields 
\bea
\lim\limits_{\theta\to\pi} B^{\varsigma_1 \varsigma_2}_{\rho_1\rho_2}  = \frac{\pi^2 \delta(\rho_1 - \rho_2)}{4\rho_1 \sinh\pi\rho_1} \left\{ \frac{\sinh\pi(\rho_1 - \varsigma_1)\sinh\pi(\rho_1 - \varsigma_2)}{\Gamma\left(\frac12 + i\rho_1 + i \varsigma_1\right)\Gamma\left(\frac12 - i\rho_1 - i\varsigma_2\right)} + \right. \nonumber\\
\left. 
+ \frac{\sinh\pi \varsigma_1 \sinh\pi \varsigma_2}{\Gamma\left(\frac12 - i\rho_ 1 + i \varsigma_1\right)\Gamma\left(\frac12 + i\rho_ 1  - i\varsigma_2\right)} 
 \right\},
\eea
and
\bea
\lim\limits_{\theta\to 0} B^{\varsigma_1 \varsigma_2}_{\rho_1\rho_2} 
= - \frac{\pi^2 \delta(\rho_1 - \rho_2)}{4\rho_1 \sinh\pi\rho_1} \left\{ \frac{1}{\Gamma\left(\frac12 + i\rho_1 + i \varsigma_1\right)\Gamma\left(\frac12 - i\rho_1 - i\varsigma_2\right)} + \right. \nonumber\\
\left. 
+ \frac{\cosh^2\pi\rho_1}{\Gamma\left(\frac12 - i\rho_ 1 + i \varsigma_1\right)\Gamma\left(\frac12 + i\rho_ 1  - i\varsigma_2\right)} 
 \right\}.
\eea

Finally, we have
\bea
\label{101_2-SOLUTION-HP10}
\left. B^{\varsigma_1 \varsigma_2}_{\rho_1\rho_2} \right|_{0}^{\pi} &=& \frac{\pi^2 \delta(\rho_1 - \rho_2)}{4 \rho_1 \sinh\pi\rho_1}
\left\{
\frac{\sinh\pi(\rho_1 - \varsigma_1)\sinh\pi(\rho_1 - \varsigma_2) + 1}{\Gamma\left(\frac12 + i\rho_1 + i \varsigma_1\right)\Gamma\left(\frac12 - i\rho_1 - i\varsigma_2\right)} 
+ \right. \nonumber \\[2mm]
&&
\left.
+ 
 \frac{\sinh\pi \varsigma_1 \sinh\pi \varsigma_2 + \cosh^2\pi\rho_1}{\Gamma\left(\frac12 - i\rho_ 1 + i \varsigma_1\right)\Gamma\left(\frac12 + i\rho_ 1  - i\varsigma_2\right)} 
\right\}.
\eea
Multiplying (\ref{02-SOLUTION-HP10}) by (\ref{00-SOLUTION-HP111}) yields
\bea
&&\int\limits_{0}^{\infty} d b\int\limits_{0}^{\pi} \frac{d\theta}{\sin^2\theta} \psi_{\rho_1 \varsigma_1}(b) \psi_{\rho_1 \varsigma_1}(\theta) \psi_{\rho_2 \varsigma_2}^{\ast}(b) \psi_{\rho_2 \varsigma_2}^{\ast}(\theta) = \nonumber \\
&&= \frac{\pi^2 e^{2\pi\rho_1}}{4\rho_1 \varsigma_1} \frac{\sinh \pi \varsigma_1}{\sinh\pi\rho_1} \left(\sinh^2\pi\rho_1 + \cosh^2\pi \varsigma_1\right)\cosh\pi(\rho_1 - \varsigma_1)\, \delta(\rho_1 - \rho_2)\delta(\varsigma_1 - \varsigma_2) - \nonumber \\
&& - \int\limits_{0}^{\infty} \frac{d b}{\sinh^2 b}\int\limits_{0}^{\pi} d\theta \psi_{\rho_1 \varsigma_1}(b) \psi_{\rho_1 \varsigma_1}(\theta) \psi_{\rho_2 \varsigma_2}^{\ast}(b) \psi_{\rho_2 \varsigma_2}^{\ast}(\theta).
\eea
Comparison with (\ref{Norm_HP}) gives 
\bea
\label{02-SOLUTION-00-HP6_A}
\Psi_{\rho \varsigma}^{\rm HP} (b, \theta) =  N_{ \rho \varsigma}^{\rm c} \sqrt{\sinh b \sin\theta}  Q_{-\frac12 + i \varsigma}^{-i\rho}(\cosh b) 
{\rm Q}_{-\frac12 + i \varsigma}^{-i\rho}(\cos\theta),
\eea
with
\be
N_{\rho \varsigma}^{\rm c} := \frac{2}{\pi R e^{\pi\rho} \Gamma^2\left(\frac12 - i \rho+ i\varsigma\right) } \sqrt{\frac{\rho \varsigma \sinh\pi\rho}{\sinh\pi \varsigma (\sinh^2 \pi\rho + \cosh^2 \pi \varsigma) \cosh\pi(\rho - \varsigma)}}.
\label{N_rho_s_c}
\ee

\subsection{Calculation of integrals  ${\cal K}^{s(1,2)}_{\rho A}$}
\label{sec:APP_1}

We introduce a new integration variable $z := - \nu$ and the parameter $t := |A/s|/4$, then the integrals (\ref{SCP-HOR-08}) will be
\bea
\label{K-1}
{\cal K}^{s(1,2)}_{\rho A} = \frac{1}{4\pi} \,
\frac{1}{\sqrt{|A| |s|}} 
\left[ I^{(+)}(t) \mp \frac{i s}{|s|} I^{(-)}(t) \right],
\eea
where we denote
\bea
\label{K-2}
I^{(+)}(t)  := \int\limits_{-\infty}^{\infty} \frac{\Gamma(3/4 + iz/2)}{\Gamma(3/4 - iz/2)} t^{- iz}
 d z, \qquad
I^{(-)}(t)  := \int\limits_{-\infty}^{\infty} \frac{\Gamma(1/4 + iz/2)}{\Gamma(1/4 - iz/2)} t^{- iz}
 d z.
\eea
Let us consider 
\bea
f^{(+)}_\epsilon(z) := \frac{\Gamma(3/4 + iz/2)\Gamma(1 - \epsilon iz/2)}{\Gamma(3/4 - iz/2)\Gamma(3 + \epsilon iz/2)} t^{- iz}, \nonumber\\
f^{(-)}_\epsilon(z) := \frac{\Gamma(1/4 + iz/2)\Gamma(1 - \epsilon iz/2)}{\Gamma(1/4 - iz/2)\Gamma(3 + \epsilon iz/2)} t^{- iz},
\eea
then 
\be
I^{(+)}(t)  = 2 \lim\limits_{\epsilon\to 0^+}  
\int\limits_{-\infty}^{\infty} f^{(+)}_\epsilon(z) dz, \quad 
I^{(-)}(t)  = 2 \lim\limits_{\epsilon\to 0^+}  \int\limits_{-\infty}^{\infty} f^{(-)}_\epsilon(z) dz.
\label{II}
\ee
For functions $ f^{(\pm)}_\epsilon(z)$ we have the asymptotics $| f^{(\pm)}_\epsilon(z)| \sim |z|^{-2}$ 
for large $|z|$, therefore is valid to apply the Residue theorem
\bea
\label{Res_theorem2}
\int\limits_{-\infty}^{\infty} f^{(\pm)}_\epsilon (z) dz = 2\pi i \sum\limits_{n = 0}^N \Res\left[f^{(\pm)}_\epsilon (z) ,z^{(\pm)}_n\right],
\eea
where $z^{(\pm)}_n$ are the poles of $f^{(\pm)}_\epsilon (z)$ in the upper complex half-plane. Note that functions $f^{(\pm)}_\epsilon (z)$ in this half-plane have only poles of the form $z^{(+)}_n = i(3/2 + 2n)$, $z^{(-)}_n = i(1/2 + 2n)$, $n = 0, 1, \dots$ and $\infty$ is a zero.

To calculate the residues, we represent $f^{(\pm)}_\epsilon$ as follows: $f^{(\pm)}_\epsilon(z) = {g^{(\pm)}(z)}/{h^{(\pm)}(z)}$,
\bea
 g^{(+)} :=   \frac{\Gamma(1 - \epsilon iz/2)\ t^{- iz}}{\Gamma(3/4 - iz/2)\Gamma(3 + \epsilon iz/2)}, \quad
h^{(+)} := \frac{1}{\Gamma(3/4 + {iz}/2)}, \\
g^{(-)} :=   \frac{\Gamma(1 - \epsilon iz/2)\ t^{- iz}}{\Gamma(1/4 - iz/2)\Gamma(3 + \epsilon iz/2)}, \quad
h^{(-)} := \frac{1}{\Gamma(1/4 + {iz}/2)}, 
\eea
then $\Res\left[f^{(\pm)}_\epsilon(z),z^{(\pm)}_n\right] = \left.g^{(\pm)}/h^{(\pm)\prime}\right|_{z = z^{(\pm)}_n}$.  Taking into account that
\bea
h^{(+)\prime} = \frac{d}{dz} \frac{1}{\Gamma(3/4 + {iz}/2)} = - \frac{i}{2} \frac{\psi\left(\frac34 + \frac{iz}{2}\right)}{\Gamma\left(\frac34 + \frac{iz}{2}\right)} \to \frac{i}{2}(-1)^n n!, \quad z \to z^{(+)}_n,\\
h^{(-)\prime} = \frac{d}{dz} \frac{1}{\Gamma(1/4 + {iz}/2)} = - \frac{i}{2} \frac{\psi\left(\frac14 + \frac{iz}{2}\right)}{\Gamma\left(\frac14 + \frac{iz}{2}\right)} \to \frac{i}{2}(-1)^n n!, \quad z \to z^{(-)}_n,
\eea
and
\bea
 \left.g^{(+)}\right|_{z = z^{(+)}_n} =  \frac{\Gamma\left(1 + \frac34\epsilon +  n\epsilon\right) t^{\frac32 + 2n}}{\Gamma(3/2 + n)\Gamma\left(3 - \frac34\epsilon -  n\epsilon\right)},
  \left.g^{(-)}\right|_{z = z^{(-)}_n} =  \frac{\Gamma\left(1 + \frac14\epsilon +  n\epsilon\right) t^{\frac12 + 2n}}{\Gamma(1/2 + n)\Gamma\left(3 - \frac14\epsilon -  n\epsilon\right)},
\eea
we obtain
\bea
\Res\left[f^{(+)}_\epsilon(z),z^{(+)}_n\right] = \frac{2}{i} \frac{(-1)^n}{n!} \frac{\Gamma\left(1 + \frac34\epsilon +  n\epsilon\right) t^{\frac32 + 2n}}{\Gamma(3/2 + n)\Gamma\left(3 - \frac34\epsilon -  n\epsilon\right)}, \\
\Res\left[f^{(-)}_\epsilon(z),z^{(-)}_n\right] = \frac{2}{i} \frac{(-1)^n}{n!} \frac{\Gamma\left(1 + \frac14\epsilon +  n\epsilon\right) t^{\frac12 + 2n}}{\Gamma(1/2 + n)\Gamma\left(3 - \frac14\epsilon -  n\epsilon\right)}.
\eea
Returning to (\ref{II}) and (\ref{Res_theorem2}),  we get
\bea
\label{K-3}
I^{(+)}(t) 
&=&  
2 \lim\limits_{\epsilon\to0^+}\lim\limits_{N\to\infty} 2\pi i \sum\limits_{n = 0}^N  \frac{2}{i} \frac{(-1)^n}{n!} \frac{\Gamma\left(1 + \frac34\epsilon +  n\epsilon\right)\ t^{\frac32 + 2n}}{\Gamma(3/2 + n)\Gamma\left(3 - \frac34\epsilon -  n\epsilon\right)} 
\nonumber\\[2mm]
&=& 
4\pi t \sum\limits_{n = 0}^\infty \frac{ (-1)^n t^{2n + \frac12}}{n! \Gamma(n + 3/2)} 
=
\pi \left|\frac{A}{s}\right|
J_{\frac12}\left(\frac{|A/s|}{2}\right)
= 2\sqrt{\pi |A/s|} \sin\frac{|A/s|}{2},
\\[2mm]
\label{K-4}
I^{(-)}(t) 
&=& 4\pi t \sum\limits_{n = 0}^\infty \frac{ (-1)^n t^{2n - \frac12}}{n! \Gamma(n + 1/2)} 
=  \pi \left|\frac{A}{s}\right| J_{-\frac12}\left(\frac{|A/s|}{2}\right) 
=  2\sqrt{\pi |A/s|} \cos\frac{|A/s|}{2}.
\eea
Thus, using equations  (\ref{K-3}), (\ref{K-4}) and (\ref{K-1}) we come to the formula  for the coefficients  of  interbasis expansions (\ref{SCP-HOR-088}).


\bibliography{bib_Georgy}

\end{document}